\documentclass[fleqn,usenatbib]{mnras}

\usepackage{mathptmx}

\usepackage[T1]{fontenc}

\DeclareRobustCommand{\VAN}[3]{#2}
\let\VANthebibliography\thebibliography
\def\thebibliography{\DeclareRobustCommand{\VAN}[3]{##3}\VANthebibliography}

\usepackage{graphicx}	
\usepackage{amssymb}	
\usepackage{amsmath}	
\usepackage{mathrsfs}

\newcommand{\Lya}{Ly$\alpha$~}

\newcommand{\LCDM}{$\Lambda$CDM~}

\newcommand{\HI}{\hbox{H$\,\rm \scriptstyle I\ $}}
 
\newcommand{\HeI}{\hbox{He$\,\rm \scriptstyle I\ $}}
\newcommand{\HeII}{\hbox{He$\,\rm \scriptstyle II\ $}}

\newcommand{\CIV}{\hbox{C$\,\rm \scriptstyle IV\ $}}

\title[Non-canonical heating and turbulence]{Limits on non-canonical
  heating and turbulence in the intergalactic medium from the low
  redshift Lyman-$\alpha$ forest}

\author[J.S. Bolton et al.] {James S. Bolton$^{1}$\thanks{E-mail: james.bolton@nottingham.ac.uk}, Prakash Gaikwad$^{2,3}$, Martin G. Haehnelt$^{2}$, Tae-Sun Kim$^{4}$, Fahad Nasir$^{5}$, \newauthor Ewald Puchwein$^{6}$, Matteo Viel$^{7,8,9,10}$ \& Bart P. Wakker$^{4}$
  \\$^1$School of Physics and Astronomy, University of Nottingham, University Park, Nottingham, NG7 2RD, UK
  \\$^{2}$Kavli Institute for Cosmology and Institute of Astronomy, Madingley Road, Cambridge, CB3 0HA, UK
 \\$^{3}$Max-Planck-Institut f{\"u}r Astronomie, K{\"o}nigstuhl 17, D-69117 Heidelberg, Germany
 \\$^{4}$Department of Astronomy, University of Wisconsin-Madison, 475 N. Charter St., Madison, WI 53706, USA
 \\$^{5}$Department of Physics and Astronomy, University of California, Riverside, CA 92521, USA
  \\$^{6}$Leibniz-Institut f\"ur Astrophysik Potsdam, An der Sternwarte 16, 14482 Potsdam, Germany
  \\$^{7}$SISSA - International School for Advanced Studies, Via Bonomea 265, I-34136 Trieste, Italy
  \\$^{8}$IFPU, Institute for Fundamental Physics of the Universe, Via Beirut 2, I-34151 Trieste, Italy
  \\$^{9}$INAF - Osservatorio Astronomico di Trieste, Via G.B. Tiepolo 11, I-34131 Trieste, Italy 
  \\$^{10}$INFN - National Institute for Nuclear Physics, Via Valerio 2, I-34127 Trieste, Italy}

\pubyear{2022}

\begin{document}
\label{firstpage}
\pagerange{\pageref{firstpage}--\pageref{lastpage}}
\maketitle

\begin{abstract}
\noindent  
We examine the column density distribution function (CDDF) and Doppler
parameter distribution from hydrodynamical simulations and
\emph{Cosmic Origins Spectrograph} (COS) observations of the \Lya
forest at redshift $0 \leq z\leq 0.2$.  Allowing for a factor of two
uncertainty in the metagalactic \HI photoionisation rate, our
hydrodynamical simulations are in good agreement ($1$--$1.5\sigma$)
with the shape and amplitude of the observed CDDF at \HI column
densities $10^{13.3}\rm\,cm^{-2}\leq N_{\rm HI}\leq
10^{14.5}\rm\,cm^{-2}$.  However, the Doppler widths of the simulated
lines remain too narrow with respect to the COS data.  We argue that
invoking AGN feedback does not resolve this discrepancy.  We also
disfavour enhanced photoheating rates as a potential solution, as this
requires an unphysically hard UV background spectrum.  If instead
appealing to a non-canonical source of heating, an additional specific
heat injection of $u \lesssim 6.9\rm\,eV\,m_{\rm p}^{-1}$ is required
at $z\lesssim 2.5$ for gas that has $N_{\rm HI}\simeq
10^{13.5}\rm\,cm^{-2}$ by $z=0.1$.  Alternatively, there may be an
unresolved line of sight turbulent velocity component of $v_{\rm
  turb}\lesssim 8.5\rm\,km\,s^{-1}(N_{\rm
  HI}/10^{13.5}\rm\,cm^{-2})^{0.21}$ for the coldest gas in the
diffuse IGM.
\end{abstract}

\begin{keywords}
methods: numerical - intergalactic medium - quasars: absorption lines 
\end{keywords}


\section{Introduction}

Hydrodynamical simulations of the \Lya forest in a \LCDM universe are
in broad agreement with a range of observational data at redshift $2.5 \lesssim z
\lesssim 4.5$
\citep{Hernquist1996,Lukic2015,Bolton2017,Rossi2020,Villasenor2021}.
Allowing for the approximately factor of two uncertainty in the
amplitude of the metagalactic UV background
\citep[e.g.][]{Bolton2005,BeckerBolton2013}, quantities such as the
\HI column density distribution \citep{Altay2011,Rahmati2013}, the
Doppler widths of the \Lya absorption lines
\citep{Schaye2000,Hiss2018}, and the distribution and power spectrum
of the transmitted flux \citep{Rorai2017_pdf,Walther2019} can be
readily reproduced.  While astrophysical processes such as feedback
\citep{Theuns2002_feedback,Viel2013,Chabanier2020} and spatial
fluctuations in the UV background
\citep{Greig2015,UptonSanderbeck2020,Molaro2022} can further modify
this picture, these effects are typically of secondary importance to
the assumed cosmological model and the \emph{average} photoionisation
and photoheating rates \citep[see][for a review]{McQuinn2016_review}.

Toward lower redshifts, however, the familiar character of the \Lya
forest changes.  Observationally, the redshifted \Lya transition
shifts from optical to UV wavelengths at $z\simeq 1.5$, and can
therefore only be observed from outside the atmosphere.  Physically,
the densities probed by the \Lya absorbers also change, from gas that
is close to the mean background density, to higher density material
that resides at the outskirts of galaxies
\citep{Theuns1998_lowz,Dave1999,Dave2010,Nasir2017,Maitra2022}.
Recent progress has been largely driven by results from the
\emph{Cosmic Origins Spectrograph} \citep[COS,][]{Green2012} on the
\emph{Hubble Space Telescope}
\citep[e.g.][]{Wakker2015,Danforth2016,Khaire2019pk,Kim2021}.
Intriguingly, the straightforward evolution of successful \Lya forest
models at $z>2$ to lower redshift does not automatically guarantee a
good match to the \Lya forest absorption lines identified with COS.
One challenge is correctly reproducing the incidence of \Lya forest
absorbers at $z\sim 0$.  \cite{Kollmeier2014} were the first to
highlight a discrepancy between the observed and simulated \HI column
density distribution function (CDDF) at $z\sim 0$ for column densities
$10^{13.6}\rm\,cm^{-2}\leq N_{\rm HI}\leq 10^{14.4}\rm\,cm^{-2}$,
finding the metagalactic \HI photoionisation rate, $\Gamma_{\rm HI}$,
required by the observed CDDF was considerably larger (by factor of
$\sim 5$) than predicted in the empirically calibrated
\citet{HaardtMadau2012} UV background model.  This implied that either
the ionising photon production rate from galaxies and quasars had been
significantly underestimated by \citet{HaardtMadau2012}, or that the
IGM simulations were missing an important physical ingredient.
Subsequent work has revised this apparent discrepancy downward to a
more manageable factor of two, either by using independent simulations
\citep{Wakker2015,Shull2015,Gaikwad2017b,Viel2017,Nasir2017}, revising
the predicted amplitude of the UV background in the empirical models
\citep{KhaireSrianand2015,KhaireSrianand2019,Puchwein2019,FaucherGiguere2020},
or by invoking black hole feedback that efficiently heats and ionises
the low density IGM \citep{Christiansen2020}.  Regardless of any
remaining tension between data and theory, however, this work has
demonstrated that the \Lya forest CDDF at $z\sim 0$ is a valuable
diagnostic of ionising photon production \emph{and} galactic feedback
\citep{Gurvich2017}.

By contrast, the direct comparison of the Doppler parameter
distribution of the \Lya forest absorption lines identified with COS
to hydrodynamical simulations has received less attention.  The few
studies that have attempted this recently have (as for the CDDF) also
struggled to reproduce the COS data at $0\leq z \leq 0.2$, finding
line widths that are too narrow compared to the observations
\citep{Gaikwad2017,Viel2017,Nasir2017}.  The reason for this
discrepancy remains unclear, but -- assuming there are no unaccounted
for instrumental systematics that systematically broaden the
absorption lines -- it suggests that the gas temperatures associated
with the simulated \Lya absorbers may be too low and/or there is
missing non-thermal broadening in the models.  It is also not certain
whether missing feedback can resolve this discrepancy.
\citet{Viel2017} and \citet{Nasir2017} found that the AGN feedback
model used in the Sherwood simulations \citep{Bolton2017} has a very
limited impact on the \Lya line widths; any additional hot gas was in
the Warm-Hot IGM (WHIM), and was too highly ionised to detect in \Lya
absorption.  On the other hand, more recently \citet{Christiansen2020}
have found that hot, highly ionised gas produced by the jet feedback
model in the SIMBA simulation \citep{Dave2019} significantly improves
agreement with the mean \Lya forest transmission at $z<0.5$.  However,
it is not obvious if this improved agreement also extends to the
Doppler parameter distribution.

Hence, the goal of this work is to present a quantitative assessment
of the additional heating and/or non-thermal contribution to the line
widths required for consistency with the \Lya forest CDDF \emph{and}
Doppler parameter distribution from COS observations at $0\leq z\leq
0.2$.  We achieve this by forward modelling the COS data, and then
fitting Voigt profiles to the simulations in the same manner as the
observations.  Our joint analysis of the CDDF and Doppler parameter
distribution is then used to simultaneously constrain the metagalactic
\HI photoionisation rate and the effective power-law spectral shape of
the UV background close to the Lyman limit.  We shall argue that
reproducing the Doppler widths of the \Lya absorbers in the COS data
by boosting the thermal line widths requires a UV background with an
unphysically hard ionising spectrum.  This implies that there is still
missing physics in the models, and that either an additional,
non-canonical heating mechanism, or a source of non-thermal broadening
that is missed by the simulations (e.g., IGM turbulence), is necessary
for achieving consistency between hydrodynamical simulations of the
\Lya forest and the COS data.

This paper is organised as follows.  In Section~\ref{sec:method} we
describe the simulations used in this work.  We then give an overview
of the physical properties of simulated \Lya forest absorbers at
$z\simeq 0.1$ in Section~\ref{sec:overview}, and perform an initial
comparison of our simulated results to the COS data.
Section~\ref{sec:heat_turb} describes the relationship between the UV
background spectral shape and the IGM temperature in our models, and
outlines the simple model we use for including an unresolved turbulent
contribution to the \Lya forest line widths.  We then present and
discuss our limits on non-canonical heating and/or the turbulent
contribution required by the COS \Lya absorbers in
Section~\ref{sec:results}, and summarise our conclusions in
Section~\ref{sec:conclude}.  The corrections to the simulated CDDF for
box size and mass resolution that we use throughout this work are
presented in Appendix~\ref{app:converge}.  An examination of the
effect that noise and spectral resolution have on the CDDF and Doppler
parameter distribution is presented in Appendix~\ref{app:sys}, along
with a test of our assumption that photoionisation equilibrium holds
in the low redshift \Lya forest. In Appendix~\ref{app:TNG}, we compare
the TNG100-1 simulation from the IllustrisTNG project
\citep{Nelson2019} to the COS data.  We confirm that the \Lya forest
line widths predicted by TNG100-1 are also narrower than observed at
$z\simeq 0.1$.

Finally, throughout this work, it may be useful to recall that a
$N_{\rm HI}\sim 10^{13.5}\rm\,cm^{-2}$ \Lya forest absorber at $z=0.1$
is typically associated with a gas overdensity of $\Delta \sim 10$ (or
equivalently $n_{\rm H}\sim 10^{-5.6}\rm\,cm^{-3}$) within our models.
Comoving and proper distance units use the prefixes ``c'' and ``p''
respectively.


\section{Numerical models}\label{sec:method}
\subsection{Hydrodynamical simulations}\label{sec:sims}

The $18$ cosmological hydrodynamical simulations used in this work are
listed in Table~\ref{tab:hydrosims}.  The simulations were performed
with a version of the Tree-PM SPH code \textsc{P-Gadget-3}
\citep{Springel2005}, modified for the Sherwood simulation project
\citep{Bolton2017,Nasir2017}.  Our fiducial box size is
$L=60h^{-1}\rm\,cMpc$ with $2\times 768^{3}$ gas dark matter and gas
particles, giving a gas (dark matter) particle mass of $M_{\rm
  gas}=6.38\times 10^{6}h^{-1}M_{\odot}$ ($M_{\rm dm}=3.44\times
10^{7}h^{-1}M_{\odot}$).  This improves our fiducial mass resolution
by a factor of $8$ compared to \citet{Nasir2017}.  The gravitational
softening length is set to $0.04$ times the mean interparticle
separation in all models.  The cosmological parameters are
$\Omega_{\rm m}=0.308$, $\Omega_{\Lambda}=0.692$, $h=0.678$,
$\Omega_{\rm b}=0.0482$, $\sigma_{8}=0.829$ and $n=0.961$
\citep{Planck2014}, with a primordial helium fraction by mass of
$Y_{\rm p}=0.24$ \citep{Hsyu2020}.

Our main simulation (AGN) incorporates energy-driven galactic outflows
and AGN feedback using the model described in detail by
\citet{PuchweinSpringel2013}.  Briefly, the star formation model is
based on \citet{SpringelHernquist2003}, but for a \citet{Chabrier2003}
rather than \citet{Salpeter1955} initial mass function and a galactic
wind velocity that is directly proportional to the escape velocity of
the galaxy.  In the black hole feedback model, for the ``quasar'' mode
when accretion rates are above $0.01$ of the Eddington rate, $0.5$ per
cent of the accreted rest mass energy is thermally coupled to the
gas. For lower accretion rates the ``radio'' mode is used instead.
This is triggered for a fractional increase in the black hole mass of
$10^{-4}$, with $2$ per cent of the rest mass energy used for
injecting hot AGN bubbles.  This is the same star formation and AGN
feedback model used in our earlier work on the \Lya forest at $z\simeq
0.1$ \citep{Viel2017,Nasir2017}.

To explore the effect of AGN feedback further, we also now consider a
second model (StrongAGN) where the fractional increase in the black
hole mass required for triggering the radio mode is increased to
$10^{-2}$, with $8$ per cent of the rest mass energy now used to
inject hot AGN bubbles.  This is the same as the ``stronger radio''
model used by \citet{Henden2018}, and it leads to less frequent but
more energetic bubble injections and substantially lower gas fractions
in the vicinity of haloes.  We use this to provide a model that
predicts a gas mass in the Warm-Hot IGM (WHIM) at low redshift that
mimics the effect of jet-mode heating in the SIMBA simulation used by
\citet{Christiansen2020} (see Section~\ref{sec:physprop} for further
details).

For the remainder of the simulations in Table~\ref{tab:hydrosims} we
ignore star formation and feedback, and instead directly convert gas
with temperature $T<10^{5}\rm\,K$ and density $\Delta=\rho/\langle
\rho \rangle>10^{3}$ into collisionless particles \citep{Viel2004}.
This ``Quick-Ly$\alpha$'' approach has been shown to be a reasonable
approximation for unsaturated absorption lines, $N_{\rm
  HI}<10^{14.5}\rm\,cm^{-2}$, in the low redshift \Lya forest
\citep{Nasir2017}.  Importantly, the Quick-Ly$\alpha$ approach is less
computationally expensive than the AGN feedback model, and for this
reason we use it to create a grid of $11$ simulations (H$00$--H$10$)
with different IGM thermal histories that we use to obtain our UV
background constraints in Section~\ref{sec:results}.

\begin{table*}
  \centering
   \caption{Summary of the \textsc{P-Gadget-3} simulations used in
     this work.  From left to right, the columns list the simulation
     name, the box size in $h^{-1}\rm cMpc$, the gas particle mass in
     $h^{-1}M_{\odot}$, the factor, $\zeta$, by which the equilibrium
     equivalent photoheating rates from \citet{Puchwein2019} are
     increased for each model, the gas temperature, $T_{0}$, at the
     mean density at $z=0.1$, the slope, $\gamma-1$, of the power-law
     temperature-density relation at $z=0.1$, the effective power-law
     spectral index, $\alpha_{\rm eff}$, for the UV background at
     $z=0.1$ (see Section~\ref{sec:T_asymp} for details), the mean
     transmission of the \Lya forest at $z=0.1$ assuming the
     \citet{Puchwein2019} photoionisation rate $\log(\Gamma_{\rm
       HI}^{\rm P19}/\rm s^{-1})=-13.04$, the photoionisation rate,
     $\Gamma_{\rm HI}^{\rm CDDF}$, required to match the amplitude of
     the CDDF at $z=0.1$, the ratio of this quantity to the
     \citet{Puchwein2019} value, and the ratio to the
     \citet{HaardtMadau2012} value, $\log(\Gamma_{\rm HI}^{\rm
       HM12}/\rm s^{-1})=-13.45$. The upper section of the table lists
     the simulations used to test the effect of AGN feedback
     \citep{PuchweinSpringel2013} and blazar heating
     \citep{Puchwein2012} on the \Lya forest absorption lines. The
     Quick-Ly$\alpha$ simulations that form our thermal history grid
     are listed in the middle section of the table.  The lower section
     of the table lists the Quick-Ly$\alpha$ simulations used to test
     the effect of simulation box size and mass resolution in
     Appendix~\ref{app:converge}.}
  \begin{tabular}{c|c|c|c|c|c|c|c|c|c|c|c}
    \hline
    Model & $L_{\rm box}$ & $N_{\rm part}$& $M_{\rm gas}$ & $\zeta$ & $T_{0}$  & $\gamma-1$ & $\alpha_{\rm eff}$  &  $\langle F \rangle$ &  $\log(\frac{\Gamma_{\rm HI}^{\rm CDDF}}{\rm s^{-1}})$ & $\frac{\Gamma_{\rm HI}^{\rm CDDF}}{\Gamma_{\rm HI}^{\rm P19}}$ &  $\frac{\Gamma_{\rm HI}^{\rm CDDF}}{\Gamma_{\rm HI}^{\rm HM12}}$  \\
    & $[\rm h^{-1} cMpc]$ & &  $ \rm [h^{-1}\,M_{\odot}]$& &$\rm [K]$ & &  &  &  &  & \\
    \hline
     AGN         & 60.0 & $2\times768^{3}$ & $6.38\times 10^{6}$ & 1.00 & 4216  & 0.58  & 1.17 & 0.981 & -13.22 & 0.66 & 1.69 \\
    StrongAGN    & 60.0 & $2\times768^{3}$ & $6.38\times 10^{6}$ & 1.00 & 4274  & 0.59  & 1.17 & 0.988 & -13.50 & 0.35 & 0.90 \\
    Blazar       & 60.0 & $2\times768^{3}$ & $6.38\times 10^{6}$ & 1.00 & 31069 & -0.85 & --   & 0.989 & -13.41 & 0.43 & 1.11\\ 
    \hline
    H00 & 60.0 & $2\times768^{3}$ & $6.38\times 10^{6}$ & 0.65 & 3262  & 0.58 & 2.89  & 0.981 & -13.22 & 0.67 & 1.71 \\
    H01 & 60.0 & $2\times768^{3}$ & $6.38\times 10^{6}$ & 0.82 & 3703  & 0.58 & 1.86  & 0.982 & -13.22 & 0.66 & 1.70 \\
    H02 & 60.0 & $2\times768^{3}$ & $6.38\times 10^{6}$ & 1.00 & 4174  & 0.58 & 1.17  & 0.983 & -13.22 & 0.67 & 1.73 \\
    H03 & 60.0 & $2\times768^{3}$ & $6.38\times 10^{6}$ & 1.20 & 4743  & 0.59 & 0.65  & 0.983 & -13.22 & 0.67 & 1.73 \\
    H04 & 60.0 & $2\times768^{3}$ & $6.38\times 10^{6}$ & 1.42 & 5129  & 0.58 & 0.24  & 0.984 & -13.24 & 0.64 & 1.64 \\
    H05 & 60.0 & $2\times768^{3}$ & $6.38\times 10^{6}$ & 1.65 & 5713  & 0.58 & -0.07 & 0.985 & -13.27 & 0.59 & 1.51 \\
    H06 & 60.0 & $2\times768^{3}$ & $6.38\times 10^{6}$ & 1.90 & 6194  & 0.59 & -0.33 & 0.985 & -13.28 & 0.58 & 1.49 \\
    H07 & 60.0 & $2\times768^{3}$ & $6.38\times 10^{6}$ & 2.16 & 6680  & 0.59 & -0.54 & 0.986 & -13.31 & 0.54 & 1.38 \\
    H08 & 60.0 & $2\times768^{3}$ & $6.38\times 10^{6}$ & 2.72 & 7797  & 0.59 & -0.86 & 0.987 & -13.35 & 0.49 & 1.25 \\
    H09 & 60.0 & $2\times768^{3}$ & $6.38\times 10^{6}$ & 3.33 & 8778  & 0.59 & -1.09 & 0.987 & -13.40 & 0.44 & 1.13 \\
    H10 & 60.0 & $2\times768^{3}$ & $6.38\times 10^{6}$ & 4.70 & 10817 & 0.59 & -1.41 & 0.989 & -13.47 & 0.38 & 0.97 \\
    \hline
    L40   & 40.0 & $2\times512^{3}$  & $6.38\times 10^{6}$ & 1.00 & 4148 & 0.58 & 1.17 & 0.983 & -13.25 & 0.62 & 1.59 \\
    L80   & 80.0 & $2\times1024^{3}$ & $6.38\times 10^{6}$ & 1.00 & 4182 & 0.58 & 1.17 & 0.984 & -13.26 & 0.60 & 1.55 \\
    N512  & 60.0 & $2\times512^{3}$  & $2.15\times 10^{7}$ & 1.00 & 4473 & 0.59 & 1.17 & 0.982 & -13.20 & 0.70 & 1.80 \\
    N1024 & 60.0 & $2\times1024^{3}$ & $2.69\times 10^{6}$ & 1.00 & 4127 & 0.59 & 1.17 & 0.983 & -13.24 & 0.64 & 1.63 \\
    \hline
  \end{tabular}
  \label{tab:hydrosims}
\end{table*}

Photoionisation and heating by a spatially uniform UV background is
included in all simulations assuming ionisation equilibrium.  We use
the equilibrium equivalent rates from the UV background model of
\citet[][hereafter P19]{Puchwein2019}.  This has the advantage of
correctly incorporating non-equilibrium ionisation effects on the
ionised fraction and gas temperature, but without the additional
computational overhead of solving a non-equilibrium thermo-chemistry
network.  At $z=0.1$, the P19 model has a photoionisation rate
$\log(\Gamma_{\rm HI}^{\rm P19}/\rm s^{-1})=-13.04$ and predicts a gas
temperature at the mean density of $T_{0}\simeq 4200\rm\,K$, and a
power-law temperature-density relation, $T=T_{0}\Delta^{\gamma-1}$,
where $\gamma \simeq 1.58$ for $\Delta=\rho/ \langle \rho \rangle
\lesssim 10$.  In the $11$ Quick-Ly$\alpha$ simulations ($\rm
H00$--$\rm H10$) we have varied the UV background photoheating rates,
$\epsilon_{\rm i}$, in the P19 UV background model.  We scale the
hydrogen and helium photoheating rates by a constant $\zeta$, where
$\epsilon_{\rm i}=\zeta\epsilon_{\rm i}^{\rm P19}$ and $i=\rm H\,\rm
\scriptstyle I$, He$\,\rm \scriptstyle I$, He$\,\rm \scriptstyle II$
\citep[cf.][]{Becker2011}.  Self-shielding of dense gas to ionising
photons is included on-the-fly in all simulations following
\citet{Rahmati2013}.  Metal line cooling is not included, although
\citet{TepperGarcia2012} found this should have a very small effect on
the column densities and Doppler parameters of low redshift \Lya
forest absorbers (see their fig. C2).  We also indirectly test this in
Appendix~\ref{app:TNG} by performing a Voigt profile analysis on \Lya
forest spectra drawn from the Illustris TNG100-1 simulation, which
does include metal line cooling.

For comparison with the canonical UV photoheating paradigm, following
\citet{Kollmeier2014} we also consider an alternative model (Blazar)
where heating in the IGM at $z<3$ is dominated by TeV emission from
blazars.  This results in a much higher temperature in the low-density
IGM, $T_{0}\simeq 31\,100\rm\,K$, compared to UV photoheating models,
with a temperature-density relation that is ``inverted''
(i.e. $\gamma<1$) due to a volumetric heating rate that is independent
of density \citep{Chang2012}.  We adopt the intermediate heating model
from \citet{Puchwein2012} for this purpose (see their eq. (1) and
table 1), and perform the simulation using the Quick-Ly$\alpha$
approximation.

Finally, we perform four more Quick-Ly$\alpha$ simulations to assess
the convergence of our results with box size and mass resolution;
these models are listed in the lower section of
Table~\ref{tab:hydrosims}.  We use these models to apply a correction
to the simulated CDDF at our fiducial mass resolution and box size.
This correction is listed in Table~\ref{tab:boxres} in
Appendix~\ref{app:converge}, along with a more detailed discussion of
the convergence properties of the Quick-Ly$\alpha$ simulations.

\subsection{Simulated and observed \Lya forest spectra}\label{sec:spectra}

Simulated \Lya forest spectra are extracted using an approach similar
to \citet{Nasir2017}.  We randomly draw $16384$ lines of sight
parallel to the box axes, where each line of sight has 2048 pixels.
The \Lya optical depths are then obtained from the particle data using
the interpolation scheme described by \citet{Theuns1998} combined with
the Voigt profile approximation from \citet{TepperGarcia2006}.

In this work we will compare the simulated spectra to observational
measurements of the \Lya forest CDDF and Doppler parameter
distribution first described in \citet{Viel2017}.  These data are
obtained from 44 AGN spectra, selected to have a signal-to-noise per
resolution element of $\rm S/N>20$ and an emission redshift of
$0.1<z_{\rm em}<0.35$, and form part of the larger data set recently
presented by \citet{Kim2021}.  Further details regarding the COS data
reduction and Voigt profile fitting can be found in \citet{Wakker2015}
and \citet{Kim2021}.  The Voigt profile fits to the COS data and
simulations have been obtained using the \Lya transition only.  The
line list we use consists of 704 \HI \Lya lines with mean redshift
$\langle z \rangle =0.09$, mean column density $\langle \log(N_{\rm
  HI}/\rm cm^{-2})\rangle =13.29$ and mean Doppler parameter $ \langle
b\rangle =36.6\rm\,km\,s^{-1}$.  The total redshift path length of the
data is $\Delta z = 4.991$, covering the \Lya forest at $0 \leq z \leq
0.2$.  As already shown in \citet{Nasir2017}, the CDDF and Doppler
parameter distribution we use are consistent with independent
measurements using COS data from \citet{Danforth2016} and
\citet{Gaikwad2017} over the range of interest for this work.

In order to approximately forward model the COS data, all the
simulated spectra are convolved with the COS line spread
function\footnote{\url{https://www.stsci.edu/hst/instrumentation/cos/performance/spectral-resolution}}
at $1341\,$\AA, for central wavelength G130M/1327 at lifetime position
LP1.  The spectra are then rebinned onto pixels of width
$0.02991\,$\AA\ (i.e 3 times the COS binning of $0.00997\,$\AA\,
following \citet{Kim2021}) and Gaussian distributed noise with a flux
independent signal-to-noise ratio of $\rm S/N=30$ per
$19\rm\,km\,s^{-1}$ resolution element (i.e. $\rm S/N\sim 17.7$ per
pixel) is added.  Voigt profile fitting to the simulated \Lya spectra
is then performed with VPFIT version 10 \citep{CarswellWebb2014},
which deconvolves the (already convolved) mock spectra with the
instrument profile to obtain the intrinsic line widths.  We emphasise
that, to ensure a fair comparison between observations and
simulations, we have considered {\it only} the \Lya lines obtained
with VPFIT in this work.  This minimises any biases that would arise
if, e.g., we had used higher order Lyman series information to perform
a curve-of-growth analysis on either the observational or simulated
data alone.

We assess the role that a different signal-to-noise ratio or line
spread function may have on the recovery of the CDDF and Doppler
parameter distribution in Appendix~\ref{app:sys}.  To summarise those
results, we find that absorption lines with $10^{13.3}\rm cm^{-2}\leq
N_{\rm HI} \leq 10^{14.5}\rm\,cm^{-2}$ and $20\rm\,km\,s^{-1} \leq b
\leq 90\rm\,km\,s^{-1}$ will remain insensitive to the expected
variations in the $\rm S/N$ or line spread function.  We will only use
the absorption lines in these ranges when obtaining UV background
constraints from the COS data.  This sub-set consists of $297$ \HI
\Lya lines with mean redshift $\langle z \rangle =0.09$, mean column
density $\langle \log(N_{\rm HI}/\rm cm^{-2})\rangle =13.70$ and mean
Doppler parameter $ \langle b\rangle =39.9\rm\,km\,s^{-1}$.


\section{The CDDF and Doppler parameter distribution} \label{sec:overview}
\subsection{Physical properties of unsaturated \Lya forest absorbers at z=0.1} \label{sec:physprop}

\begin{figure*}
  \begin{minipage}{1.0\textwidth}
    \includegraphics[width=0.48\textwidth]{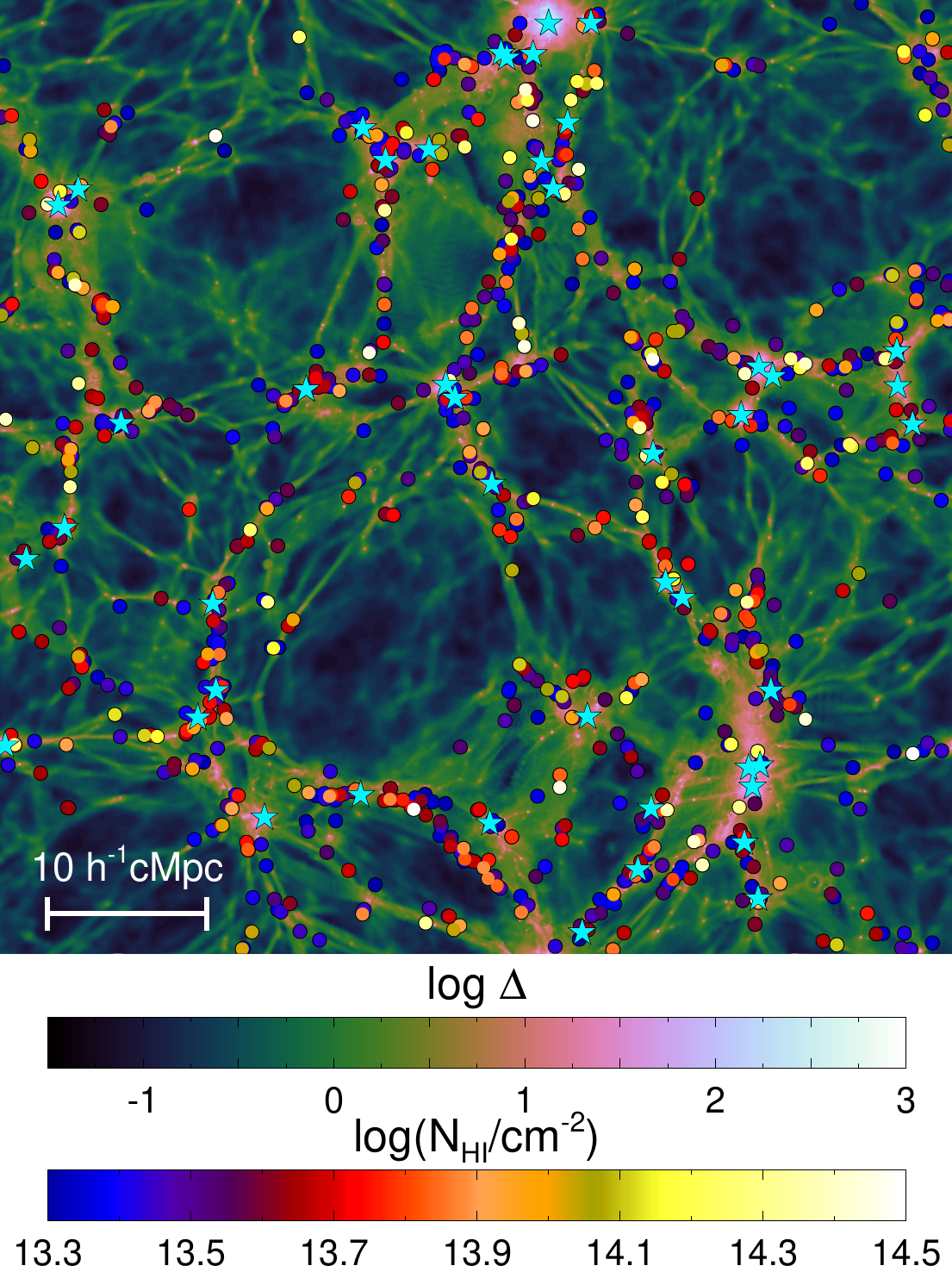}
    \includegraphics[width=0.48\textwidth]{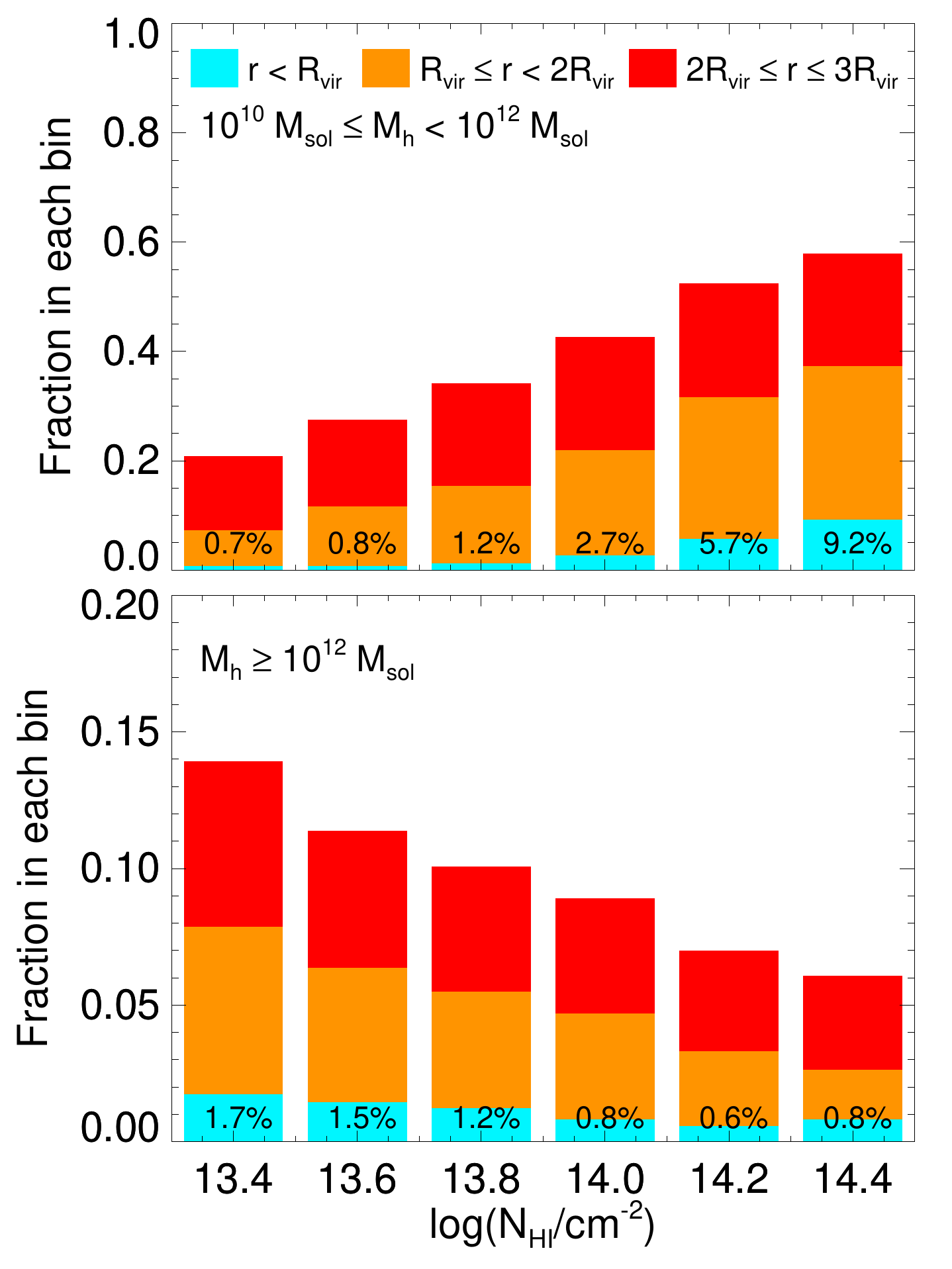}
  \end{minipage}
 
  \caption{{\it Left:} A square slice ($60 \,h^{-1}\rm cMpc$ on each
    side) displaying the logarithm of the gas density, $\log
    \Delta=\log(\rho/\langle \rho \rangle)$, in the AGN simulation at
    $z=0.1$, projected over a distance of 500 $h^{-1}\rm ckpc$.  The
    \Lya absorbers in the slice with \HI column densities
    $10^{13.3}\rm\,cm^{-2}\leq N_{\rm HI} \leq 10^{14.5}\rm\,cm^{-2}$
    are over-plotted as coloured circles.  The cyan stars show the
    location of haloes in the slice with total mass $M_{\rm h}\geq
    10^{12}M_{\odot}$.  {\it Right:} The fraction of \Lya absorbers in
    the full simulation volume at redshift $z=0.1$, in bins of width
    $\Delta \log(N_{\rm HI}/\rm cm^{-2})=0.2$, located within one
    (cyan bars), two (orange bars) or three (red bars) virial radii,
    $R_{\rm vir}$, of haloes with total mass $ 10^{10}M_{\odot} \leq
    M_{\rm h}< 10^{12}M_{\odot}$ (upper panel) or $M_{\rm h}\geq
    10^{12}M_{\odot}$ (lower panel).  Note the different scale on the
    vertical axis of the lower panel.  The percentages at the base of
    each bar give the fraction of absorbers within $r<R_{\rm vir}$.}
  \label{fig:map}
\end{figure*}

\begin{figure*}
  \begin{minipage}{1.0\textwidth}
    \centering
    \includegraphics[width=0.75\textwidth]{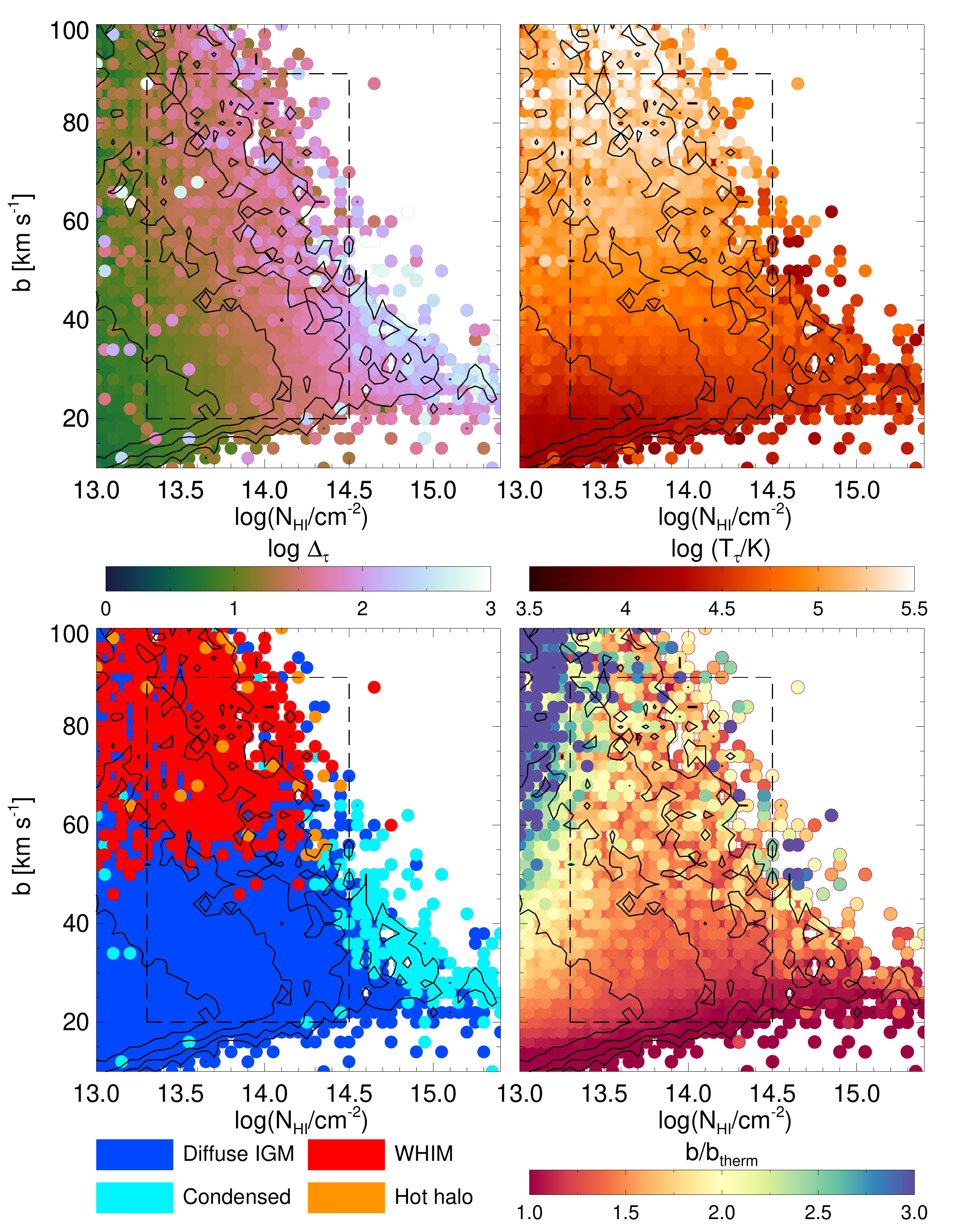}
  \end{minipage}
  \vspace{-0.3cm}
  \caption{{\it Upper left:} The Doppler parameter, $b$, against the
    logarithm of the \HI column density, $\log N_{\rm HI}$, for \Lya
    absorption lines at redshift $z=0.1$ in the AGN simulation.  The
    number of absorption lines increases by $0.5$ dex within each
    contour, where the colour indicates the logarithm of the optical
    depth weighted gas density, $\log \Delta_{\tau}$, associated with
    each absorber.  The boxed region corresponds to the column
    densities and Doppler parameters used to constrain the UV
    background in this work; these represent $28$ per cent of all the
    identified lines. {\it Upper right:} As for the upper left panel,
    except the colour now indicates the logarithm of the optical depth
    weighted gas temperature for each absorber. {\it Lower left:} As
    for the upper left panel, except the four colours now indicate the
    baryon phase associated with the absorbers, following
    \citet{Dave2010}.  The phases are: diffuse IGM ($T<10^{5}\rm\,K$,
    $\Delta<97.2$), WHIM ($T\geq 10^{5}\rm\,K$, $\Delta<97.2$), hot
    halo ($T\geq 10^{5}\rm\,K$, $\Delta\geq 97.2$) and condensed ($T<
    10^{5}\rm\,K$, $\Delta\ \geq 97.2$).  The fraction of absorbers in
    the boxed region within each phase are: $83$ per cent diffuse IGM,
    $15$ per cent WHIM, $1$ per cent hot halo and $1$ per cent
    condensed (see also Table~\ref{tab:phases}).  {\it Lower right:}
    As for the upper left panel, except now showing the ratio of the
    Doppler parameters to the expected thermal line widths, $b/b_{\rm
      therm}$, where $b_{\rm therm}=(2k_{\rm B}T_{\tau}/m_{\rm
      H})^{1/2}$.}
  \label{fig:bN}  
\end{figure*}

\begin{table*}
  \centering
  \caption{The percentage of \Lya absorbers with $20\rm\,km\,s^{-1}
    \leq b \leq 90\rm\,km\,s^{-1} $ and $10^{13.3}\rm\,cm^{-2}\leq
    N_{\rm HI} \leq 10^{14.5}\rm\,cm^{-2}$ in the AGN, StrongAGN, H02
    and Blazar simulations that are associated with the four
    temperature-density phases defined by \citet{Dave2010}.  For
    comparison, the number in parentheses gives the baryon mass
    fraction associated with each phase, where we also include stars
    in the condensed phase.  The mass fraction in the
    condensed$+$stars phase is artificially high in the H02 and Blazar
    models as a result of using the Quick-Ly$\alpha$ scheme for
    converting gas to star particles.  Note also that while the
    relative fraction of \Lya absorbers in each phase remains similar
    across all models, the mass fraction in the WHIM (shown in
    parentheses) increases significantly in the StrongAGN simulation.}
  \begin{tabular}{c|c|c|c|c}
    \hline
    Model & Diffuse IGM & WHIM & Hot halo & Condensed(+stars) \\
    & $T<10^{5}\rm\,K$, $\Delta<97.2$ & $T\geq 10^{5}\rm\,K$, $\Delta<97.2$ & $T \geq 10^{5}\rm\,K$, $\Delta \geq 97.2$ & $T<10^{5}\rm\,K$, $\Delta \geq 97.2$\\
    \hline
    AGN & 83.4 (36.7) & 15.2 (37.9) & 0.7 (16.1) & 0.7 (9.3) \\
    StrongAGN & 82.1 (20.1)  & 16.8 (68.4) & 0.6 (3.0) & 0.5 (8.5) \\
    H02 (No SF/feedback) & 87.3 (35.5) & 10.2 (21.2) & 2.0 (13.9) & 0.5 (29.4) \\
    Blazar (No SF/feedback) & 82.9 (23.2) & 14.9 (34.3) & 1.9 (13.5) & 0.3 (29.0) \\
    \hline
  \end{tabular}
  \label{tab:phases}
\end{table*}

\begin{figure*}
  \begin{minipage}{1.00\textwidth}
    \includegraphics[width=0.49\textwidth]{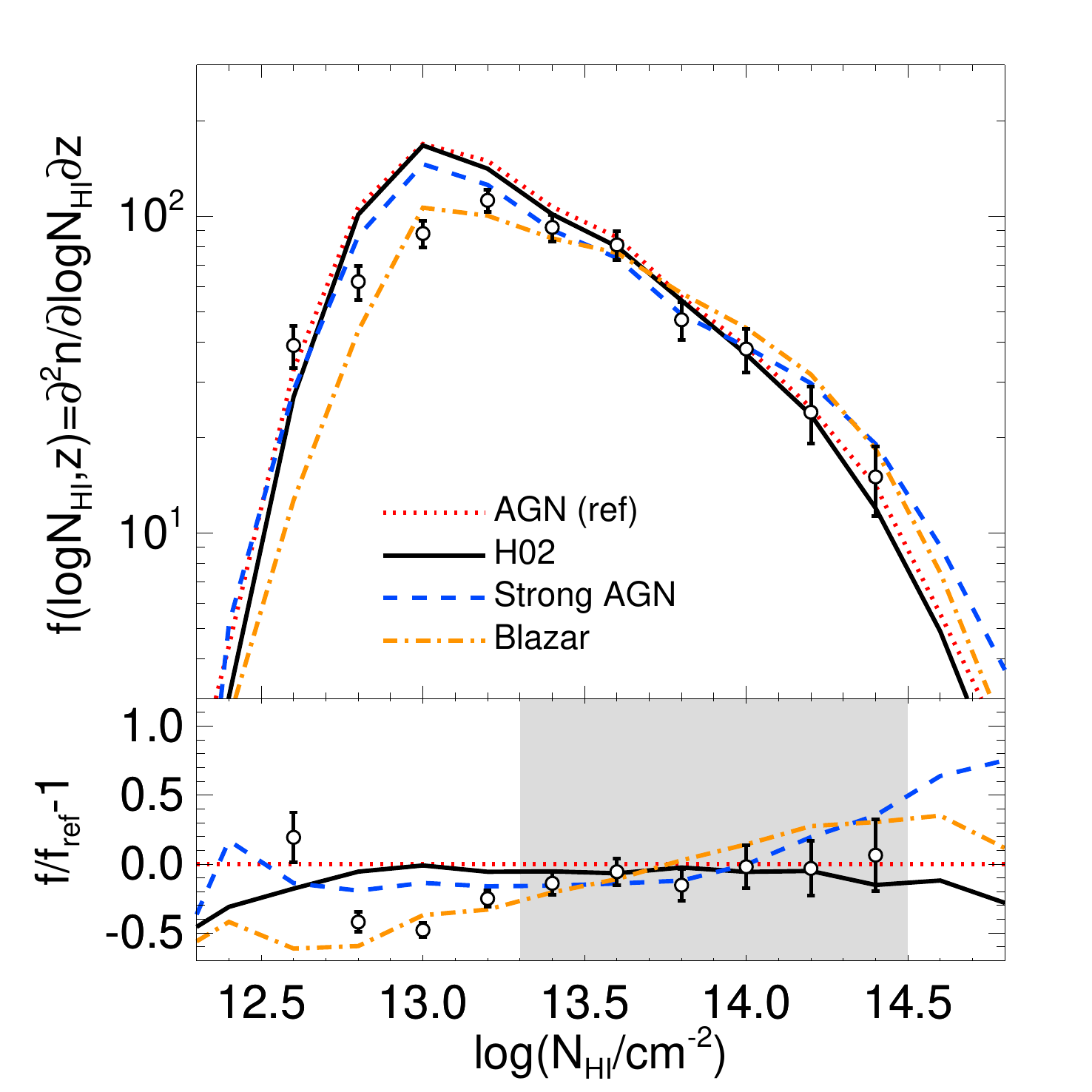}
    \includegraphics[width=0.49\textwidth]{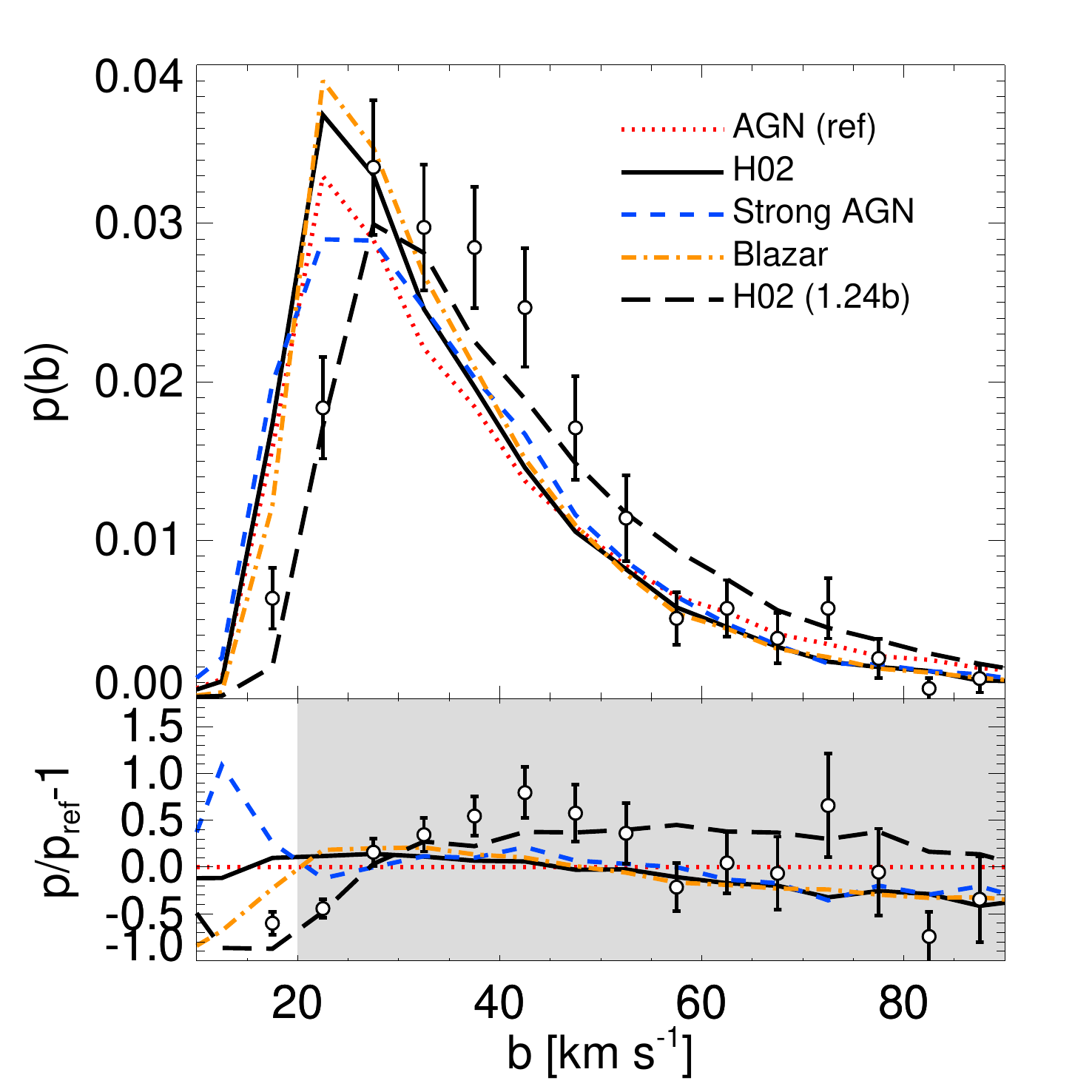}
  \end{minipage}
  \vspace{-0.3cm}
  \caption{{\it Left:} The column density distribution function (CDDF)
    measured from COS data \citep[open circles,][]{Viel2017} for \Lya
    absorbers with Doppler parameters $20\rm \,km\,s^{-1}\leq b \leq
    90\rm \,km\,s^{-1} $, compared to the CDDF obtained from the AGN
    (red dotted curve), StrongAGN (blue dashed curve), Blazar (orange
    dot-dashed curve) and H02 (black solid curve) models.  Note the
    CDDF is incomplete at $N_{\rm HI}\lesssim 10^{13.3}\rm\,cm^{-2}$.
    Following \citet{Viel2017}, the column densities obtained from the
    mock spectra have been linearly rescaled to achieve good agreement
    with the observed CDDF.  A correction for box size and mass
    resolution has also been applied to the simulations (see
    Appendix~\ref{app:converge} for details).  The lower panel
    displays the residuals with respect to the AGN model.  {\it
      Right:} The corresponding Doppler parameter probability
    distribution for \Lya absorbers with column densities
    $10^{13.3}\rm\,cm^{-2}\leq N_{\rm HI} \leq 10^{14.5}\rm\,cm^{-2}$.
    The simulated line widths are narrower than the observational
    measurements.  The black dashed curve shows the H02 simulation
    after the Doppler widths have been increased by a factor of
    $1.24$.  The grey shading in the lower panels display the $N_{\rm
      HI}$ and $b$ range where we judge the comparison between
    observations and simulations to be reliable (see
    Appendix~\ref{app:sys} for details).}
  \label{fig:phystest}  
\end{figure*}

First, it is instructive to briefly recap the general properties of
the low redshift \Lya forest predicted in cosmological hydrodynamical
simulations.  In Fig.~\ref{fig:map} we show a $60\,h^{-1}\rm \,cMpc\,
\times\, 60\,h^{-1}\rm\,cMpc$ slice through the AGN simulation volume,
displaying the gas density at $z=0.1$ projected over a distance of
$500\, h^{-1}\rm\,ckpc$.  The \Lya absorbers with \HI column densities
$10^{13.3}\rm \, cm^{-2}\leq N_{\rm HI} \leq 10^{14.5}\rm \, cm^{-2}$
in the slice are over-plotted as coloured circles.  Rather than fit
Voigt profiles -- as we do in the rest of this work -- following
\citet{Gurvich2017}, these \HI column densities are obtained by
integrating the \HI number densities in the model over a scale of
$50\rm\,km\,s^{-1}$.  The cyan stars show the position of haloes with
total mass $M_{\rm h}\geq 10^{12}\rm\,M_{\odot}$.

On visual inspection of Fig.~\ref{fig:map}, we observe the \Lya forest
absorbers are located in the overdense filaments and nodes in the
web-like distribution of intergalactic gas \citep[see
  also][]{Tonnesen2017}.  This is quantified further in the right
panel of Fig.~\ref{fig:map}, which shows the fraction of these
absorbers within $1$, $2$ or $3$ virial radii\footnote{In this work we
identify the virial radius, $R_{\rm vir}$, as the radius of a sphere
centred at the halo centre of mass that has mean density $\Delta_{\rm
  c}$ times the critical density, where $\Delta_{\rm c}$ is given by
eq. (6) of \citet{BryanNorman1998}.} of haloes with masses
$10^{10}\,M_{\odot}\leq M_{\rm h}<10^{12}\,M_{\odot}$ (upper panel)
and $M_{\rm h}\geq 10^{12}\,M_{\odot}$ (lower panel).  We find that
only a few per cent of \Lya forest absorbers with $ N_{\rm HI}\leq
10^{14}\rm\,cm^{-2}$ are within $r<R_{\rm vir}$, with a fraction that
increases with $N_{\rm HI}$ for halo masses $10^{10}\,M_{\odot}\leq
M_{\rm h}<10^{12}\,M_{\odot}$.

For comparison, using the $\rm CGM^{2}$ (Cosmic Origins Spectrograph
and Gemini Mapping the Circumgalactic Medium) survey,
\citet{Wilde2021} have recently found that -- while \HI absorption
lines with $N_{\rm HI}<10^{14}\rm\,cm^{-2}$ are not strongly
associated with massive galaxies -- absorbers with $N_{\rm
  HI}>10^{14}\rm\,cm^{-2}$ have a high probability (i.e. $>50$ per
cent) of residing within $1.2 R_{\rm vir}$ ($2.4 R_{\rm vir}$) of
galaxies with stellar masses
$10^{9.2}\rm\,M_{\odot}<M_{\star}<10^{9.9}\,M_{\odot}$
($M_{\star}>10^{9.9}\,M_{\odot}$).  As a rough comparison, for halo
masses $10^{10}\,M_{\odot}\leq M_{\rm h}<10^{12}\,M_{\odot}$, we find
the fraction of \HI absorbers within $r<3R_{\rm vir}$ in our
simulations increases with $N_{\rm HI}$, from $\sim 20$ per cent at
$\log(N_{\rm HI}/\rm cm^{-2})=13.4\pm 0.1$ up to almost $60$ per cent
at $\log(N_{\rm HI}/\rm cm^{-2})=14.4\pm 0.1$.  The qualitative
picture is therefore one where the majority of the strongest absorbers
with $N_{\rm HI}>10^{14}\rm\,cm^{-2}$ occur close to haloes, while the
majority of weaker absorbers are associated with intergalactic gas and
are at $r>3R_{\rm vir}$ \citep[see
  also][]{ChenMulchaey2009,Tejos2014,Keeney2018}.  In total, $29.2$
($1.5$) per cent of the identified \HI absorbers with
$10^{13.3}\rm\,cm^{-2}\leq N_{\rm HI}\leq 10^{14.5}\rm\,cm^{-2}$ are
within $3R_{\rm vir}$ ($R_{\rm vir}$) of a halo with
$10^{10}\,M_{\odot}\leq M_{\rm h}<10^{12}\,M_{\odot}$.

By contrast, for the more massive haloes with $M_{\rm h}\geq
10^{12}\,M_{\odot}$ shown in the lower right panel of
Fig.~\ref{fig:map}, only $11.7$ ($1.4$) per cent of identified \HI
absorbers with $10^{13.3}\rm\,cm^{-2}\leq N_{\rm HI}\leq
10^{14.5}\rm\,cm^{-2}$ are within $r<3R_{\rm vir}$ ($r<R_{\rm vir}$)
of a halo.  Interestingly, the fraction of gas with $r<3R_{\rm vir}$
now has the opposite trend with $N_{\rm HI}$ for the most massive
haloes, where there are now relatively \emph{fewer} absorbers in the
vicinity of haloes as $N_{\rm HI}$ increases.  This is due to the
presence of hot, collisionally ionised gas around the more massive
haloes, associated with shocks from gravitational infall and AGN
feedback.

The physical properties of the \Lya forest absorbers are explored
further in Fig.~\ref{fig:bN}, where the Doppler parameters, $b$, and
\HI column densities, $N_{\rm HI}$, obtained from fitting mock spectra
with Voigt profiles are displayed in the $b$--$N_{\rm HI}$ plane.  In
each panel, the colour scale shows the (optical depth weighted) gas
density, gas temperature, baryon phase, and the ratio $b/b_{\rm
  therm}$ (where $b_{\rm therm}=(2k_{\rm B}T/m_{\rm H})^{1/2}$ is the
thermal line width) associated with the absorbers.  The baryon phase
definitions follow those introduced by \cite{Dave2010} (see
Table~\ref{tab:phases} for details).  The absorption lines enclosed by
the dashed lines (i.e. for $10^{13.3}\rm\,cm^{-2}\leq N_{\rm HI}\leq
10^{14.5}\rm\,cm^{-2}$, $20\rm\,km\,s^{-1}\leq b \leq
90\rm\,km\,s^{-1}$) are associated with predominantly photoionised gas
with a median gas density $\log \Delta = 1.14$, median temperature
$T=38\,380\rm\,K$ and median $b/b_{\rm therm}=1.28$. Note that
$b/b_{\rm therm}$ is consistent with curve-of-growth analyses that use
higher order Lyman series lines to separate unresolved \Lya
components.  For example, \citet{Danforth2010} perform a
curve-of-growth analysis on 164 \Lya absorbers using HST/STIS data,
finding a median ratio $b_{\rm Ly\alpha}/b_{\rm
  cog}=1.26^{+0.49}_{-0.25}$ \citep[see
  also][]{Shull2000,Danforth2008}.  In general, the larger column
densities and Doppler parameters are associated with higher gas
densities and temperatures, respectively.  The majority of the
absorbers ($90$ per cent) with $10^{13.3}\rm\,cm^{-2}\leq N_{\rm
  HI}\leq 10^{14.5}\rm\,cm^{-2}$ exhibit suprathermal line widths and
$83$ per cent are associated with the diffuse IGM.  Only the absorbers
close to the lower boundary of the $b$--$N_{\rm HI}$ plane have
thermal widths; most of the other \Lya lines are also broadened by
peculiar motions and the blending of unresolved \Lya components, gas
pressure (Jeans) smoothing and expansion with the Hubble flow.

Finally, in Table~\ref{tab:phases} we summarise the percentage of \Lya
absorbers associated with the four gas phases (diffuse IGM, WHIM, hot
halo, and condensed) defined by \cite{Dave2010} are listed for the
AGN, StrongAGN, H02 (i.e.  the Quick-\Lya simulation with no feedback
or star formation) and Blazar models.  Note that a similar fraction of
the \Lya absorbers are associated with each baryon phase in all four
simulations.  This may be contrasted with the baryon mass fraction in
each phase, shown in the parentheses in Table~\ref{tab:phases}.  In
particular, the mass fraction associated with the WHIM increases
dramatically in the StrongAGN model.  However, this has a very limited
effect on the fraction of \emph{absorbers} in each phase, since most
of this gas is too hot and/or low density to produce \Lya absorption
lines \citep{Viel2017}.

\subsection{Comparison of the models to COS data} \label{sec:obs}

We now perform an initial comparison of the simulated spectra to
measurements of the column density distribution function (CDDF) and
Doppler parameter distribution.  The CDDF and Doppler parameter
distribution observations were first presented in \cite{Viel2017}.  In
this work, the $1\sigma$ uncertainties on the measurements are
obtained using $10^{3}$ bootstrap samples with replacement.

The AGN, StrongAGN, Blazar and H02 simulations are compared directly
to the COS data in Fig.~\ref{fig:phystest}.  Following
\citet{Viel2017}, the \HI column densities from each model have been
rescaled by a constant to match the amplitude of the observed CDDF at
$10^{13.3}\rm\,cm^{-2}\leq N_{\rm HI}\leq 10^{14.5}\rm\,cm^{-2}$.
Assuming the \Lya absorbers are in photoionisation equilibrium, this
is equivalent to a rescaling of the \HI photoionisation rate
$\Gamma_{\rm HI}$, since $N_{\rm HI} \propto \Gamma_{\rm HI}^{-1}$.
We verify in Appendix~\ref{app:sys} that this approximation is a good
one for modelling the CDDF at $10^{13.3}\rm\,cm^{-2}\leq N_{\rm
  HI}\leq 10^{14.5}\rm\,cm^{-2}$ (although see also
\citet{Khaire2019pk}, who find this is not the case for the \Lya
forest power spectrum at $z<0.5$.  We speculate this may be due to
including absorbers with $N_{\rm HI}>10^{14.5}\rm\,cm^{-2}$, and/or
the influence of hot gas that produces absorption near the continuum
that is too weak and/or broad to be identified reliably with VPFIT).
The photoionisation rate required to match the observed
CDDF\footnote{We choose to match the CDDF over a limited range in
$N_{\rm HI}$, rather than matching the mean or distribution of the
\Lya forest transmission, as the CDDF is less susceptible to
systematic uncertainties associated with signal-to-noise and the
uncertain properties of high density gas in the CGM.  Nevertheless, we
note the mean transmission associated with the rescaled CDDF -- given
in Table~\ref{tab:hydrosims} -- is in excellent agreement with
$\langle F \rangle=0.983\pm 0.009$ at $\langle z \rangle = 0.08$ from
\citet{Kim2021} and the flux decrement $D_{\rm A}(z)=1- \langle F
\rangle = 0.014(1+z)^{2.2\pm 0.2}$ from \citet{Shull2015}.},
$\Gamma_{\rm HI}^{\rm CDDF}$, is a factor of $0.4$--$0.6$ times the
P19 UV background model value at $z=0.1$, suggesting that the P19
model may overproduce the UV background at $z\sim 0.1$.  The
simulations with a higher fraction of hot gas, either from additional
physics (StrongAGN and Blazar) or from artificially increased
photoheating rates (e.g. H10) require the smallest $\Gamma_{\rm
  HI}^{\rm CDDF}$ as a result of the reduced recombination rate in the
hotter IGM (see the second last column of Table~\ref{tab:hydrosims}).

For comparison to the literature, we also calculate the ratio of
$\Gamma_{\rm HI}^{\rm CDDF}$ to the photoionisation rate at $z=0.1$
from the \citet[][hereafter HM12]{HaardtMadau2012} UV background
model; we list this in the final column of Table~\ref{tab:hydrosims}.
For absorbers with $10^{13.6}\rm\,cm^{-2}\leq N_{\rm HI}\leq
10^{14.4}\rm\,cm^{-2}$, \citet{Kollmeier2014} found that $\Gamma_{\rm
  HI}^{\rm CDDF}$ differed from the HM12 model by a factor of $\sim 5$
within their simulations.  \citet{Christiansen2020} recently revised
this discrepancy downward by a factor of $\sim 2$ by invoking IGM
heating associated with jet feedback in the SIMBA simulation.  This
revised estimate is in better agreement with the smaller factor of
$\sim 2$ discrepancy noted by other independent studies
\citep{Wakker2015,Shull2015,Gaikwad2017b,Viel2017}, although for
different reasons, as these models did not include jet feedback. In
this work we find a difference of at most a factor of $\sim 1.7$
compared to HM12 for our fiducial AGN feedback \emph{and} no feedback
models (H02), which is within the factor of $\sim 2$ uncertainty
expected in UV background models
\citep[e.g.][]{KhaireSrianand2015,FaucherGiguere2020}. The additional
hot gas in the StrongAGN model decreases the $\Gamma_{\rm HI}^{\rm
  CDDF}$ by a further factor of $\sim 2$ relative to our fiducial AGN
and H02 simulations, which is also in good agreement with the relative
change found by \citet{Christiansen2020} when comparing the SIMBA jet
and no-jet feedback models.\footnote{Note the mass fraction in the
diffuse IGM and WHIM for the StrongAGN (AGN) models are very similar
to the Jet (No-Jet) models presented by \citet{Christiansen2020} which
have 16.4 (38.8) per cent and 70.5 (28.7) per cent of the baryon mass
in the diffuse IGM and WHIM, respectively (cf. Table~\ref{tab:phases}
in this work).}  Sufficiently potent AGN feedback that heats the low
density IGM can therefore relax the requirement on the number of
ionising photons that was first pointed out by \cite{Kollmeier2014},
although in contrast to \citet{Christiansen2020}, we find this
additional heating is not necessarily required.  Blazar heating has a
qualitatively similar effect on the required $\Gamma_{\rm HI}^{\rm
  CDDF}$ to strong AGN feedback, as was also noted by
\cite{Kollmeier2014} (see their fig. 3).

Following the rescaling of the column densities, all four models in
the left panel of Fig.~\ref{fig:phystest} are in good agreement
(i.e. within $\sim 1$--$1.5 \sigma$) with the observed CDDF at
$10^{13.3}\rm\,cm^{-2}\leq N_{\rm HI}\leq 10^{14.5}\rm\,cm^{-2}$.
Note again that at column densities $N_{\rm HI}\lesssim
10^{13.3}\rm\,cm^{-2}$ the CDDF is incomplete and is dependent on the
assumed signal-to-noise (see Appendix~\ref{app:sys}).  Comparing the
H02 and StrongAGN models, we find that strong AGN feedback can alter
the shape and amplitude of the CDDF.  This is in agreement with
earlier work by \citet{Gurvich2017}, although differences between the
models are within $\sim 1 \sigma$ at the column densities, $N_{\rm
  HI}\geq 10^{13.3}\rm\,cm^{-2}$, where the COS data are complete.
With a higher precision measurement of the CDDF, it may therefore be
possible to use the slope of the CDDF around $N_{\rm HI}\sim
10^{14}\rm\,cm^{-2}$ to distinguish between different feedback models.
We note, however, that the differences between H02 (no feedback) and
our fiducial AGN model are very small at $N_{\rm HI}\geq
10^{13.3}\rm\,cm^{-2}$, and are comparable to those found by
\citet{Nasir2017} and \citet{Viel2017}, reiterating their finding that
AGN feedback plays a negligible role in changing the amplitude of the
CDDF \emph{for our fiducial AGN feedback model}. We therefore conclude
that the extent to which AGN feedback and galaxy formation impact on
the CDDF will depend on the specifics of the (uncertain) sub-grid
modelling.  In contrast our earlier work \citep{Viel2017,Nasir2017},
however, we find better agreement between the simulated and observed
CDDF at $10^{14}\rm\,cm^{-2}<N_{\rm HI}\leq 10^{14.5}\rm\,cm^{-2}$.
The main reason for this improvement is that the simulations in this
work have higher mass resolution and include a correction to the CDDF
for both box size and mass resolution (see
Appendix~\ref{app:converge}).

Finally, in the right panel of Fig.~\ref{fig:phystest} we show the
Doppler parameter distribution.  As has been noted previously, the
simulated line widths are narrower than the observations
\citep{Gaikwad2017,Nasir2017}. The observed Doppler parameter
distribution peaks in the bin at $b=27.5\pm 2.5 \rm\,km\,s^{-1}$
\citep[see also][]{Danforth2010,Danforth2016}, whereas the simulated
distributions all peak at $b=22.5\pm 2.5\rm\,km\,s^{-1}$,
corresponding to thermally broadened lines with $T\simeq
10^{4.5}\rm\,K$.  More quantitatively, the AGN (H02) simulations
exceed the observed distribution by $4.6\sigma$ ($5.8\sigma$) at
$b=22.5\pm 2.5\rm\,km\,s^{-1}$.  Interestingly, this discrepancy also
holds for the StrongAGN and Blazar heating models.  The explanation
for this was outlined in \citet{Viel2017}; the additional hot gas in
the WHIM does not produce \Lya absorption at the necessary column
densities, since that gas is typically hot ($T\sim 10^{6}\rm\,K$) and
collisionally ionised. Hence, while strong AGN feedback that
efficiently heats the low density IGM can alleviate differences
between the observed and simulated CDDF and the mean transmission
$\langle F \rangle$ \citep[in agreement with][]{Christiansen2020},
unless this feedback is tuned to produce gas at just the right density
and temperature, it will not reproduce the number of lines with
$b\lesssim 30\rm\,km\,s^{-1}$ in the \Lya forest at $z=0.1$
\citep{Viel2017}.  Lastly, the black dashed curve in
Fig.~\ref{fig:phystest} provides a crude estimate of the additional
line broadening required for consistency with the COS data at
$b\lesssim 30\rm\,km\,s^{-1}$, where we show that a factor $1.24$
increase to the H02 model line widths achieves much better agreement.
A $1.5\sigma$ discrepancy at $b\sim 42.5 \pm 2.5 \rm\,km\,s^{-1}$
remains, possibly indicating there is also a lack of $T\sim
10^{5}\rm\, K$ gas in the simulations.  As we demonstrate later,
however, this difference does not preclude a statistically acceptable
fit to the data.


\section{Thermal and turbulent line broadening} \label{sec:heat_turb}
\subsection{The thermal asymptote at $z \leq 0.5$} \label{sec:T_asymp}

\begin{figure*}

  \begin{minipage}{1.0\textwidth}
\centering
\includegraphics[width=0.495\textwidth]{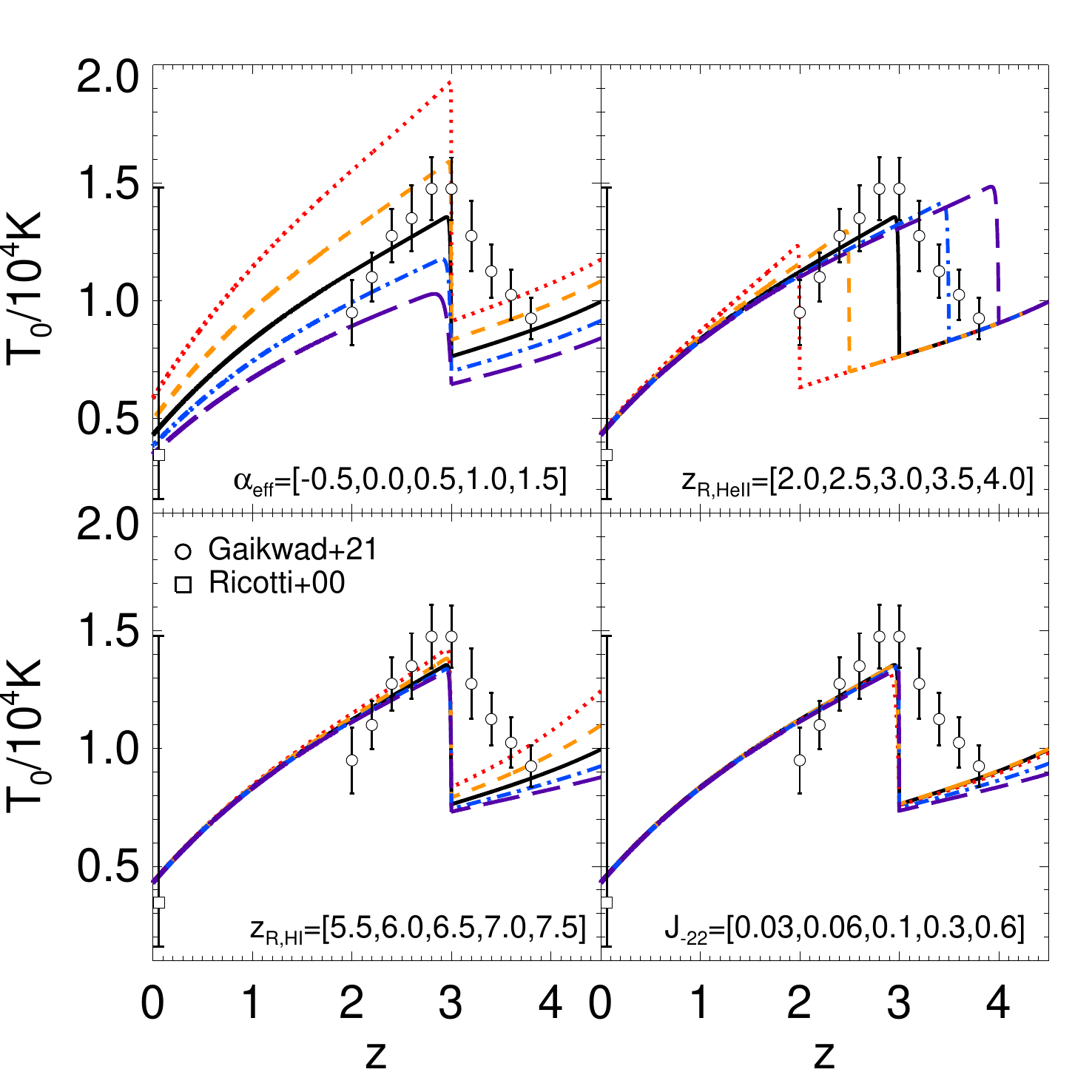}
\includegraphics[width=0.495\textwidth]{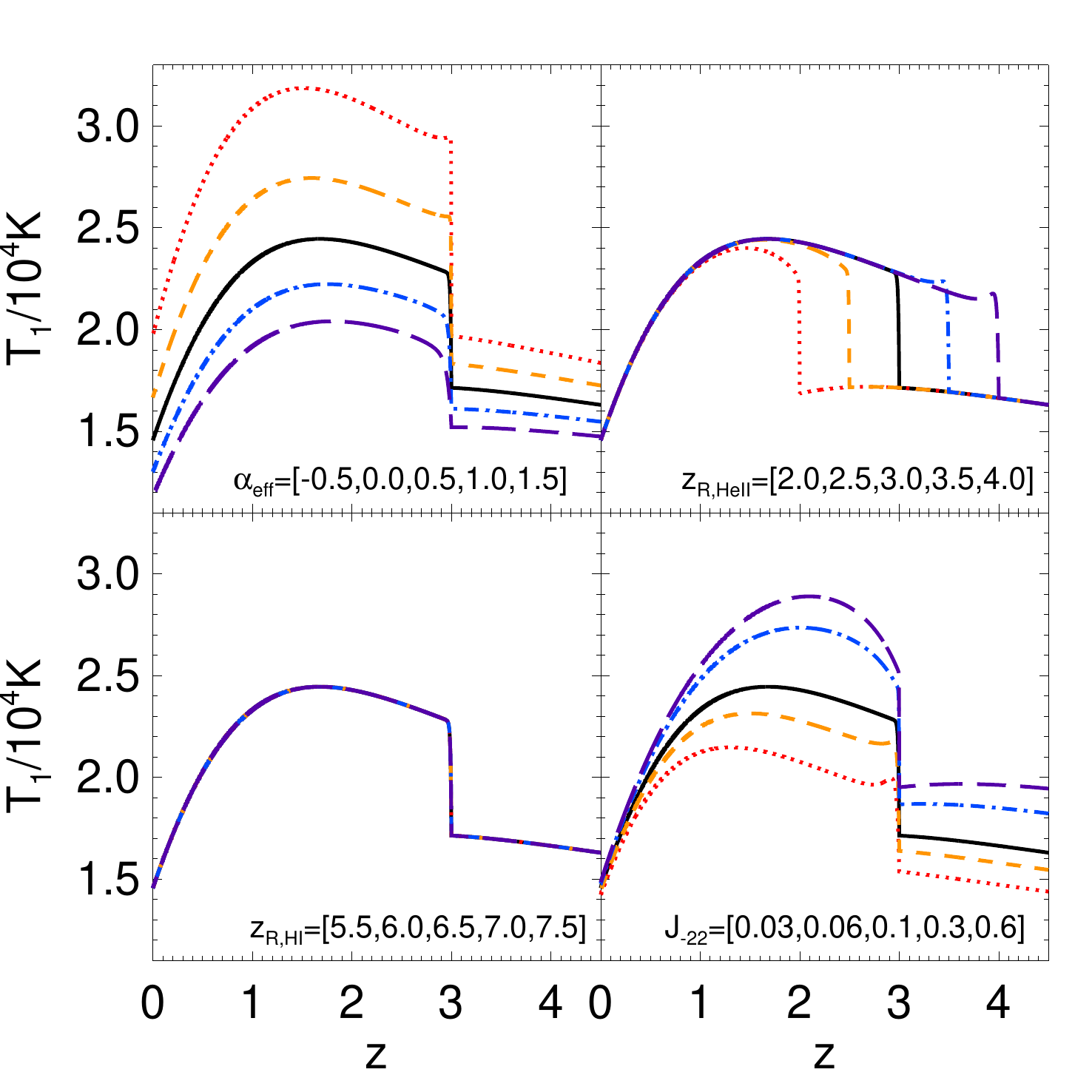}
  \vspace{-0.3cm}
\end{minipage}
  \caption{\emph{Left:} Temperature evolution with redshift for a gas
    parcel at the mean density, $\Delta=1$, that has been photoionised
    and heated by a power-law UV spectrum, $J_{\rm E}\propto
    E^{-\alpha_{\rm eff}}$.  Individual panels show the effect of
    varying different parameters on $T_{0}$. Clockwise from the top
    left, these are: the effective power-law spectral index,
    $\alpha_{\rm eff}$, the redshift of \HeII reionisation, $z_{\rm R,
      HeII}$, the specific intensity at the Lyman limit,
    $J_{-22}=J_{13.6\rm\,eV}/10^{-22}\rm\,erg\,s^{-1}\,cm^{-2}\,sr^{-1}\,Hz^{-1}$,
    and the redshift of \HI reionisation, $z_{\rm R, HI}$.  The
    central parameter values, corresponding to the black curves in
    each sub-panel, are $\alpha_{\rm eff}=0.5$, $z_{\rm R,HI}=6.5$,
    $z_{\rm R,HeII}=3.0$ and $J_{-22}=0.1$.  Note that the temperature
    at $z\leq 0.5$ depends only on the spectral index.  For
    comparison, temperature measurements from the \Lya forest at
    $2\leq z\leq 3.8$ are shown as open circles \citep{Gaikwad2021},
    and at $z=0.06$ by an open square
    \citep{Ricotti2000}. \emph{Right:} As for the left panel, but now
    for a gas parcel with $\Delta=10$, similar to the densities
    typically probed by the $z=0.1$ \Lya forest.  Note the different
    scale on the vertical axis compared to the left panel.}
  \label{fig:tevol}
\end{figure*}

\begin{figure}
\begin{center}
  \includegraphics[width=0.47\textwidth]{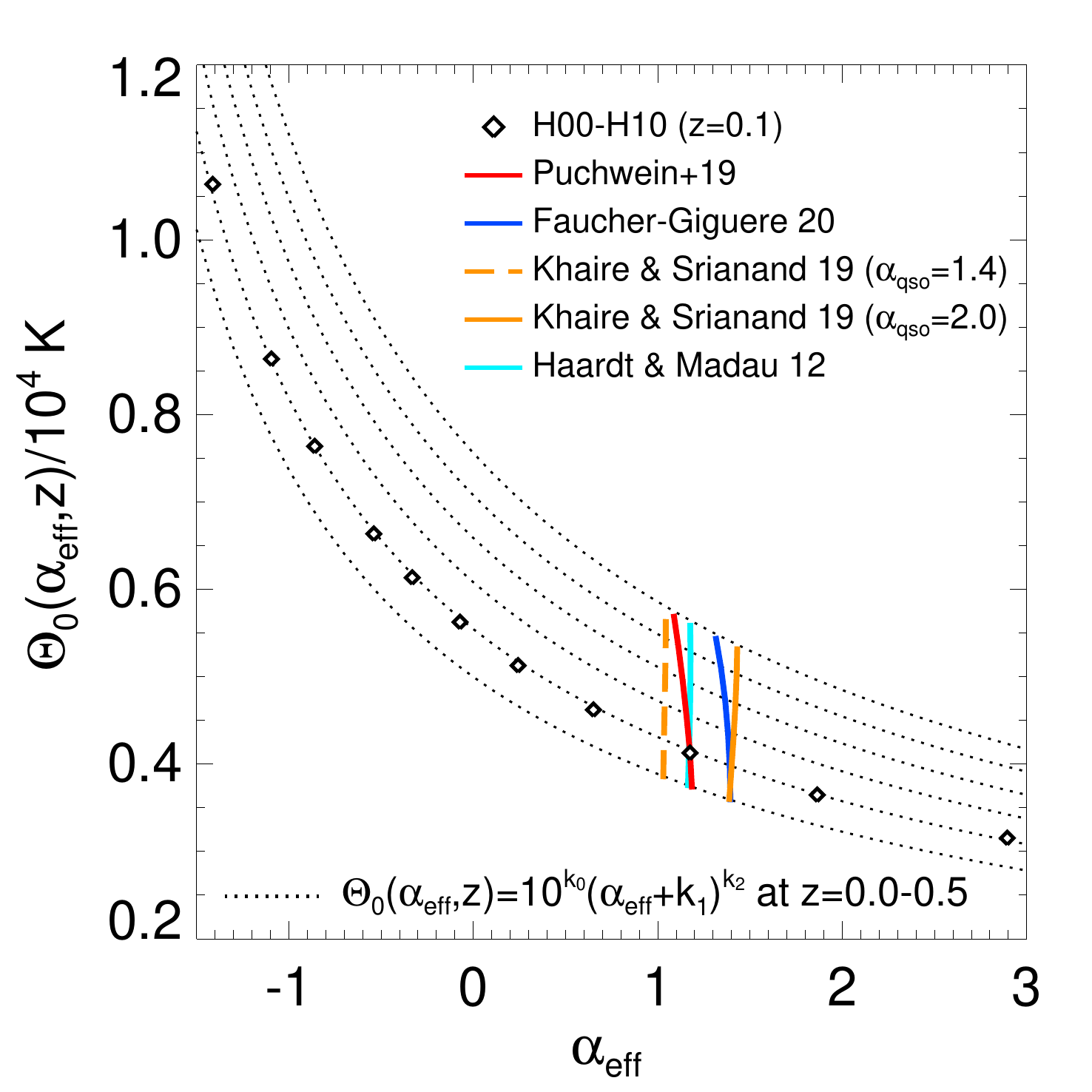}
  \vspace{-0.3cm}
  \caption{The temperature at the mean density along the thermal
    asymptote, $\Theta_{0}$, as a function of the effective power-law
    spectral index, $\alpha_{\rm eff}$, for a hydrogen and helium gas
    parcel heated by a power-law ionising spectrum with specific
    intensity $J_{\rm E}\propto E^{-\alpha_{\rm eff}}$.  The dotted
    curves are obtained at intervals of $\Delta z =0.1$ from $z=0.0$
    (lower curve) to $z=0.5$ (upper curve) using
    Eq.~(\ref{eq:T0fit}). The black diamonds assume $\Theta_{0}=T_{0}$
    at $z=0.1$ for the hydrodynamical simulations $\rm H00$--$\rm H10$
    listed in Table~\ref{tab:hydrosims}.  The solid and dashed curves
    show the thermal asymptote predicted by various UV background
    models: \citet{Puchwein2019} (solid red curve),
    \citet{FaucherGiguere2020} (solid blue curve),
    \citet{KhaireSrianand2019} for a quasar contribution with spectral
    index $\alpha_{\rm qso}=1.4$ (dashed orange curve) or $\alpha_{\rm
      qso}=2.0$ (solid orange curve), and \citet{HaardtMadau2012}
    (solid cyan curve).  The empirically calibrated UV background
    models all predict thermal asymptotes in a narrow range,
    consistent with $\alpha_{\rm eff}=1.0$--$1.4$.}
  \label{fig:alphabk}
\end{center}
\end{figure}

Since we argue that our AGN feedback models fail to explain the
observed line widths in the \Lya forest at $z\simeq 0.1$, we now turn
to consider alternatives.  One possible explanation for the
discrepancy between the simulated and observed Doppler parameter
distribution is increased photoheating associated with a hardening of
the UV background spectrum.  The resulting enhancement to the IGM
temperature produces additional thermal broadening in the \Lya forest.
Indeed, additional photoheating is typically invoked to explain the
IGM temperature boost inferred from the \Lya forest at $z \simeq 3$
\citep{Schaye2000,Ricotti2000,Becker2011,Walther2019}, when the rising
contribution to the ionising emissivity from quasars is thought to
harden the UV background spectrum and drive \HeII reionisation
\citep{Theuns2002,Bolton2009,Puchwein2015,UptonSanderbeck2016}.  An
interesting question is then: how hard would the UV background
spectrum need to be to achieve consistency between the observed COS
line widths and theoretical models?  More importantly, is the required
spectral hardening physically plausible?

In the canonical model, the IGM is expected to follow a power-law
temperature-density relation following reionisation,
$T=T_{0}\Delta^{\gamma-1}$, with some additional scatter around this
relation due to shock heating \citep{HuiGnedin1997,McQuinn2016}.  This
arises due to photoheating and adiabatic cooling for densities
$\Delta\lesssim 10$.  At late times the slope of this relation
approaches $T\propto \Delta^{1/1.72}\simeq \Delta^{0.58}$, where the
exponent $\gamma-1=0.58$ arises through the temperature dependence of
the case-A recombination coefficient for hydrogen, $\alpha_{\rm
  A}\propto T^{-0.72}$ \citep[see
  e.g.][]{MiraldaEscudeRees1994,McQuinn2016}.  The temperature-density
relation furthermore retains no memory of its earlier reionisation and
heating history $\Delta z\sim 1$--$2$ after \HI reionisation at
$z\simeq 6$ or \HeII reionisation at $z\simeq 3$.  Once this ``thermal
asymptote'' is reached and the IGM is in photoionisation equilibrium,
the IGM temperature is set only by the spectral shape of the UV
background \citep{HuiHaiman2003}, and is independent of the UV
background intensity.

This is further illustrated in the Fig.~\ref{fig:tevol}, where we
perform a non-equilibrium ionisation calculation to obtain the
temperature of a hydrogen and helium gas parcel that is reionised by a
time independent UV background.  The results are displayed at two gas
densities, $\Delta=1$ (left panels) and $\Delta=10$ (right panels),
i.e. at the mean density and the density where most of the \Lya forest
absorption at $z=0.1$ occurs for $10^{13.3}\rm\,cm^{-2}\leq N_{\rm
  HI}\leq 10^{14.5}\rm\,cm^{-2}$.  We include the effect of
photoionisation and photoheating, collisional ionisation, radiative
cooling, adiabatic cooling and secondary ionisations by fast
electrons, following the equations given in appendix B of
\cite{Soltinsky2021}.  The spectrum of the ionising radiation is
treated as a single power-law, $J_{\rm E}\propto E^{-\alpha_{\rm
    eff}}$, with effective\footnote{We refer to this as an effective
spectral index, as in reality the spectral shape of the UV background
will be modified by intervening absorption and recombination emission
from the IGM, even if the intrinsic spectrum emitted by the ionising
source population is a pure power-law.  See e.g. the right panel of
Fig.~\ref{fig:obsvals}, which we discuss later. } spectral index
$\alpha_{\rm eff}$.  For the photons with energies $13.6\rm\,eV\leq E
< 54.4\rm\,eV$ that are capable of \HI and \HeI ionisation, we assume
a specific intensity
\begin{equation} J_{\rm E}=
  \begin{cases} J_{13.6\rm\,eV}\left(\frac{E}{13.6\rm\,eV}\right)^{-\alpha_{\rm eff}} & \mbox{for } z<z_{\rm R,HI}, \\
   0 & \mbox{for } z\geq z_{\rm R,HI}, \\
\end{cases}
\label{eq:JE}
\end{equation}
where $z_{\rm R,HI}$ is the redshift of (instantaneous) \HI and \HeI
reionisation.  For the higher energy photons, $E\geq 54.4\rm\,eV$,
responsible for \HeII reionisation, we instead assume
\begin{equation} J_{\rm E}=
\begin{cases} J_{13.6\rm\,eV}\,4^{-\alpha_{\rm eff}} \left(\frac{E}{54.4\rm\,eV}\right)^{-\alpha_{\rm eff}} & \mbox{for } z<z_{\rm R,HeII},\\
  0 & \mbox{for } z\geq z_{\rm R,HeII}, \\
\end{cases}
\label{eq:JE2}
\end{equation}
\noindent
where $z_{\rm R,HeII}$ is the redshift of (instantaneous) \HeII
reionisation.  

Clockwise from the top left in each sub-panel of Fig.~\ref{fig:tevol},
we vary the spectral index $\alpha_{\rm eff}$, the redshift of \HeII
reionisation $z_{\rm R,HeII}$, the intensity of the UV background at
the \HI Lyman limit, $J_{-22}=J_{13.6\rm\, eV}/10^{-22}
\rm\,erg\,s^{-1}\,cm^{-2}\,Hz^{-1}\,sr^{-1}$, and the redshift of \HI
reionisation, $z_{\rm R,HI}$.  We adopt ranges for each parameter that
bracket the plausible values: the bulk of \HI and \HeII reionisation
should be complete by $z\simeq 6$ \citep{Yang2020} and $z\simeq 3$
\citep{Worseck2019}, respectively, while typical values in UV
background models at $z=0.1$ are $\alpha_{\rm eff}\simeq 1.0$--$1.4$
(see Fig.~\ref{fig:alphabk}) and $J_{-22}\simeq 0.2$
\citep{Puchwein2019,KhaireSrianand2019,FaucherGiguere2020}.

The results for $\Delta=1$ in Fig.~\ref{fig:tevol} (left panel)
encompass the recent IGM temperature measurements at the mean density,
$T_{0}$, from \citet{Gaikwad2021}, who use a variety of different
statistical measures of the \Lya forest transmitted flux to obtain
joint constraints on $T_{0}$ and $\gamma$ at $2\leq z\leq 3.8$.  The
models are also consistent with the only IGM temperature measurement
to date from the \Lya forest at $z\simeq 0$ \citep{Ricotti2000},
although the error bars on this measurement are large (see their
fig. 12).  Note, however, our aim is not to match these data points; a
single zone reionisation model will in any case not correctly capture
the volume averaged gas temperature during inhomogeneous \HeII
reionisation at $z\simeq 3$ \citep[see
  e.g.][]{UptonSanderbeck2016,UptonSanderbeck2020}.  The important
point, as we now discuss below, is that this approach captures the
\emph{late time behaviour} of the IGM temperature when it tracks the
thermal asymptote \citep[cf.][]{HuiHaiman2003}.

First, note that Fig.~\ref{fig:tevol} shows the gas temperatures in
the low density IGM at $z<0.5$ follow a single thermal asymptote, as
expected.  This implies a temperature sensitive statistic, such as the
Doppler parameter distribution, should provide an excellent constraint
on the spectral shape of the UV background at $z\simeq 0$ if the low
column density \Lya forest is primarily thermally broadened by
photoheating.\footnote{Note, however, that in photoionisation
equilibrium the gas temperature at densities $\Delta \gtrsim 10$ will
become increasingly sensitive to the spectral shape \emph{and}
specific intensity of the UV background, as the dominant gas cooling
mechanism transitions from adiabatic to radiative cooling.  In this
regime, lowering $J_{-22}$ produces colder gas temperatures, since the
\HI collisional excitation cooling rate scales as $\sim n_{\rm HI}$.
For gas with $\Delta=100$ at $z\sim 0.1$ ($N_{\rm HI}\sim
10^{15}\rm\,cm^{-2}$) the range of temperatures that result from
$J_{-22}=0.03$--$0.6$ in our single zone model becomes comparable to
the range of temperatures for $\alpha_{\rm eff}=-0.5$--$1.5$.}  It is
convenient to provide a fit to the temperature at the thermal
asymptote at the mean density, $\Theta_{0}$, as a function of
$\alpha_{\rm eff}$ at $0\leq z\leq 0.5$.  Our numerical results from
the non-equilibrium ionisation calculations are well approximated by
\begin{equation} \Theta_{0}(\alpha_{\rm eff},z) = 10^{k_{0}(z)}\rm\,K\, [\alpha_{\rm eff}+k_{1}(z)]^{k_{2}(z)}, \label{eq:T0fit} \end{equation}
\noindent
where the redshift dependent coefficients $k_{0}$, $k_{1}$ and $k_{2}$
are listed in Table~\ref{tab:fit}.  Eq.~(\ref{eq:T0fit}) reproduces
our numerical calculation to within $1$ per cent for $-1.5\leq
\alpha_{\rm eff}\leq 3.0$.  The thermal asymptote at $\Delta=10$ is
furthermore similarly well approximated by $\Theta_{1}(\alpha_{\rm
  eff})=3.27 \Theta_{0}(\alpha_{\rm eff})$ at $z=0.1$.

\begin{table}
  \centering
  \caption{Coefficients for the best fit to the thermal asymptote,
    $\Theta_{0}$, obtained using Eq.~(\ref{eq:T0fit}) at redshifts
    $0.0\leq z \leq 0.5$ for a power-law UV background spectrum with
    $J_{\rm E}\propto E^{-\alpha_{\rm eff}}$.  The fits reproduce the
    numerical calculation to within $1$ per cent for $-1.5\leq
    \alpha_{\rm eff}\leq 3.0$.}
  \begin{tabular}{c|c|c|c}
    \hline
    Redshift, $z$ & $k_{0}$ & $k_{1}$ & $k_{2}$ \\
    \hline
    $0.0$ &  $3.976$ &  $2.412$ & $-0.725$ \\ 
    $0.1$ &  $4.022$  & $2.413$ & $-0.727$ \\
    $0.2$ &  $4.063$ &  $2.414$ & $-0.729$ \\ 
    $0.3$ &  $4.099$  & $2.415$ & $-0.732$ \\ 
    $0.4$ &  $4.131$ &  $2.416$ & $-0.734$ \\
    $0.5$ &  $4.161$ &  $2.417$ & $-0.737$ \\
    \hline
  \end{tabular}
  \label{tab:fit}
\end{table}

We can therefore use Eq.~(\ref{eq:T0fit}) to estimate the effective
power-law spectra indices, $\alpha_{\rm eff}$, that are
\emph{equivalent} to the spectral shape assumed in different synthesis
models for the metagalactic UV background
\citep{HaardtMadau2012,KhaireSrianand2019,Puchwein2019,FaucherGiguere2020}.
In Fig.~\ref{fig:alphabk}, the black dotted curves show
Eq.~(\ref{eq:T0fit}) at $z=0$--$0.5$.  The $\Theta_{0}$ values
predicted at $z=0.0$--$0.5$ by the UV background synthesis models are
shown by the (roughly vertical) solid and dashed curves in
Fig.~\ref{fig:alphabk}, where we have again calculated the temperatures
using our non-equilibrium ionisation code.  For comparison, the black
diamonds show $\Theta_{0}\simeq T_{0}$ at $z=0.1$ for the
hydrodynamical simulations H00-H10 listed in
Table~\ref{tab:hydrosims}.

Note that the gas temperatures at $\Delta=1$ predicted by the UV
background models are all within a very narrow temperature range,
$\Theta_{0}\simeq 3950\rm\,K$--$4250\rm\,K$ ($\Theta_{1}\simeq
12920\rm\,K$--$13900\rm\,K$) at $z=0.1$, which from
Eq.~(\ref{eq:T0fit}) is equivalent to a spectrum with an effective
power-law slope of $\alpha_{\rm eff}=1.0$--$1.4$.  This is expected,
as at least half (and possibly all, depending on the assumed spectral
shape) of the contribution to the UV background at $z\simeq 0$ is
thought to be from quasars and active galactic nuclei
\citep[e.g.,][]{Kulkarni2019,Shen2020}.  The UV background models from
\citet{KhaireSrianand2019} adopt the hardest (softest)
\emph{intrinsic} quasar spectrum of $\alpha_{\rm qso}=1.4$
($\alpha_{\rm qso}=2.0)$, shown by the dashed (solid) orange curves in
Fig.~\ref{fig:alphabk}.  We find the gas temperatures predicted by
these models correspond to hardened\footnote{This hardening is due to
the combined effect of the \HeII opacity of the IGM and the rapidly
declining UVB emissivity at $z<1$.  The mean free path at the \HeII
ionisation edge ($E=54.4\rm\,eV$) is $\sim 500 \rm\,pMpc$ by $z\sim 1$
\citep[see fig. 1 in][]{Puchwein2019}, so a non-negligible fraction of
the $z\sim 0$ UVB at $E\sim 54.4\rm\,eV$ is produced by redshifted
photons emitted at higher energies.  The average excess energy per
\HeII photoionisation is therefore significantly increased above that
expected for an optically thin IGM at $z\sim 0$ \citep[see fig. 3
  in][]{Puchwein2019}.  By contrast, the UVB spectrum between
$13.6\rm\,eV<E<54.4\rm\,eV$ can instead slightly soften as a result of
IGM processing \citep{Shull2020}.}  effective spectral indices of
$\alpha_{\rm eff}=1.03$ and $\alpha_{\rm eff}=1.40$, respectively.
For comparison, the \citet{Puchwein2019} and
\citet{FaucherGiguere2020} models both assume $\alpha_{\rm qso}=1.7$,
but predict temperatures consistent with $\alpha_{\rm eff}=1.17$ and
$\alpha_{\rm eff}=1.40$, respectively; the precise amount of hardening
is dependent on the different IGM opacity models used in these
synthesis UV background models.  Finally, \citet{HaardtMadau2012}
assume $\alpha_{\rm qso}=1.57$, but their UVB model predicts
temperatures consistent with $\alpha_{\rm eff}=1.17$.  Overall, this
implies a modest spectral hardening of $\alpha_{\rm eff}-\alpha_{\rm
  qso} \simeq -0.5$ associated with the propagation of the ionising
photons through the IGM.

For comparison, \citet{Lusso2015} measure an extreme-UV spectral index
of $\alpha_{\rm qso}=1.70\pm 0.61$ at $600\,$\mbox{\AA }$\leq \lambda
\leq 912\,$\mbox{\AA } from a stacked spectrum of 53 luminous quasars
at $z=2.4$, while \citet{Stevans2014} obtain $\alpha_{\rm qso}=1.41\pm
0.15$ at $500\,$\mbox{\AA }$\leq \lambda \leq 1000\,$\mbox{\AA}, from
159 AGN with $\langle z \rangle=0.37$.  However, in an analysis of
$11$ AGN at $1.45\leq z\leq 2.14$, \citet{Tilton2016} find a harder
spectral index $\alpha_{\rm qso}=0.72\pm 0.26$ at $450\,$\mbox{\AA
}$\leq \lambda \leq 770\,$\mbox{\AA } \citep[see also][]{Scott2004}.
Hence, if taking the intrinsic quasar spectral index from
\citet{Stevans2014} (consistent with the $\alpha_{\rm qso}=1.4$ model
from \cite{KhaireSrianand2019}, a plausible hardness limit for quasar
dominated UV background models is approximately $\alpha_{\rm
  eff}\simeq 1$, yielding $\Theta_{0}\simeq 4300\rm\,K$
($\Theta_{1}\simeq 14\,050\rm\,K$) for the thermal asymptote at
$z=0.1$.  A more extreme case is possible if instead adopting the
harder \citet{Tilton2016} composite spectrum and once again assuming
some spectral hardening at the \HeII edge, such that $\alpha_{\rm
  eff}-\alpha_{\rm qso} \simeq -0.5$.  If taking the lower bound of
the \citet{Tilton2016} $1\sigma$ measurement, we obtain $\alpha_{\rm
  eff}\simeq 0$, yielding $\Theta_{0}\simeq 5550\rm\,K$
($\Theta_{1}\simeq 18\,150\rm\,K$).  We caution, however, the
\citet{Tilton2016} composite spectrum may be less reliable than
\citet{Stevans2014} due their use of low resolution COS/G140L spectra
and their small sample size.  In either case, as we will demonstrate
in Section~\ref{sec:results}, an IGM heated by a UV background
spectrum with $0<\alpha_{\rm eff}< 1$ would still be too cold to
reproduce the COS \Lya line width distribution in our simulations in
the absence of any additional non-thermal broadening.

\subsection{Unresolved non-thermal broadening} \label{sec:turb}

An alternative to increasing the thermal widths of the \Lya absorbers
is the introduction of (unresolved) non-thermal broadening in
hydrodynamical simulations of the IGM
\citep[e.g.][]{OppenheimerDave2009,Gaikwad2017}.  We can make a crude
estimate of the turbulent contribution needed to reproduce the COS
Doppler parameter distribution using the approach introduced by
\citet{OppenheimerDave2009}.  Let there be an unresolved
(i.e. sub-grid) turbulent component in the simulated \Lya absorbers,
where
\begin{equation} b_{\rm turb}^{2}=b_{\rm obs}^{2}-b_{\rm noturb}^{2}. \label{eq:bturb1} \end{equation}
\noindent
Here $b_{\rm turb}$ is the turbulent contribution to the Doppler
parameters, $b_{\rm obs}$ are the observed Doppler parameters in the
COS data, and $b_{\rm noturb}$ are the Doppler parameters obtained
from our simulated spectra.  Defining $\xi = b_{\rm obs}/b_{\rm
    noturb}$, Eq.~(\ref{eq:bturb1}) then becomes
  \begin{equation} b_{\rm turb} = (\xi^{2}-1)^{1/2}b_{\rm noturb}. \label{eq:bturb1b} \end{equation}
\noindent
We may now estimate the $b_{\rm turb}$ required to match the COS line
widths by taking the photoheating in our fiducial UVB model
($\alpha_{\rm eff}=1.17$) as an effective prior on the gas
temperature.  From the black dashed curve in the right panel of
Fig.~\ref{fig:phystest}, we will assume a boost to the line widths of
$\xi = b_{\rm obs}/b_{\rm noturb}\simeq 1.24$ will allow the AGN and
H02 simulations to approximately match the COS data.  From
Eq.~(\ref{eq:bturb1b}), this corresponds to a turbulent Doppler
parameter contribution of $b_{\rm turb}\simeq 0.73b_{\rm noturb}$.

It is also instructive to have an estimate for $b_{\rm turb}$ as a
function of column density, $N_{\rm HI}$.  Here we make use of the
fact that the \emph{narrowest} \Lya line widths associated with gas on
the temperature-density relation in the diffuse IGM at $z\simeq 0.1$
are thermally broadened (see Fig.~\ref{fig:bN} and associated
discussion).  Assuming $b_{\rm noturb}\simeq b_{\rm therm}$ for these
absorbers, Eq.~(\ref{eq:bturb1}) then becomes
\begin{equation} b_{\rm turb}  = (\xi^{2}-1)^{1/2}\left(\frac{2k_{\rm B}T}{m_{\rm H}}\right)^{1/2}. \label{eq:bturb} \end{equation}
\noindent
Next, assuming the size of \Lya forest absorbers is set by the local
Jeans scale in the IGM \citep{Schaye2001,Garzilli2015}, a power law
relationship between $N_{\rm HI}$ and density will hold,
\begin{equation} N_{\rm HI}  = N_{\rm HI,0}\Delta^{\beta}, \label{eq:NH1Delta} \end{equation}
\noindent
where $\beta=1.72-0.22\gamma$ for $T=T_{0}\Delta^{\gamma-1}$ and a
case-A recombination coefficient\footnote{The power-law approximation
for $\alpha_{\rm A}$ given here reproduces the more accurate fit from
\citet{VernerFerland1996} to within 10 per cent at $10^{3}\rm\,K \leq
T \leq 10^{5}\rm\,K$.} $\alpha_{\rm A}\simeq 4.06\times
10^{-13}\rm\,cm^{3}\,s^{-1}(T/10^{4}\rm\,K)^{-0.72}$.  For our AGN
simulation, we find a best fit value of $N_{\rm
  HI,0}=10^{12.12}\rm\,cm^{-2}$, assuming $\beta=1.37$ for
$\gamma=1.58$.  Combining Eq.~(\ref{eq:NH1Delta}) and
Eq.~(\ref{eq:bturb}), and once again assuming a power-law temperature
density relation, we obtain
\begin{equation} b_{\rm turb}= (\xi^{2}-1)^{1/2}\left(\frac{2k_{\rm B}T_{0}}{m_{\rm H}}\right)^{1/2}\left(\frac{N_{\rm HI}}{N_{\rm HI,0}}\right)^{(\gamma-1)/2\beta}, \label{eq:bturb2} \end{equation}
\noindent
Evaluating Eq.~(\ref{eq:bturb2}) for the AGN model at $N_{\rm
  HI}=10^{13.5}\rm\,cm^{-2}$ assuming $\xi=1.24$ then gives a density
dependent turbulent contribution of
\begin{equation} b_{\rm turb} \simeq 12.1\rm\,km\,s^{-1} \left(\frac{N_{\rm HI}}{10^{13.5}\rm\,cm^{-2}}\right)^{0.21}, \end{equation}
\noindent
for the \Lya absorbers with $10^{13.3}\rm\,cm^{-2}\leq N_{\rm HI}\leq
10^{14.5}\rm\,cm^{-2}$ associated with gas on the power-law
temperature-density relation. The required turbulent velocity
component along the line of sight is then $v_{\rm turb}=b_{\rm
  turb}/\sqrt{2}\simeq 8.5\rm\,km\,s^{-1}(N_{\rm
  HI}/10^{13.5}\rm\,cm^{-2})^{0.21}$. This gives an approximate upper
limit on the turbulent contribution to the narrowest, thermally
broadened \Lya lines arising from the coldest gas in the diffuse IGM
at $z\sim 0.1$.


\section{Results} \label{sec:results}

\subsection{Best fit $\Gamma_{\rm HI}$ and $\alpha_{\rm eff}$ for the UV background at $z=0.1$} \label{sec:bestfit}

\begin{figure*}
  \begin{minipage}{1.00\textwidth}
    \includegraphics[width=1.0\textwidth]{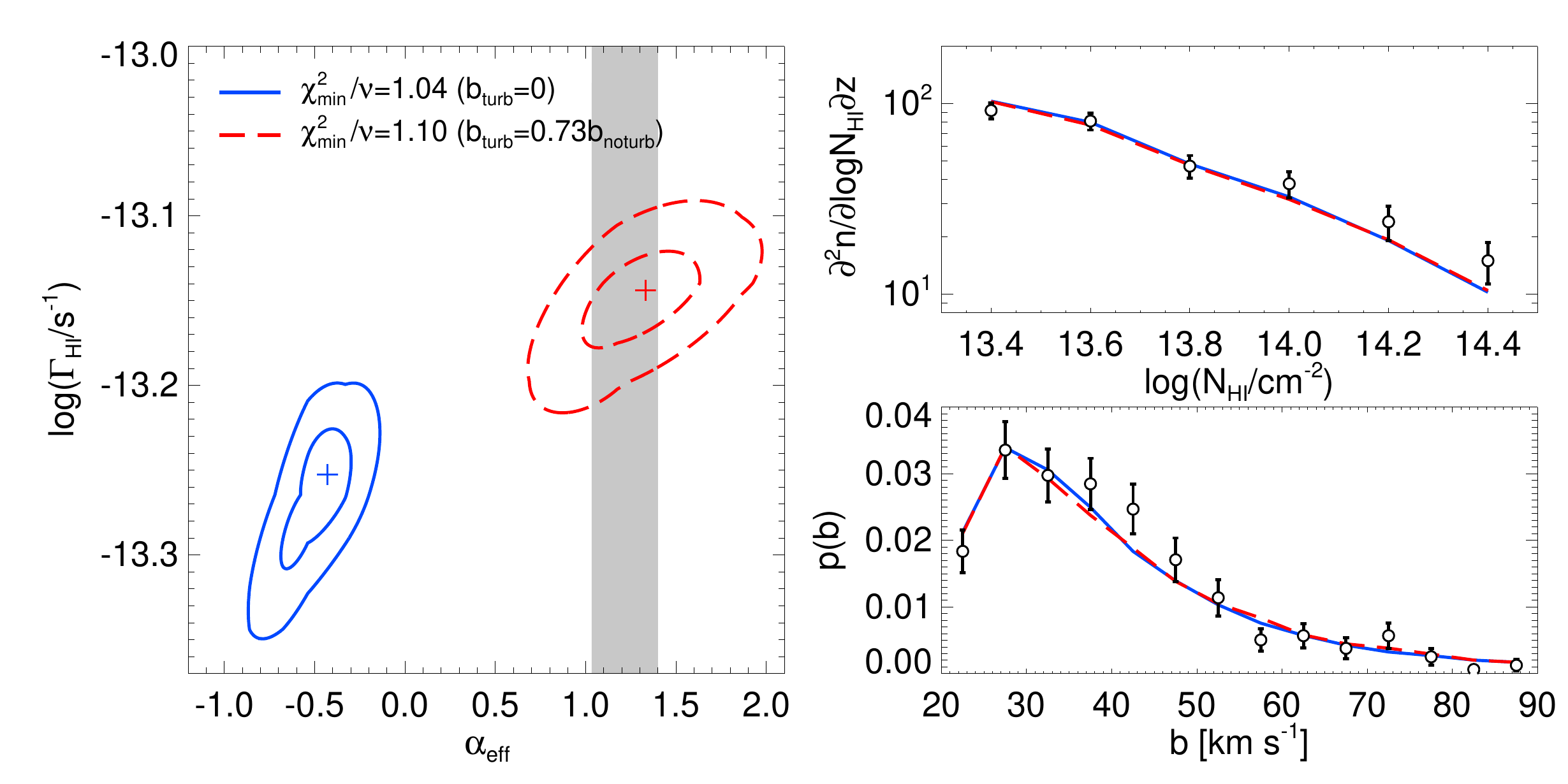}
  \end{minipage}
  \vspace{-0.4cm}
  \caption{{\it Left:} The projection of $\chi^{2}$ for the joint fit
    to the CDDF and Doppler parameter distribution including (red
    dashed curves) and excluding (blue solid curves) an additional
    turbulent contribution to the line widths (see
    Section~\ref{sec:turb} for details).  The contours show $\Delta
    \chi^{2}=\chi^{2}-\chi^{2}_{\min}=1$ and $4$, corresponding to the
    $1\sigma$ and $2\sigma$ confidence intervals for the individual
    parameters $\Gamma_{\rm HI}$ and $\alpha_{\rm eff}$.  The best fit
    model parameters (shown by the crosses) and $1\sigma$
    uncertainties are $\log(\Gamma_{\rm
      HI}/\rm\,s^{-1})=-13.25^{+0.03}_{-0.06}$ ($\log(\Gamma_{\rm
      HI}/\rm\,s^{-1})=-13.14^{+0.02}_{-0.03}$) and $\alpha_{\rm
      eff}=-0.43^{+0.13}_{-0.26}$ ($\alpha_{\rm
      eff}=1.33^{+0.30}_{-0.35}$) when excluding (including) the
    turbulent contribution to the line widths.  For comparison, the
    grey shaded region shows the $\alpha_{\rm eff}$ range consistent
    with the selection of UV background models shown in
    Fig.~\ref{fig:alphabk}, $\alpha_{\rm eff}\simeq 1.0$--$1.4$.  Note
    that \emph{the turbulent contribution has been deliberately
    calibrated} for consistency with the grey band.  {\it Right:} The
    best fitting models to the CDDF and Doppler width distribution,
    compared to the COS data.  The best fits have $\chi^{2}/\nu=1.04$
    ($1.10$) for $\nu=18$ degrees of freedom excluding (including) the
    turbulent contribution, with a probability that the $\chi^{2}$ may
    be exceeded randomly of $p=0.41$ ($p=0.35$).}
  \label{fig:chisqfit} 
\end{figure*}

\begin{figure*}
  \begin{minipage}{1.00\textwidth}
    \includegraphics[width=0.49\textwidth]{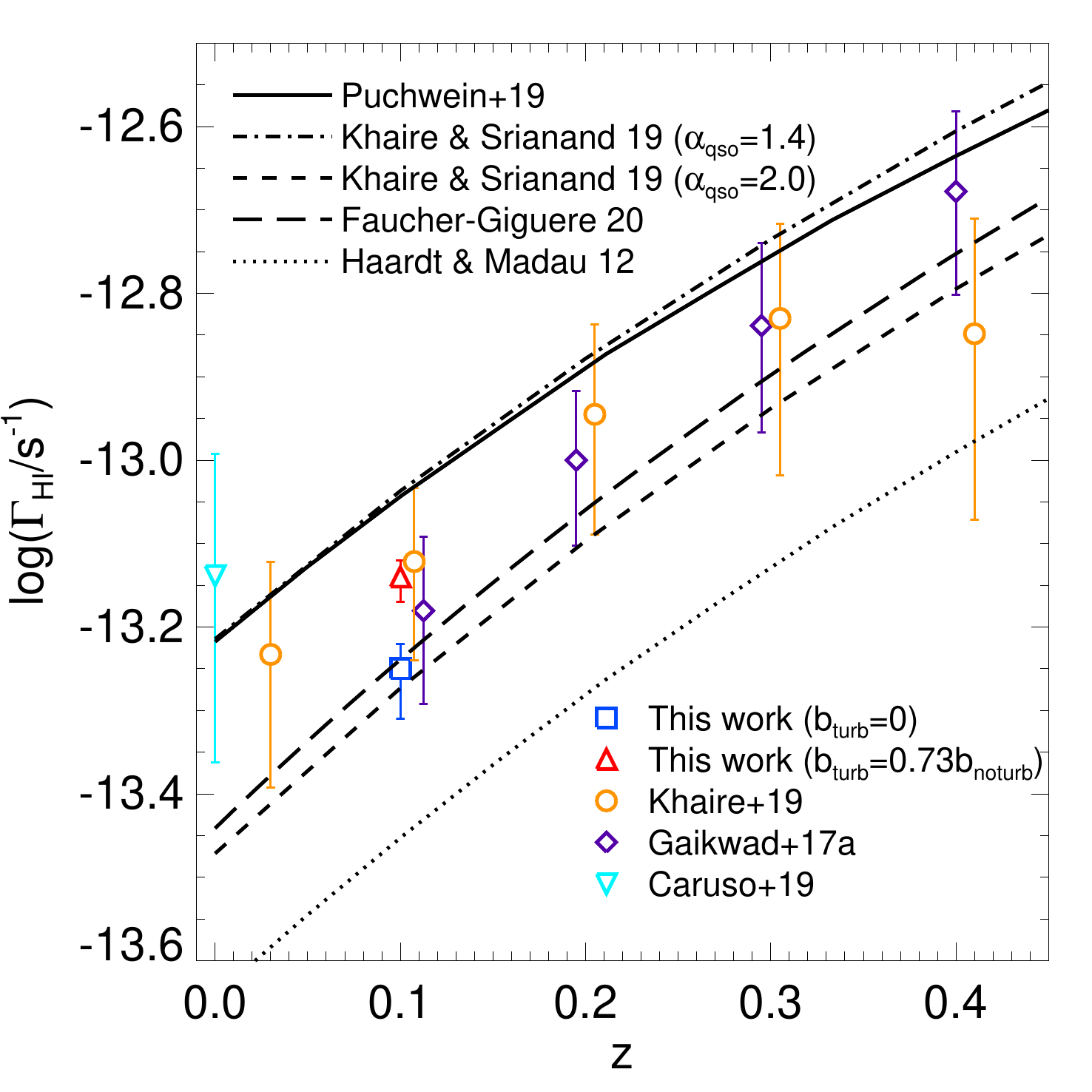}
    \includegraphics[width=0.49\textwidth]{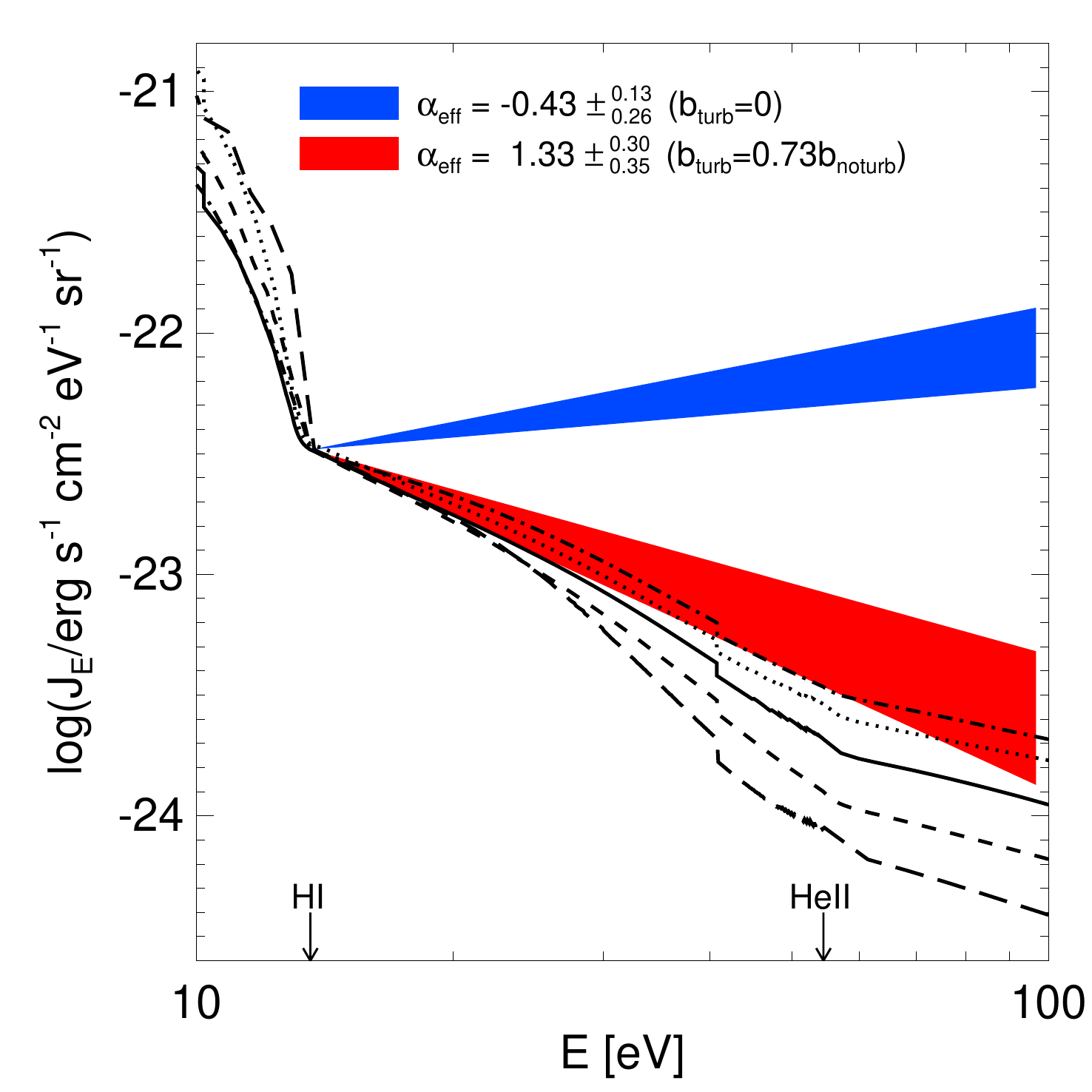}
  \end{minipage}
    \vspace{-0.3cm}
  \caption{{\it Left:} The metagalactic \HI photoionisation rate,
    $\Gamma_{\rm HI}$, at $z<0.5$.  The results of this work are shown
    by the blue square (red triangle) excluding (including) an
    additional unresolved turbulent contribution to the \Lya line
    widths, $b_{\rm turb}=0.73b_{\rm noturb}$. The curves show
    predictions for $\Gamma_{\rm HI}$ from the UV background synthesis
    models of \citet{Puchwein2019} (solid curve),
    \citet{KhaireSrianand2019} for $\alpha_{\rm qso}=1.4$ (dot-dashed
    curve) and $\alpha_{\rm qso}=2.0$ (short dashed curve),
    \citet{FaucherGiguere2020} (long dashed curve) and
    \citet{HaardtMadau2012} (dotted curve).  Independent $\Gamma_{\rm
      HI}$ measurements are shown from a joint analysis of the power
    spectrum and PDF of the \Lya forest transmitted flux
    \citep[][purple diamonds]{Gaikwad2017b}, the power spectrum only
    \citep[][orange circles]{Khaire2019pk}, and from H$\alpha$
    fluorescence in a galactic disc \citep[][cyan inverted
      triangle]{Caruso2019}.  {\it Right:} The specific intensity,
    $J_{\rm E}$, of the metagalactic UV background.  The black curves
    show the UV background models displayed in the left panel, while
    the shaded blue (red) regions show our constraint on $\alpha_{\rm
      eff}$ excluding (including) turbulent broadening.  Note that
    \emph{the model with the turbulent contribution (red shading) has
      been deliberately calibrated} for consistency with the spectral
    shape of the UV background models at the Lyman limit.  To
    facilitate the comparison, all models have been normalised to match
    the amplitude of the \citet{Puchwein2019} spectrum at
    $E=13.6\rm\,eV$.  Vertical arrows show the location of the \HI and
    \HeII ionisation edges at $13.6\rm\,eV$ and $54.4\rm\,eV$.}
  \label{fig:obsvals}  
\end{figure*}

We now proceed to perform a joint fit of our simulated spectra to the
COS measurements of the CDDF and Doppler parameter distribution.  We
vary two model parameters in our analysis: the metagalactic \HI
photoionisation rate, $\Gamma_{\rm HI}$, and the effective power-law
spectral index of the UV background, $\alpha_{\rm eff}$.  As discussed
earlier, $\alpha_{\rm eff}$ is directly related to the thermal
asymptote at $z<0.5$.

A grid of models is used for this fitting procedure.  The line fits
are obtained by performing a Voigt profile analysis on hydrodynamical
simulations with $11$ different $\alpha_{\rm eff}$ values (H00-H10 in
Table~\ref{tab:hydrosims}).  For each of these simulations we also
assume $7$ different photoionisation rates ($0.4-1.0\Gamma_{\rm
  HI,P19}$ in steps of $0.1$).  This gives a total of $77$ separate
sets of mock \Lya forest spectra for Voigt profile fitting.  The
redshift path length of each set is $\Delta z=210.1$.

We furthermore consider two different cases in our analysis (i.e. we
actually fit $2 \times 77$ sets of mocks).  These two cases assume
$b_{\rm turb}=0$ (i.e no additional turbulent broadening, $\xi=1$) and
$b_{\rm turb}=0.73b_{\rm noturb}$ (i.e. $\xi=1.24$), where for the
latter we have followed the argument in Section~\ref{sec:turb} and
assumed a prior limit on the thermal broadening of the \Lya
absorbers.\footnote{In principle, we could also find a best fit
    sub-grid turbulent contribution to the line broadening by treating
    $\xi$ as a free parameter.  However, this would require fitting an
    additional grid of models in $\xi$ for each of the 77
    ($\alpha_{\rm eff}$, $\Gamma_{\rm HI}$) pairs, which would greatly
    increase the cost of the already time consuming Voigt profile
    fitting process.  We leave this to future work.}

A $\chi^{2}$ minimisation is then performed on the COS data, such that
\begin{equation} \chi^{2}= \sum_{i}\sum_{j}[x_{\rm i}(\Gamma_{\rm HI},\alpha_{\rm eff})-\mu_{\rm i}]C_{\rm ij}^{-1}[x_{\rm j}(\Gamma_{\rm HI},\alpha_{\rm eff})-\mu_{\rm j}], \label{eq:chi2} \end{equation}
\noindent
where the vector $\vec {\bf x}$ is the Doppler parameter distribution
and CDDF from the simulations, the vector $\vec {\bf \mu}$ is the COS
data, and $C_{\rm ij}$ is the COS data covariance matrix.  However,
the off-diagonal terms in the covariance matrix for the COS data are
noisy, making the inversion of the covariance matrix difficult.  We
therefore assume the COS data and the AGN model have similar
covariance properties.  If the simulated covariance matrix is $C_{\rm
  ij}^{\rm s}$, the correlation coefficients are $r_{\rm ij}^{\rm s} =
\rm C_{\rm ij}^{\rm s}/[\rm C_{\rm ii}^{\rm s}C_{\rm jj}^{\rm
    s}]^{1/2}$.  The off-diagonal terms in the covariance matrix for
the COS data can then be estimated from the observed diagonal elements
and the simulated correlation coefficients, such that $\rm C_{\rm ij}
= r_{\rm ij}^{\rm s}[C_{\rm ii}C_{\rm jj}]^{1/2}$ \citep{Lidz2006}.

The resulting best fit parameters are displayed in the left panel of
Fig.~\ref{fig:chisqfit}, where the $\Delta \chi^{2}$ contours are
shown for no additional turbulent contribution to the \Lya forest
(blue solid curves) and including our simple estimate for an
additional, unresolved turbulent velocity (red dashed curves).  For
comparison, the grey shaded region show the effective spectral index,
$\alpha_{\rm eff}$, predicted by various UV background models
\citep{HaardtMadau2012,KhaireSrianand2019,Puchwein2019,FaucherGiguere2020}.
Note again \emph{the turbulent contribution has been chosen by hand}
so that the grey region and red contours in Fig.~\ref{fig:chisqfit}
intersect.

There is a degeneracy line between $\Gamma_{\rm HI}$ and $\alpha_{\rm
  eff}$ in Fig.~\ref{fig:chisqfit}, where a softer spectrum leads to
larger $\Gamma_{\rm HI}$, and vice versa.  This is because, in
photoionisation equilibrium, the \HI column density scales as $N_{\rm
  HI} \propto T_{0}^{-0.22}/\Gamma_{\rm HI}$ \citep{Schaye2001} and,
from Eq.~(\ref{eq:T0fit}), the IGM temperature decreases for a softer
ionising spectrum (this is also exemplified by the behaviour of
$\Gamma_{\rm HI}^{\rm CDDF}$ in Table~\ref{tab:hydrosims}).  The best
fit models excluding (including) a turbulent contribution are shown in
the right panel of Fig.~\ref{fig:chisqfit}, and have
$\chi^{2}/\nu=1.04$ ($\chi^{2}/\nu=1.10$) for $\nu=18$ degrees of
freedom, with a probability that the $\chi^{2}$ may be exceeded
randomly of $p=0.41$ ($p=0.35$).  The corresponding best fit
parameters and $1\sigma$ (statistical) uncertainties are $\log
(\Gamma_{\rm HI}/\rm s^{-1})=-13.25^{+0.03}_{-0.06}$ ($\log
(\Gamma_{\rm HI}/\rm s^{-1})=-13.14^{+0.02}_{-0.03}$) and $\alpha_{\rm
  eff}=-0.43^{+0.13}_{-0.26}$ ($\alpha_{\rm
  eff}=1.33^{+0.30}_{-0.35}$). From Eq.~(\ref{eq:T0fit}), this is
equivalent to a thermal asymptote with
${\Theta}_{0}=6390^{+680}_{-290}\rm\,K$ ($\Theta_{\rm
  0}=4030^{+300}_{-220}\rm\,K$), or for gas at $\Delta=10$,
$\Theta_{1}=20900^{+2220}_{-950}\rm\,K$
($\Theta_{1}=13180^{+980}_{-720}\rm\,K$).

In Fig.~\ref{fig:obsvals}, these results are compared to the
predictions from UV background models and independent observational
constraints on the metagalactic photoionisation rate, $\Gamma_{\rm
  HI}$.  Our constraints on $\Gamma_{\rm HI}$ in the left panel of
Fig.~\ref{fig:obsvals} are consistent with the measurements from
\citet{Gaikwad2017b} and \citet{Khaire2019pk} at $z=0.1$ from the
probability distribution function (PDF) and/or power spectrum of the
\Lya forest transmission.  Our best fit values are a factor of
$0.6$--$0.8$ times the \citet{Puchwein2019} model and a factor of
$1.6$--$2.0$ times larger than \citet{HaardtMadau2012} (see also the
discussion of $\Gamma_{\rm HI}^{\rm CDDF}$ in Section~\ref{sec:obs}).
However, the $1\sigma$ uncertainties in this work are a factor of
$\sim 3$--$4$ smaller than the earlier studies.  This is partly
because we have not included systematic uncertainties from continuum
fitting and the assumed cosmology, that may (conservatively) double
the size of the error bar on $\Gamma_{\rm HI}$ \citep[see e.g. table 8
  in][]{Gaikwad2017b}.  However, another reason is that the joint
analysis of the CDDF and Doppler parameter distribution is effective
at breaking the degeneracy between $\Gamma_{\rm HI}$ and the thermal
state of the IGM.


Our constraints on $\alpha_{\rm eff}$ are shown in the right hand
panel of Fig.~\ref{fig:obsvals}, and are compared to various UV
background models.  By design, the model with turbulent broadening,
$b_{\rm turb}=0.73 b_{\rm noturb}$, is in very good agreement with the
spectral shape of UV background models close to the Lyman limit.
However, the constraint on $\alpha_{\rm eff}$ for $b_{\rm turb}=0$
yields an unphysically hard spectral shape for the UV background and
differs from our fiducial $\alpha_{\rm eff}=1.17$ model
\citep{Puchwein2019} by $\sim 6$--$7\sigma$.  It is possible that
existing UV background models have spectral shapes that are still
slightly too soft, although given the wide range of observables at
different redshifts these models calibrate to, as well as the good
agreement between independent groups, we regard this as unlikely.  Our
statistical error bars may also underestimate the true uncertainty on
$\alpha_{\rm eff}$.  If, as previously discussed, we instead take
$\alpha_{\rm eff}=0$ as a conservative limit on the hardness of the UV
background spectrum at $z=0.1$ (i.e. assuming it is dominated by
emission from quasars with extreme UV spectral indices of $\alpha_{\rm
  qso}\simeq 0.5$ \citep{Scott2004,Tilton2016}), this differs by $\sim
2\sigma$ from our constraint of $\alpha_{\rm
  eff}=-0.43^{+0.13}_{-0.26}$.  This still suggests that photoheating
by a hard UV background is disfavoured as the sole explanation for the
observed line widths in the \Lya forest at $z=0.1$.

\subsection{Discussion}

We now discuss the implications of our results.  First, if -- as we
have argued -- matching the COS line widths in the \Lya forest through
photoheating by a hard UV background is unlikely, then if assuming a
negligible turbulent component, what other (if any) heating processes
might provide the requisite injection of energy?
\citet{UptonSanderbeck2016} demonstrated that the IGM thermal history
is already well described by the photoheating expected from \HI and
\HeII reionisation at $z \geq 2.5$.  Interestingly, however, they
noted there may be room for an additional $\sim 1\rm\,eV$ per proton
at $\Delta\simeq 5$ by $z=2$ (see their fig. 8), based on the IGM
temperature measurements at $1.5\leq z \leq 2.5$ from
\citet{Boera2014}.  This suggests that any \emph{additional} IGM
heating process must be sub-dominant to the \HI and \HeII photoheating
before $z\simeq 2.5$.

We therefore take a similar approach to \citet{UptonSanderbeck2016},
where -- based on the observed COS line widths -- we estimate an upper
limit for the additional energy injection into the IGM required at
$z\leq 2.5$.  The specific energy deposited into the IGM at $z_{0}\leq
z \leq z_{\rm R,HI}$ from photoheating at overdensity
$\Delta=\rho/\langle \rho \rangle$ is

\begin{equation} u = \int_{z_{0}}^{z_{\rm R,HI}}\frac{\mathscr H}{\rho}\frac{dz^{\prime}}{H(z^{\prime})(1+z^{\prime})}, \label{eq:u0} \end{equation}
\noindent
where $\rho=\rho_{\rm crit}\Omega_{\rm b}\Delta (1+z)^{3}$, and for
photoheating ${\mathscr H}= \sum_{i}n_{i}\epsilon_{i}$, where $n_{\rm
  i}$ and $\epsilon_{\rm i}$ are respectively the proper number
density and photoheating rate for $i=[\rm H{\,\rm \scriptstyle I}, \rm
  He{\,\rm \scriptstyle I}, \rm He{\,\rm \scriptstyle II}]$
\citep[e.g.][]{Nasir2016}.  Taking the \citet{Puchwein2019} UV
background model (which is already calibrated to match existing IGM
temperature measurements at $z>2$ and has $\alpha_{\rm eff}=1.17$ at
$z=0.1$), we find boosting these model photoheating rates by a factor
of $2$ at $z\leq 2.5$ gives good agreement with the $\alpha_{\rm
  eff}=-0.43^{+0.13}_{-0.26}$ (or equivalently the thermal asymptote
temperature $\Theta_{1}=20900^{+2220}_{-950}\rm\,K$) we infer for
$b_{\rm turb}=0$ in Section~\ref{sec:bestfit}.  From
Eq.~(\ref{eq:u0}), this boost corresponds to an additional $u\simeq
6.9\rm\,eV\,m_{\rm p}^{-1}$ ($u\simeq 1.6\rm\,eV\,m_{\rm p}^{-1}$)
injected into the IGM\footnote{Note this is consistent with a density
scaling of $u\propto \Delta^{1-0.72(\gamma-1)} \sim \Delta^{0.58}$ for
$\gamma=1.58$, which is the expectation for photoheating assuming the
IGM is in photoionisation equilibrium.}  at $\Delta=10$ ($\Delta=1$)
by $z_{0}=0.1$, \emph{in excess} of that already provided by
photoheating in the \citet{Puchwein2019} model.  Note again, however,
that this level of enhanced photoheating would require an unphysically
hard UV background.

Any non-canonical heating process would therefore need to inject
$\lesssim 6.9\rm\,eV\,m_{\rm p}^{-1}$ into $N_{\rm HI}\simeq
10^{13.5}\rm\,cm^{-2}$ absorbers by $z=0.1$, while also having a
negligible effect on the IGM temperature at $z>2.5$.  We now speculate
on which processes are plausible.  As already discussed, we find a
volumetric heating process like blazar heating \citep{Puchwein2012}
with $u\propto \Delta^{-1}$ will not heat the IGM sufficiently at
$\Delta\gtrsim 10$. A similar situation likely holds for Compton
heating of the IGM by X-rays \citep{MadauEfstathiou1999}, where the
heating rate ${\mathscr H}_{\rm C} \propto n_{\rm e}$ and $u\propto
\Delta^{0}$.  Heating from dark matter annihilations
\citep[e.g.][]{Mapelli2006,Cirelli2009,Liu2021} would need to be
fine-tuned to avoid a substantial injection of energy into the IGM at
$z>2.5$.  Cosmic rays can introduce significant non-thermal pressure
in the IGM \citep{Lacki2015,Butsky2021}, but they are not expected to
directly increase the temperature of the low density IGM unless they
can efficiently couple to the gas \citep{NathBiermann1993,Samui2018}.
Further study will be necessary for confirming or ruling out these
possibilities, however.

In this work we instead focus on photoelectric emission by dust grains
\citep{Nath1999,WeingartnerDraine2001,InoueKamaya2003}, by virtue of
the fact this heating rate should naturally increase toward lower
redshift as the IGM is enriched with heavy elements.  Large, high
velocity dust grains with sizes $\gtrsim 0.1 \rm {\mu}m$ and
velocities $v\geq 100\rm\,km\,s^{-1}$ may be able to escape into the
low density IGM \citep{BianchiFerrara2005}, where the destruction
timescale due to thermal sputtering will exceed a Hubble time
\citep{Draine2011}.  Smaller, slower grains are instead more likely to
be eroded within hot halo gas where the sputtering timescale is much
shorter.  Following \citet{InoueKamaya2010}, the specific energy from
dust heating scales as $u\propto \Delta^{1/3}T^{-1/6}\propto
\Delta^{1/3 - (\gamma-1)/6}$, where the heating rate is proportional
to the dust-to-gas mass ratio, ${\mathscr D}$, and depends on the
uncertain grain size distribution.  The density dependence of $u$ also
implies that dust heating should flatten the power law
temperature-density relation in the low redshift IGM
(i.e. $\gamma<1.6$), in addition to raising the gas temperature.
\cite{InoueKamaya2010} provide an approximate expression for the dust
heating rate (their eq. 16) assuming the grain size distribution from
\citet{Mathis1977}. Taking the \citet{Puchwein2019} UV background
model and adding the \cite{InoueKamaya2010} heating rate at $z<2.5$,
we find a constant dust-to-gas mass ratio of ${\mathscr D}= 1.5\times
10^{-3}$ gives an additional $\simeq 6.5\rm\,eV$ ($\simeq 4.0\rm\,eV$)
per proton at $\Delta=10$ ($\Delta=1$) by $z=0.1$.  For comparison,
${\mathscr D}\sim 10^{-2}$ for the Milky Way \citep{Draine2011}, while
observations of nearby galaxies exhibit a large scatter ($2$--$3$ dex)
in ${\mathscr D}$ at fixed metallicity
\citep{RemyRuyer2014,DeVis2019}.  If we adopt a naive extrapolation of
the observed dust-to-gas and metallicity relation from local galaxies
\citep[e.g.][]{DeVis2019} to the $z \sim 0$ IGM, ${\mathscr D}\sim
10^{-3}$ is broadly consistent with $Z\gtrsim 0.1 Z_{\odot}$.  The
requisite heating is therefore only possible if the low density IGM is
highly enriched by a viable dust transport mechanism (e.g. galactic
winds or radiation pressure).  Further investigation of the expected
heating rates using a more detailed dust model
\citep[e.g.][]{Popping2017,McKinnon2017,Hou2019,Li2019}, along with
updated assessment of whether or not dust can be effectively
transported into the IGM without being eroded by hot halo gas may be
of interest.

Alternatively, if non-canonical heating is negligible, how plausible
is our crude upper limit on the (density dependent) line of sight
turbulent velocity, $v_{\rm turb}\lesssim 8.5\rm\,km\,s^{-1}(N_{\rm
  HI}/10^{13.5}\rm\,cm^{-2})^{0.21}$?  Assuming the IGM has kinematic
viscosity $\nu \simeq 5\times 10^{24}\rm\,cm^{2}\,s^{-1}$
\citep{Evoli2011}\footnote{For comparison, adopting representative
values for the temperature and density in the \Lya forest at $z=0.1$,
$T=10^{4.5}\rm\,K$ and $n_{\rm H}=10^{-5.5}\rm\,cm^{-3}$, the
kinematic viscosity of fully ionised hydrogen is $\nu=2.9\times
10^{24}\rm\,cm^{2}\,s^{-1}$ \citep{Chapman1954}.  The suggests
$\nu\sim 10^{24}\rm\,cm^{2}\,s^{-1}$ provides a reasonable
order-of-magnitude estimate for the IGM kinematic viscosity at
$z\simeq 0.1$.}, if the typical flow speed in the \Lya forest is the
speed of sound, $U=c_{\rm s}\sim 25\rm\,km\,s^{-1}$, and the
characteristic length scale of \Lya forest absorbers at $z\simeq 0.1$
is the Jeans scale, $L \simeq L_{\rm Jeans}\simeq 250\rm\,pkpc$
\citep{Schaye2001}, the Reynolds number for the low redshift \Lya
forest is $\rm Re=UL/\nu \simeq 3.9\times 10^{5}$.  Although only an
order of magnitude estimate, $\rm Re \gg 10^{3}$ suggests that, if an
appropriate mechanism for continuously generating vorticity is present
(e.g. feedback, shocks or magnetic fields), the IGM should indeed be
turbulent on small scales
\citep{Evoli2011,Gregori2012,Iapichino2013,Zhu2013}.  However, given
that the Doppler widths of the \Lya absorbers in our simulations
\emph{decrease} with increasing mass resolution (see
Appendix~\ref{app:converge}), this implies that -- if present -- any
turbulence must be injected below a spatial resolution scale of
roughly $l_{\rm res} \simeq L_{\rm box}(N_{\rm
  part}/2)^{-1/3}\Delta^{-1/3}$ in the simulations (i.e. $l_{\rm res}
\sim 40 \rm\,ckpc$ at $\Delta=10$ for the N$1024$
simulation). Alternatively, it could be that the gas responsible for
the bulk of the low column density \Lya absorption is just not
sufficiently agitated by the shocks and/or outflows within our models,
or there is some other non-thermal broadening missing in the simulated
spectra.  Regarding the latter possibility, we note again, however,
that most of the \Lya absorbers in our fiducial AGN simulation with
$10^{13.3}\rm\,cm^{-2}\leq N_{\rm HI}\leq 10^{14.5}\rm\,cm^{-2}$ are
already suprathermal with a median $b/b_{\rm therm}=1.28$.  If adding
a turbulent component by hand, such that $b_{\rm turb}=0.73b_{\rm
  noturb}$, we instead obtain a median $b/b_{\rm therm}=1.48$.  Both
of these values are consistent with the curve-of-growth analysis
presented by \citet{Danforth2010}, who find $b_{\rm Ly\alpha}/b_{\rm
  cog}=1.26^{+0.49}_{-0.25}$.

Nevertheless, if taking our turbulent velocity estimate at face value,
it is consistent with the results from the galactic outflow driven IGM
turbulence model of \citet{Evoli2011} at the higher redshift of $z=1$,
where from their fig. 5, $b_{\rm turb}=8\pm 4\rm\,km\,s^{-1}$ for
$N_{\rm HI}\simeq 10^{13.6}\rm\,cm^{-2}$.  Observationally,
non-thermal broadening can also be constrained by measuring the
Doppler parameters of species with different masses in the same gas
phase \citep[e.g.][]{Rauch1996}.  However, while there is some
evidence for turbulence in the low redshift CGM from well aligned $\rm
O\,\rm \scriptstyle VI$, \CIV and \HI absorbers at $z<0.5$
\citep{Tripp2008,ThomChen2008,Savage2014,Werk2016,Manuwal2021}, the
picture for the lower density diffuse IGM is arguably less clear.  The
few existing constraints instead come from observations of the \Lya
forest at $z\simeq 3$.  This has been attempted with \HI and \HeII
\Lya absorption, where \citet{Zheng2004} found evidence for purely
turbulent broadening from the Doppler parameter ratio of aligned \HI
and \HeII \Lya lines.  However, in an independent analysis,
\citet{FechnerReimers2007} found that just under half of aligned \HeII
and \HI \Lya absorbers are consistent with purely turbulent
broadening.  This confusion arises in part because, at $z\simeq 3$,
the Hubble broadening of \Lya absorbers across the Jeans smoothing
scale, $b_{\rm Jeans}\simeq H(z)L_{\rm Jeans}$, becomes comparable to
the thermal widths of the lines
\citep[e.g.][]{Peeples2010,Garzilli2015}.  Hence, without a
self-consistent hydrodynamical model for the IGM density field, this
``Jeans smoothing'' can easily be confused with turbulence.

\citet{Rauch2001Lya} side-stepped this problem by measuring
\emph{transverse} correlations between the \Lya absorption in
gravitationally lensed quasar images separated by $\sim 0.3 \rm\,ckpc$
at $z\sim 3$. They found no evidence for turbulence on this scale,
although this may not be surprising: the \Lya forest at $z\sim 3$ is
sensitive to gas close to the mean density \citep{Becker2011} and is
therefore unlikely to be disturbed by shocks or feedback
\citep{Theuns2002_feedback,Viel2013,Chabanier2020}.  By contrast, in a
companion study of \CIV absorbers at $z\sim 2$--$3$ in three lensed
quasars (where the typical gas densities probed are more like
$\Delta\sim 10$--$100$, \citet{BoltonViel2011}), \citet{Rauch2001CIV}
found that a turbulent velocity component of $v_{\rm turb}\sim
4.7\rm\,km\,s^{-1}$ was required at a scale of $\sim 0.3 \rm\,ckpc$,
which is consistent with our (line of sight) estimate of $\lesssim
8.5\rm\,km\,s^{-1}$ for $N_{\rm HI}\sim 10^{13.5}\rm\,cm^{-2}$.  At a
minimum, this suggests that our crude upper limit on the turbulent
contribution to the coldest \Lya forest absorbers at $z=0.1$ is at
least plausible.  If a suitably lensed background source could be
identified for the \Lya forest at $z<0.5$, repeating the
\citet{Rauch2001Lya} experiment at $z\simeq 0.1$ would be valuable for
testing this possibility further.


\section{Conclusions} \label{sec:conclude}

We have performed a Voigt profile analysis of the column density
distribution function (CDDF) and Doppler parameter distribution
measured from hydrodynamical simulations and \emph{Cosmic Origins
  Spectrograph} (COS) observations of the low redshift \Lya forest at
$z\simeq 0.1$ \citep{Viel2017,Kim2021}.  We re-examine the tension
between the observations and theoretical predictions for the widths of
the \Lya forest absorption lines, where the \Lya absorber Doppler
parameters in hydrodynamical simulations are too narrow with respect
to the COS data \citep{Nasir2017,Gaikwad2017}.  We also assess the
level of agreement between the COS \Lya forest CDDF and simulations, a
statistic that is sensitive to both the UV background amplitude and
(sufficiently strong) AGN feedback
\citep{Kollmeier2014,Shull2015,KhaireSrianand2015,Gurvich2017,Christiansen2020}.
Our primary conclusions are as follows.

\begin{itemize}

\item We focus on absorption lines with column densities
  $10^{13.3}\rm\,cm^{-2}\leq N_{\rm HI}\leq 10^{14.5}\rm\,cm^{-2}$ and
  Doppler parameters $20\rm\,km\,s^{-1}\leq b \leq 90\rm\,km\,s^{-1}$
  at $z\simeq 0.1$.  We show these absorption lines will be minimally
  impacted by systematic uncertainties in the signal-to-noise and
  spectral resolution of the data.  In this range, the majority of the
  absorbers (83 per cent) we identify in our fiducial simulation (AGN)
  reside in the diffuse IGM (i.e. gas with $T<10^{5}\rm\,K$ and
  $\Delta<97.2$).  Strong absorbers with $N_{\rm HI}\geq
  10^{14}\rm\,cm^{-2}$ are preferentially located close to haloes,
  with over half of these within $r<3R_{\rm vir}$ of haloes of (total)
  mass $10^{10}M_{\odot}\leq M_{\rm h}\leq 10^{12}M_{\odot}$
  \citep[cf.][]{ChenMulchaey2009,Tejos2014,Keeney2018}.  By contrast,
  fewer than 10 per cent of $N_{\rm HI}\geq 10^{14}\rm\,cm^{-2}$
  absorbers are within $r<3R_{\rm vir}$ of haloes of mass $M_{\rm
    h}\geq 10^{12}\,M_{\odot}$.  Hot, collisionally ionised gas from
  shocks and feedback reduces the incidence of strong absorbers around
  the most massive haloes at $z=0.1$.\\

\item After applying a small correction for box size and mass
  resolution, our fiducial AGN and Quick-\Lya (H02) simulations are
  both in good agreement (within $\sim 1$--$1.5\sigma$) with the shape
  of the CDDF measured from the COS data at $10^{13.3}\rm\,cm^{-2}\leq
  N_{\rm HI}\leq 10^{14.5}\rm\,cm^{-2}$.  Adopting an \HI
  photoionisation rate, $\Gamma_{\rm HI}$, that is $\sim 0.6$ times
  the \citet{Puchwein2019} model (or $\sim 1.7$ times the
  \citet{HaardtMadau2012} model) provides a good match to the
  amplitude of the COS CDDF.  We confirm that potent AGN feedback
  and/or blazar heating models that produce a substantial fraction of
  the low density, warm-hot IGM (WHIM) further lower the $\Gamma_{\rm
    HI}$ required for consistency with the CDDF amplitude
  \citep[cf.][]{Christiansen2020}, and will also flatten the CDDF at
  $N_{\rm HI}\sim 10^{14}\rm\,cm^{-2}$ \citep{Gurvich2017}.\\

\item The simulated Doppler width distribution for lines with
  $10^{13.3}\rm\,cm^{-2}\leq N_{\rm HI}\leq 10^{14.5}\rm\,cm^{-2}$ is
  inconsistent with the COS measurements; the number of narrow lines
  with $b=22.5\pm 2.5\rm\,km\,s^{-1}$ are over-predicted by
  $4.6\sigma$ in our fiducial AGN model.  We show that introducing
  additional hot gas into the low density IGM by invoking strong AGN
  feedback or blazar heating does not resolve this discrepancy.  As
  already noted by \citet{Viel2017}, this is because this hot gas is
  primarily in the WHIM with temperatures $T\simeq 10^{6}\rm\,K$.
  While this changes the average ionisation level of the IGM (and
  hence induces a shift the amplitude and shape of the CDDF), it does
  not produce additional \Lya absorption at the necessary density and
  temperature for resolving the discrepancy in the line widths.  We
  argue this implies the presence of additional heating or turbulence
  in the low density IGM.\\

\item We perform a joint analysis of the CDDF and Doppler parameter
  distribution to find the best fit values for the metagalactic \HI
  photoionisation rate, $\Gamma_{\rm HI}$, and the effective power-law
  spectral index, $\alpha_{\rm eff}$ (where $J_{\rm E}\propto
  E^{-\alpha_{\rm eff}}$), of the UV background close to the Lyman
  limit.  Assuming there is no missing non-thermal broadening in the
  simulations (e.g., from turbulence), the best fit values are
  $\log(\Gamma_{\rm HI}/\rm s^{-1})=-13.25^{+0.03}_{-0.06}$ and
  $\alpha_{\rm eff}=-0.43^{+0.13}_{-0.26}$ for $\chi^{2}/\nu=1.04$.
  While this photoionisation rate is consistent with previous
  constraints \citep{Shull2015,Gaikwad2017b,Khaire2019pk,Caruso2019},
  the inferred value of $\alpha_{\rm eff}$ is unphysically hard and is
  inconsistent by $6$--$7\sigma$ (statistical) with the much softer
  spectral shape, $\alpha_{\rm eff}\simeq 1$--$1.4$, predicted by
  state-of-the-art UV background synthesis models that use intrinsic
  power-law spectral indices of $\alpha_{\rm qso}=1.4$--$2.0$
  \citep{HaardtMadau2012,KhaireSrianand2019,
    Puchwein2019,FaucherGiguere2020}.  Even if allowing for a rather
  extreme UV background with $\alpha_{\rm eff}\simeq 0$, as might be
  expected for intrinsic quasar extreme-UV spectral indices of
  $\alpha_{\rm qso}=0.5$ \citep{Scott2004,Tilton2016} combined with
  some spectral hardening by the IGM at the \HeII ionisation edge,
  this remains a $\sim 2\sigma$ discrepancy. We conclude that enhanced
  UV background photoheating rates in the low density IGM that
  increase the thermally broadened line components are disfavoured as
  the only solution to the discrepancy between the observed and
  simulated \Lya forest line widths at $z=0.1$.\\

 \item If taking the UV background heating rates from
   \citet{Puchwein2019} as a prior (with $\alpha_{\rm eff}=1.17$), we
   may instead appeal to a non-canonical source of heating in the IGM
   (i.e. heating that is not due to photoheating by the UV
   background).  We then find a specific energy injection of
   $u\lesssim 6.9\rm\,eV\,m_{\rm p}^{-1}$ in addition to that expected
   from UV photoheating is required for gas with $\Delta=10$ at
   $z<2.5$.  We briefly discuss the likelihood that other physical
   processes could contribute the additional energy, including heating
   by dark matter annihilations
   \citep{Mapelli2006,Cirelli2009,Liu2021}, Compton heating by X-rays
   \citep{MadauEfstathiou1999}, cosmic rays
   \citep{NathBiermann1993,Samui2018} and photoelectric emission by
   dust grains \citep{Nath1999,WeingartnerDraine2001,InoueKamaya2003}.
   We speculate on the role of dust heating in particular, as the
   specific energy injected into the IGM scales as $\sim
   \Delta^{1/3}T^{-1/6}$, and (unlike other mechanisms) the heating
   rate should naturally increase toward lower redshift as the IGM is
   enriched with heavy elements.  On adopting the dust heating rates
   from \citet{InoueKamaya2010}, an additional $\simeq
   6.5\rm\,eV\,m_{\rm p}^{-1}$ at $z<2.5$ requires a constant dust to
   gas ratio of ${\mathscr D}= 1.5\times 10^{-3}$ for mildly overdense
   IGM gas with $\Delta=10$.  This is broadly consistent with a
   metallicity of $Z\gtrsim 0.1 Z_{\odot}$, based on a naive
   extrapolation of the dust-to-gas and metallicity relation in local
   galaxies, and would therefore imply a highly enriched IGM.
   However, it remains an open question as to whether or not
   sufficient quantities of dust can survive passage through hot halo
   gas \citep[but see][]{BianchiFerrara2005}.  A combination of
   several non-canonical heating mechanisms along with some
   non-thermal line broadening may also provide a plausible route for
   reconciling the COS \Lya line widths and the simulations. \\

\item Alternatively, the additional line broadening may be entirely
  due to non-thermal broadening that is missing in the hydrodynamical
  simulations.  If again adopting a prior limit on the thermal widths
  of the \Lya absorbers using the \citet{Puchwein2019} UV background
  model, we obtain a crude upper limit on a possible additional
  turbulent contribution to the \Lya forest line widths.  For an
  assumed line width ratio of $b_{\rm turb}/b_{\rm noturb}=0.73$, the
  best fit UV background parameters are instead $\log(\Gamma_{\rm
    HI}/\rm s^{-1})=-13.14^{+0.02}_{-0.03}$ and $\alpha_{\rm
    eff}=1.33^{+0.30}_{-0.35}$, where $\alpha_{\rm eff}$ is now
  consistent with \citet{Puchwein2019} by design.  For the coldest gas
  in the diffuse IGM at $z\simeq 0.1$, the ratio $b_{\rm turb}/b_{\rm
    noturb}=0.73$ translates to an upper limit of $v_{\rm
    turb}\lesssim 8.5\rm\,km\,s^{-1}(N_{\rm
    HI}/10^{13.5}\rm\,cm^{-2})^{0.21}$ for the additional turbulent
  velocity component along the line of sight.  This estimate is
  comparable to theoretical estimates at $z=1$ \citep{Evoli2011} and
  observational estimates of turbulence from \CIV absorbers at $z\sim
  3$ \citep{Rauch2001CIV}, and would suggest that the stirring of the
  low density IGM is widespread by $z\simeq 0$.\\
   
 \end{itemize}

\noindent

\noindent
In summary, we reaffirm that the low redshift \Lya forest provides a
powerful diagnostic of complex and poorly understood physical
processes in low density \emph{intergalactic} gas.  It would be
interesting to assess how well numerical models that are anchored to
these data reproduce the observed relationship between galaxies and
gas at higher densities and on smaller scales at $z<0.5$.

\section*{Acknowledgements}

We thank the referee, Mike Shull, for a constructive report, and Peng
Oh, Sebastiano Cantalupo and George Becker for useful discussions
during the early stages of this work, during the ``What matter(s)
between galaxies'' conference held at Abbazia di Spineto in 2019.  Our
thanks also to Raghunathan Srianand and Sowgat Muzahid for comments on
the draft version of the manuscript.  The hydrodynamical simulations
were performed using the Cambridge Service for Data Driven Discovery
(CSD3), part of which is operated by the University of Cambridge
Research Computing on behalf of the STFC DiRAC HPC Facility
(www.dirac.ac.uk). The DiRAC component of CSD3 was funded by BEIS
capital funding via STFC capital grants ST/P002307/1 and ST/R002452/1
and STFC operations grant ST/R00689X/1. This work also used the
DiRAC@Durham facility managed by the Institute for Computational
Cosmology on behalf of the STFC DiRAC HPC Facility. The equipment was
funded by BEIS capital funding via STFC capital grants ST/P002293/1
and ST/R002371/1, Durham University and STFC operations grant
ST/R000832/1. DiRAC is part of the National e-Infrastructure.  We
thank Volker Springel for making \textsc{P-Gadget-3} available. JSB is
supported by STFC consolidated grant ST/T000171/1.  For the purpose of
open access, the author has applied a creative commons attribution (CC
BY) to any author accepted manuscript version arising.


\section*{Data Availability}

All data and analysis code used in this work are available from the
first author on reasonable request. 


\appendix

\section{Box size and mass resolution} \label{app:converge}

\begin{figure*}
  \begin{minipage}{1.00\textwidth}
    \includegraphics[width=0.49\textwidth]{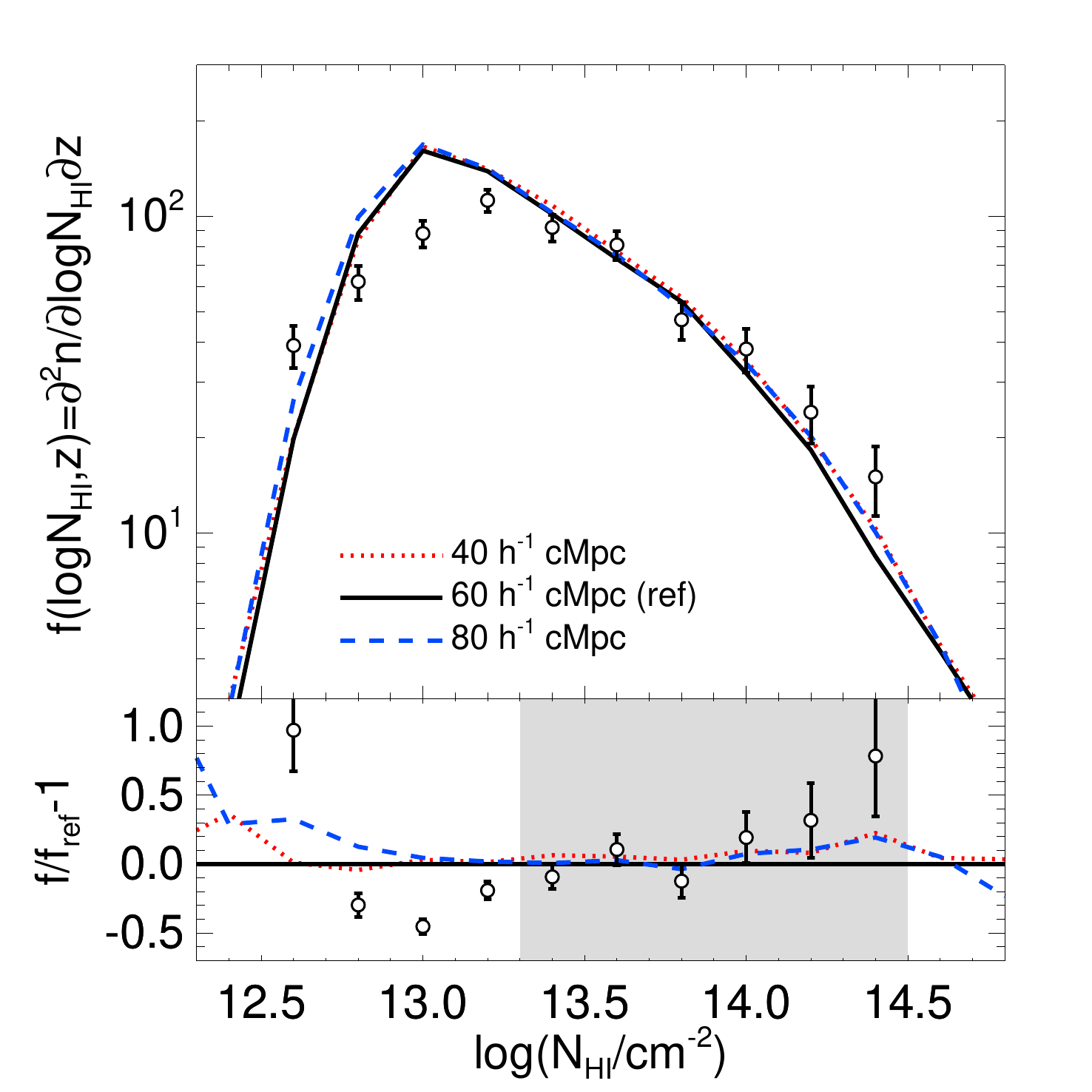}
    \includegraphics[width=0.49\textwidth]{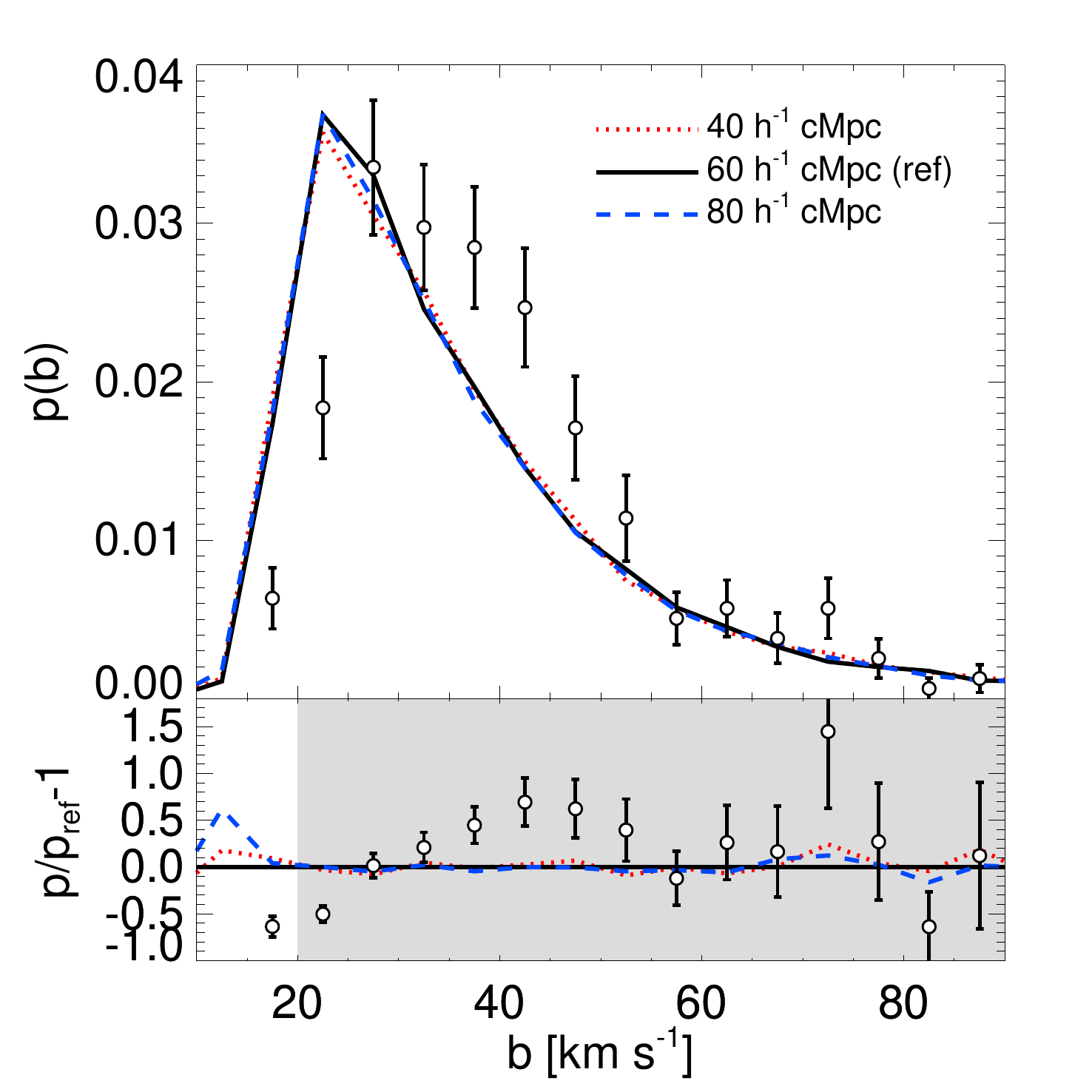}
  \end{minipage}
  \vspace{-0.3cm}
  \caption{The effect of box size for a fixed gas particle mass.  {\it
      Left:} The CDDF measured from COS data for \Lya absorbers with
    Doppler parameters $20\rm\,km\,s^{-1}\leq b \leq 90
    \rm\,km\,s^{-1}$, compared to the CDDF obtained from the L40 (red
    dotted), H02 (black solid) and L80 (blue dashed) simulations.  The
    figure is otherwise the same as Figure~\ref{fig:phystest}, but
    excludes the box size and mass resolution correction to the
    simulated CDDF.  {\it Right:} The corresponding Doppler parameter
    probability distribution for \Lya absorbers with column densities
    $10^{13.3}\rm\,cm^{-2}\leq N_{\rm HI} \leq
    10^{14.5}\rm\,cm^{-2}$.}
  \label{fig:boxtest}  
\end{figure*}

\begin{figure*}
  \begin{minipage}{1.00\textwidth}
    \includegraphics[width=0.49\textwidth]{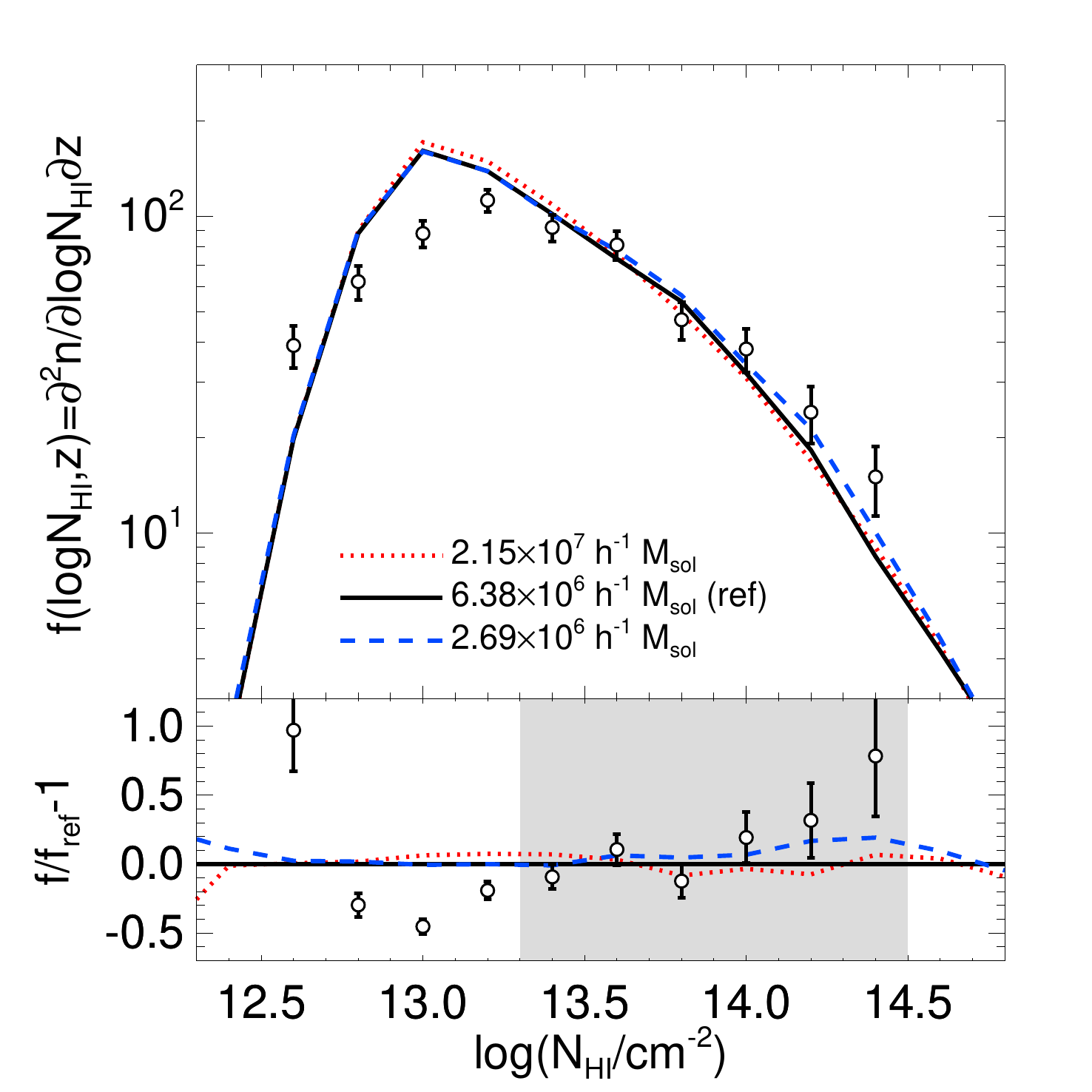}
    \includegraphics[width=0.49\textwidth]{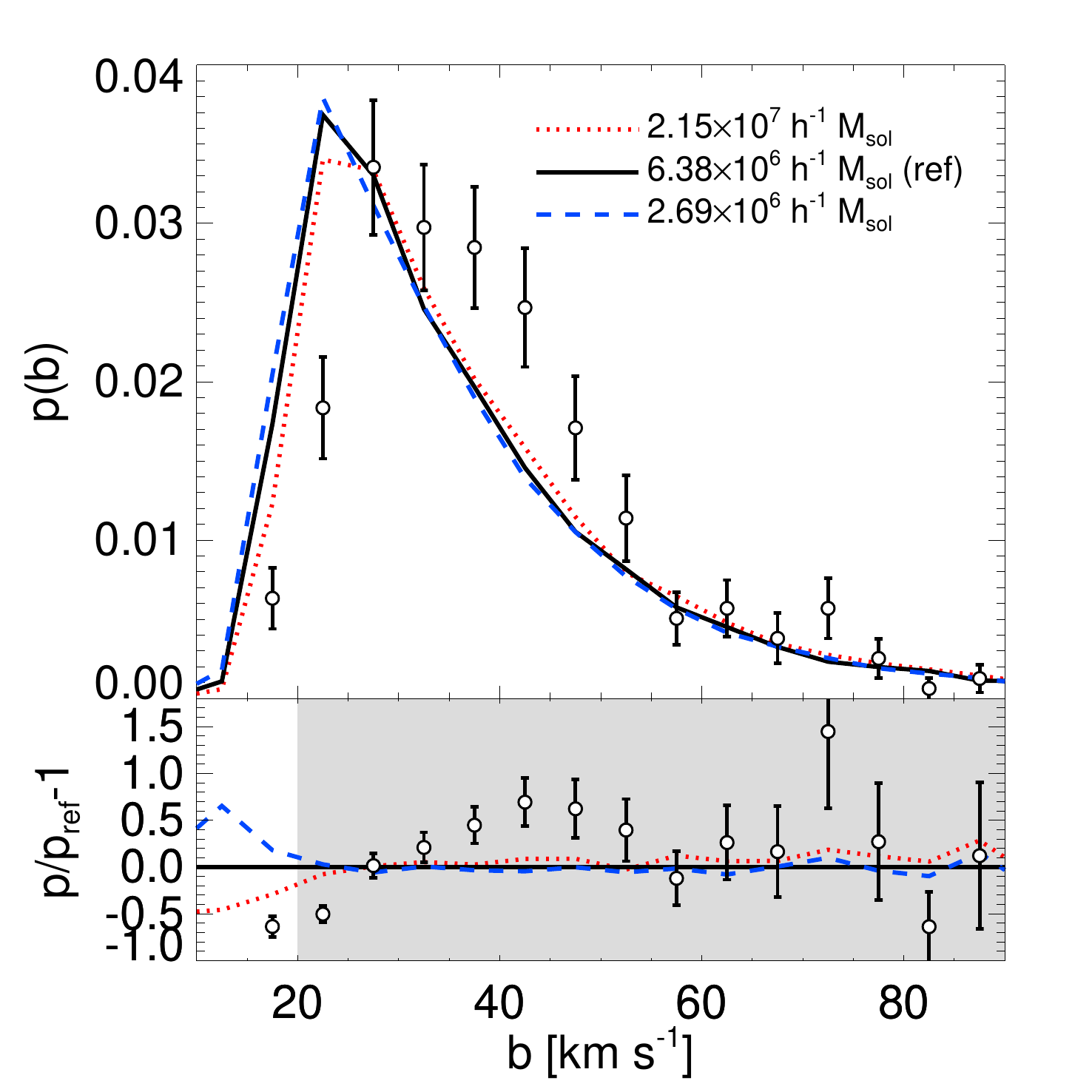}
  \end{minipage}
  \vspace{-0.3cm}
  \caption{The effect of gas particle mass for fixed box size.  {\it
      Left:} The CDDF measured from COS data for \Lya absorbers with
    Doppler parameters $20\rm\,km\,s^{-1} \leq b \leq
    90\rm\,km\,s^{-1}$, compared to the CDDF obtained from the N512
    (red dotted), H02 (black solid) and N1024 (blue dashed)
    simulations.  {\it Right:} The corresponding Doppler parameter
    probability distribution for \Lya absorbers with column densities
    $10^{13.3}\rm\,cm^{-2}\leq N_{\rm HI} \leq
    10^{14.5}\rm\,cm^{-2}$. The figure is otherwise the same as
    Figure~\ref{fig:phystest}.}
  \label{fig:restest}  
\end{figure*}

Numerical convergence tests of the \Lya forest CDDF and Doppler
parameter distribution with simulation box size and gas particle mass
are presented in Fig.~\ref{fig:boxtest} and Fig.~\ref{fig:restest},
respectively.  The Doppler parameter distribution is well converged
with respect to the $1\sigma$ uncertainties on the COS data for our
fiducial values of $L=60h^{-1}\rm\,cMpc$ and $M_{\rm gas}=6.38\times
10^{6}\,h^{-1}\,M_{\odot}$.  However, a $\sim 10$--$30$ per cent
correction to the CDDF -- particularly for strong \Lya absorbers with
$N_{\rm HI}\geq 10^{14}\rm\,cm^{-2}$ -- is required on comparing our
fiducial model to simulations with $L=80h^{-1}\rm\,cMpc$ or $M_{\rm
  gas}=2.69\times 10^{6}h^{-1}\,M_{\odot}$ (the blue dashed curves in
Fig.~\ref{fig:boxtest} and Fig.~\ref{fig:restest}).  This correction
is typically comparable to the $1\sigma$ uncertainties on the COS CDDF
measurement.  The combined correction for box size and mass resolution
that we apply to the CDDF is given in Table~\ref{tab:boxres}.

\begin{table}
\centering
   \caption{The correction we apply to each CDDF bin of width $\Delta
     \log N_{\rm HI}=0.2$ to account for the convergence with
     simulation box size and mass resolution at our fiducial box size
     and mass resolution of $L=60h^{-1}\rm\,cMpc$ and $M_{\rm
       gas}=6.38\times 10^{6}h^{-1}\,M_{\odot}$.  The corrected CDDF
     is given by $\kappa_{\rm boxres}f(\log N_{\rm HI},z)$.}
   \begin{tabular}{c|c}
     \hline
     $\log(N_{\rm HI}/\rm cm^{-2})$ & $\kappa_{\rm boxres}$ \\
     \hline
     $13.4$ & $1.00$ \\ 
     $13.6$ & $1.09$ \\
     $13.8$ & $1.01$ \\ 
     $14.0$ & $1.15$ \\ 
     $14.2$ & $1.29$ \\
     $14.4$ & $1.42$ \\
     \hline
 \end{tabular}
  \label{tab:boxres}
\end{table}

\section{Systematics} \label{app:sys}

\begin{figure*}
  \begin{minipage}{1.00\textwidth}
    \includegraphics[width=0.49\textwidth]{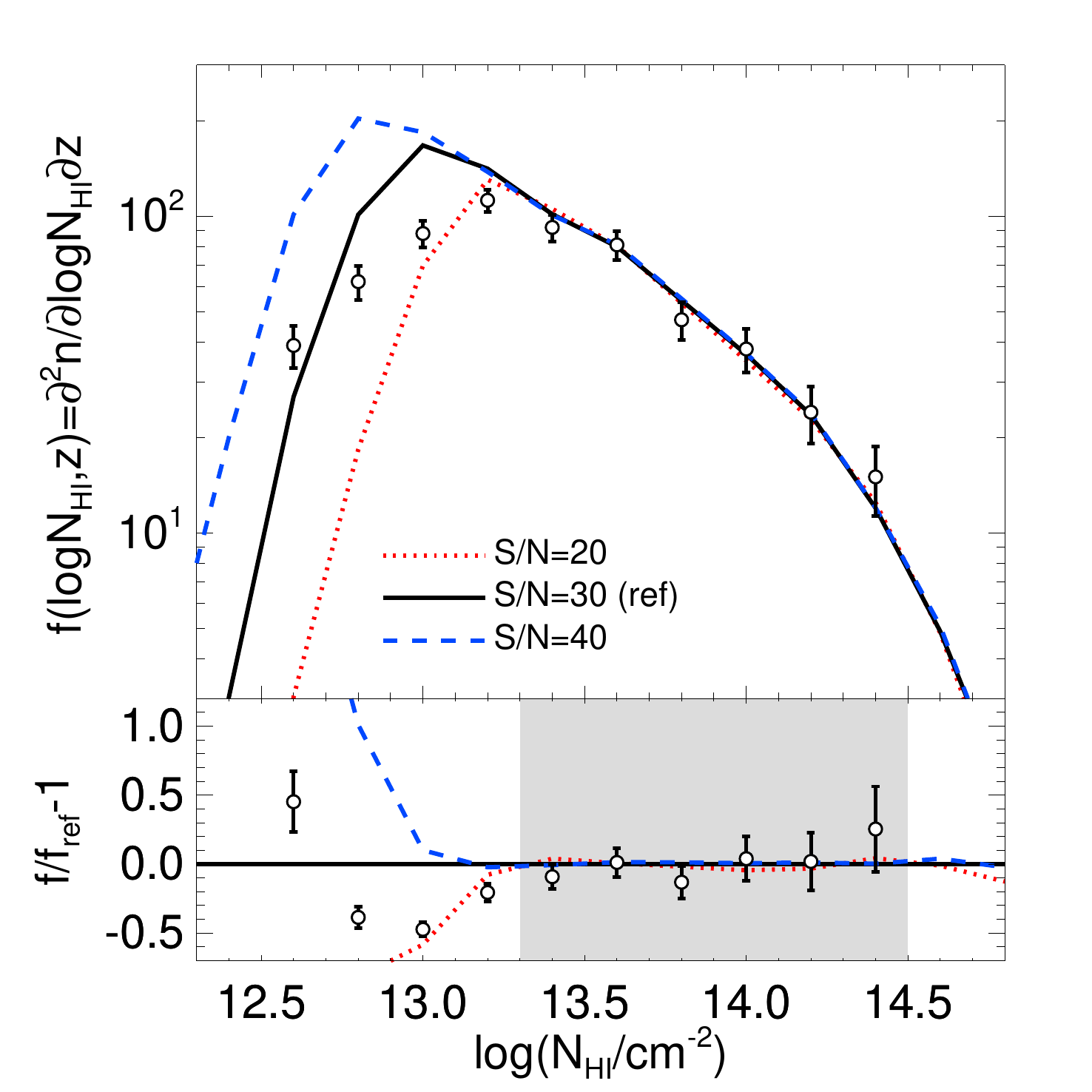}
    \includegraphics[width=0.49\textwidth]{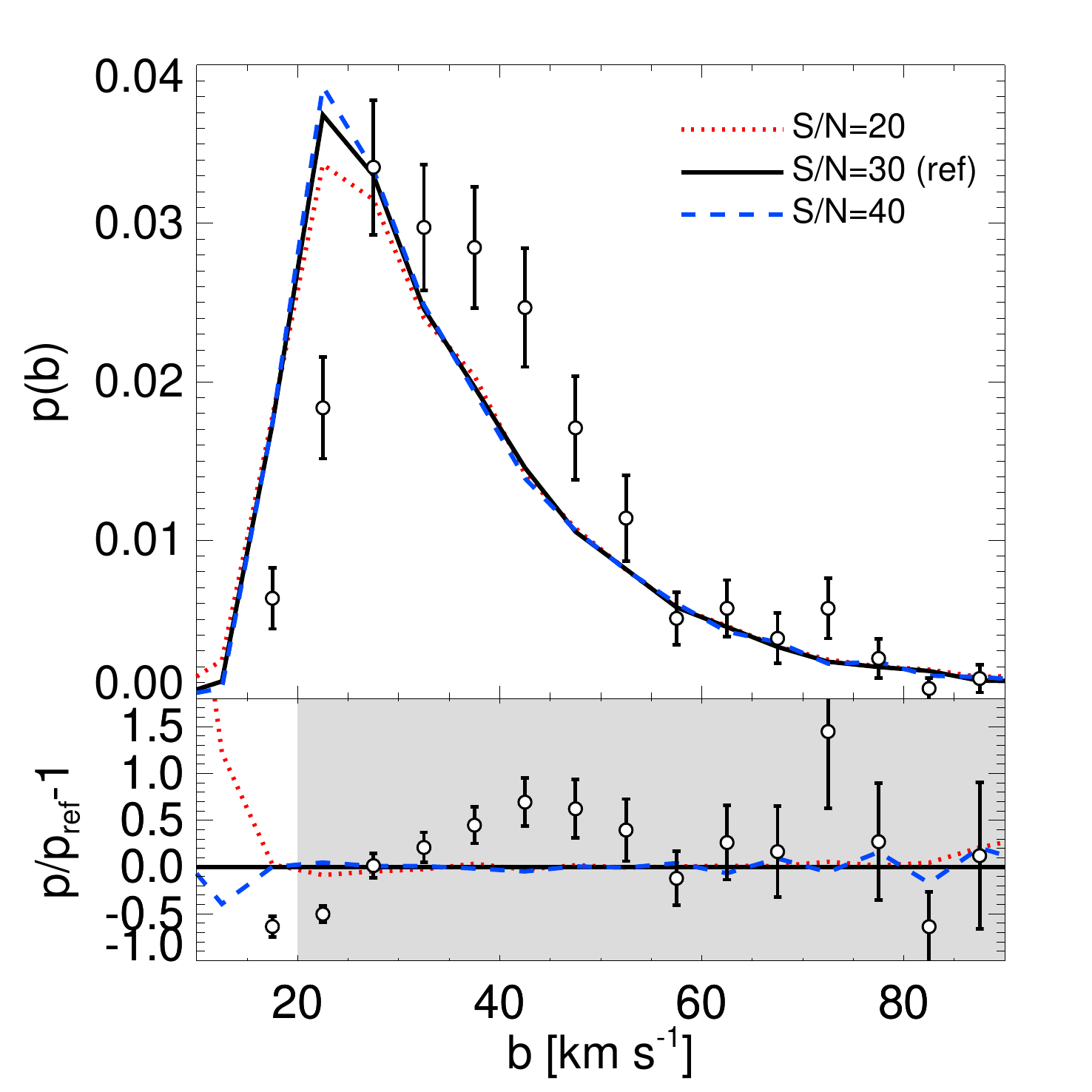}
  \end{minipage}
  \vspace{-0.3cm}
  \caption{The effect of the signal-to-noise per resolution element.
    {\it Left:} The CDDF measured from COS data for \Lya absorbers
    with Doppler parameters $20\rm km\,s^{-1} \leq b \leq 90 \rm
    km\,s^{-1}$, compared to the CDDF obtained from the H0 simulation
    for mock spectra with a flux independent signal-to-noise per
    resolution element of $\rm S/N=20$ (red dotted), $\rm S/N=30$
    (black solid) and $\rm S/N=40$ (blue dashed).  {\it Right:} The
    corresponding Doppler parameter probability distribution for \Lya
    absorbers with column densities $10^{13.3}\rm\, cm^{-2} \leq
    N_{\rm HI}\leq 10^{14.5}\rm\, cm^{-2}$.}
  \label{fig:SNtest}  
\end{figure*}

\begin{figure*}
  \begin{minipage}{1.00\textwidth}
    \includegraphics[width=0.49\textwidth]{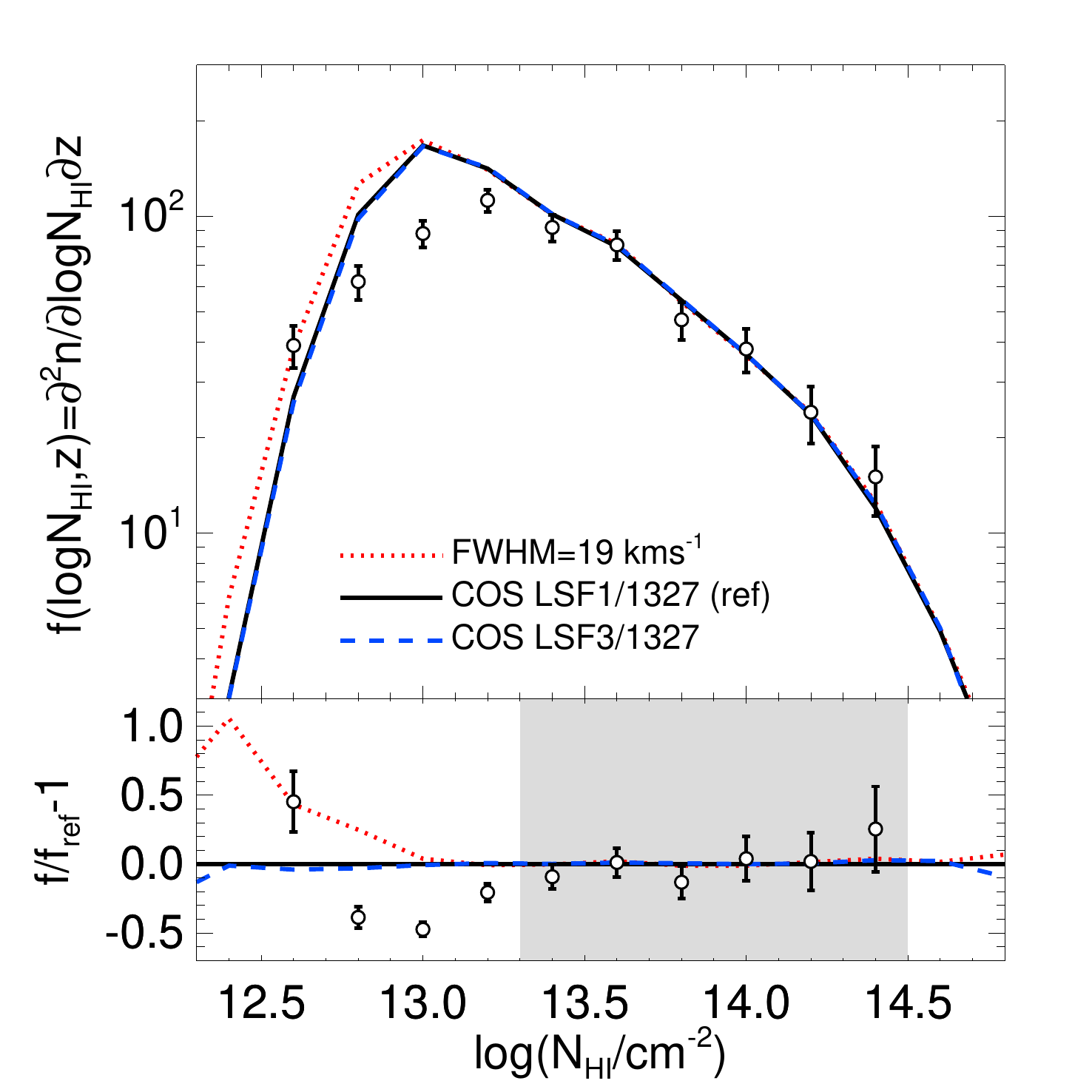}
    \includegraphics[width=0.49\textwidth]{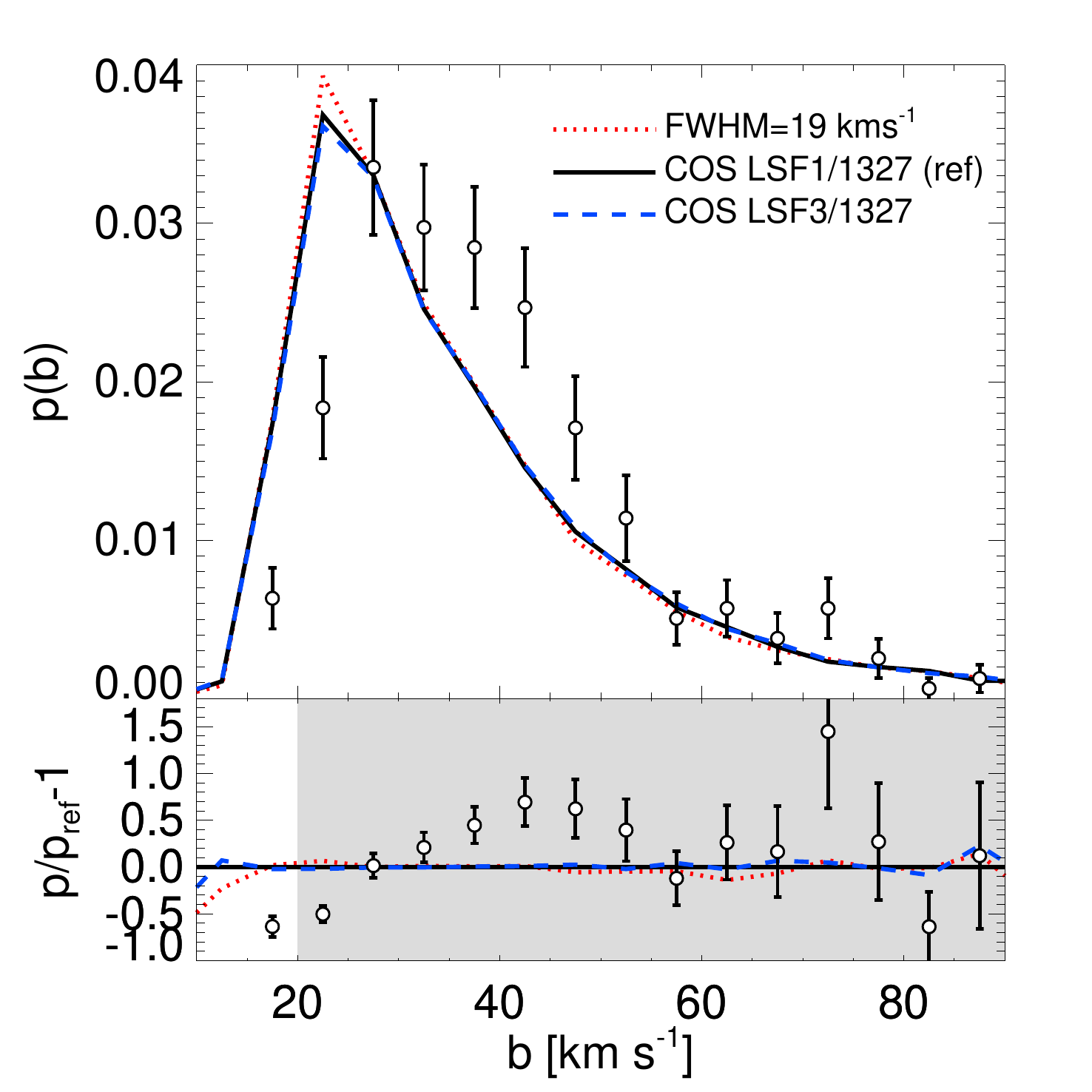}
  \end{minipage}
  \vspace{-0.3cm}
  \caption{The effect of the COS line spread function.  {\it Left:}
    The CDDF measured from COS data for \Lya absorbers with Doppler
    parameters $20\rm km\,s^{-1} \leq b \leq 90\rm km\,s^{-1}$,
    compared to the CDDF obtained from the H02 simulation after
    convolving the mock spectra with a Gaussian line profile with
    FWHM$=19\rm\,km\,s^{-1}$ (red dotted), the COS LSF for central
    wavelength G130M/1327 at LP1 (black solid) and at LP3 (blue
    dashed).  {\it Right:} The corresponding Doppler parameter
    probability distribution for \Lya absorbers with column densities
    $10^{13.3}\rm\, cm^{-2} \leq N_{\rm HI} \leq 10^{14.5}\rm\,
    cm^{-2}$. The figure is otherwise the same as
    Figure~\ref{fig:phystest}.}
  \label{fig:LSFtest}  
\end{figure*}

\begin{figure*}
  \begin{minipage}{1.00\textwidth}
    \includegraphics[width=0.49\textwidth]{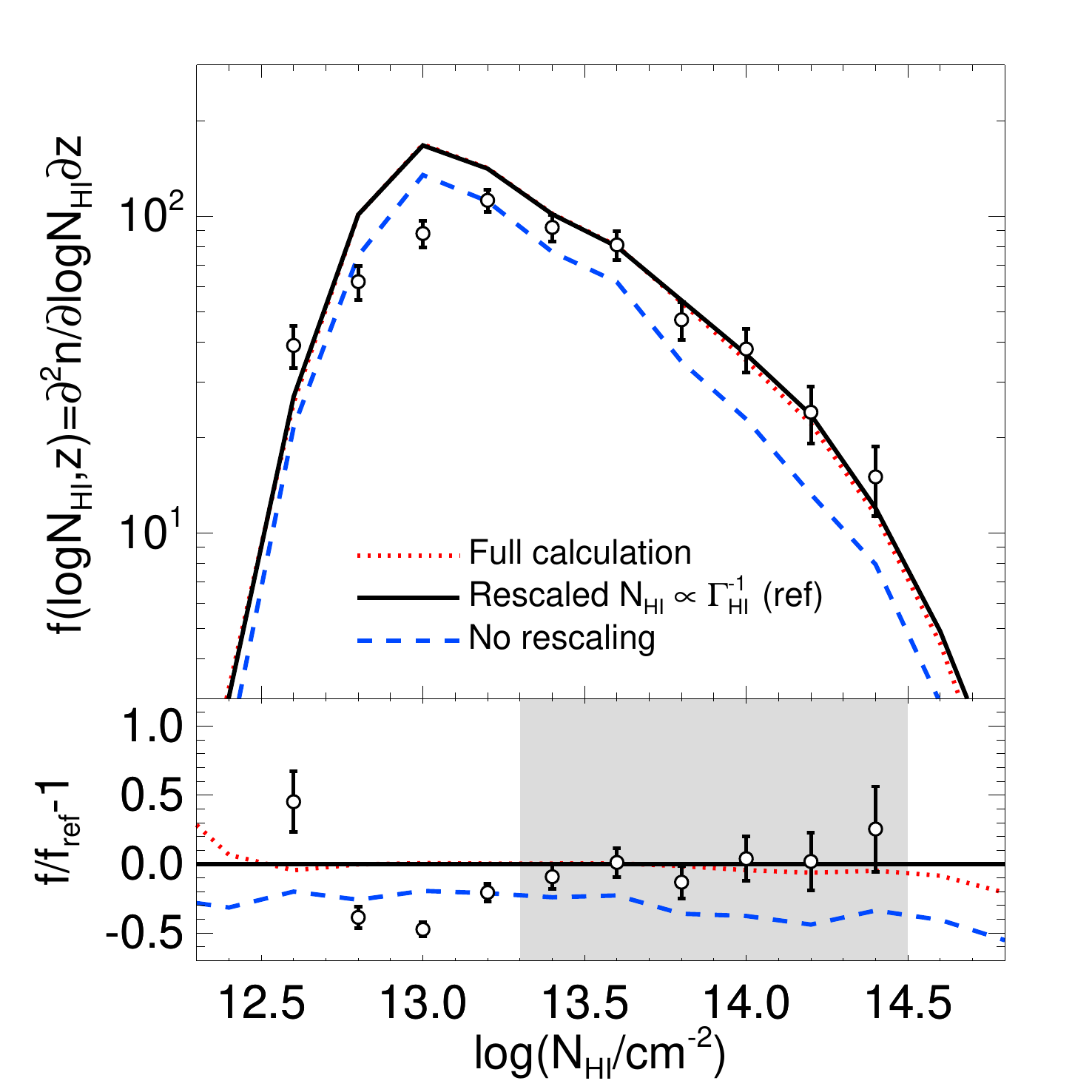}
    \includegraphics[width=0.49\textwidth]{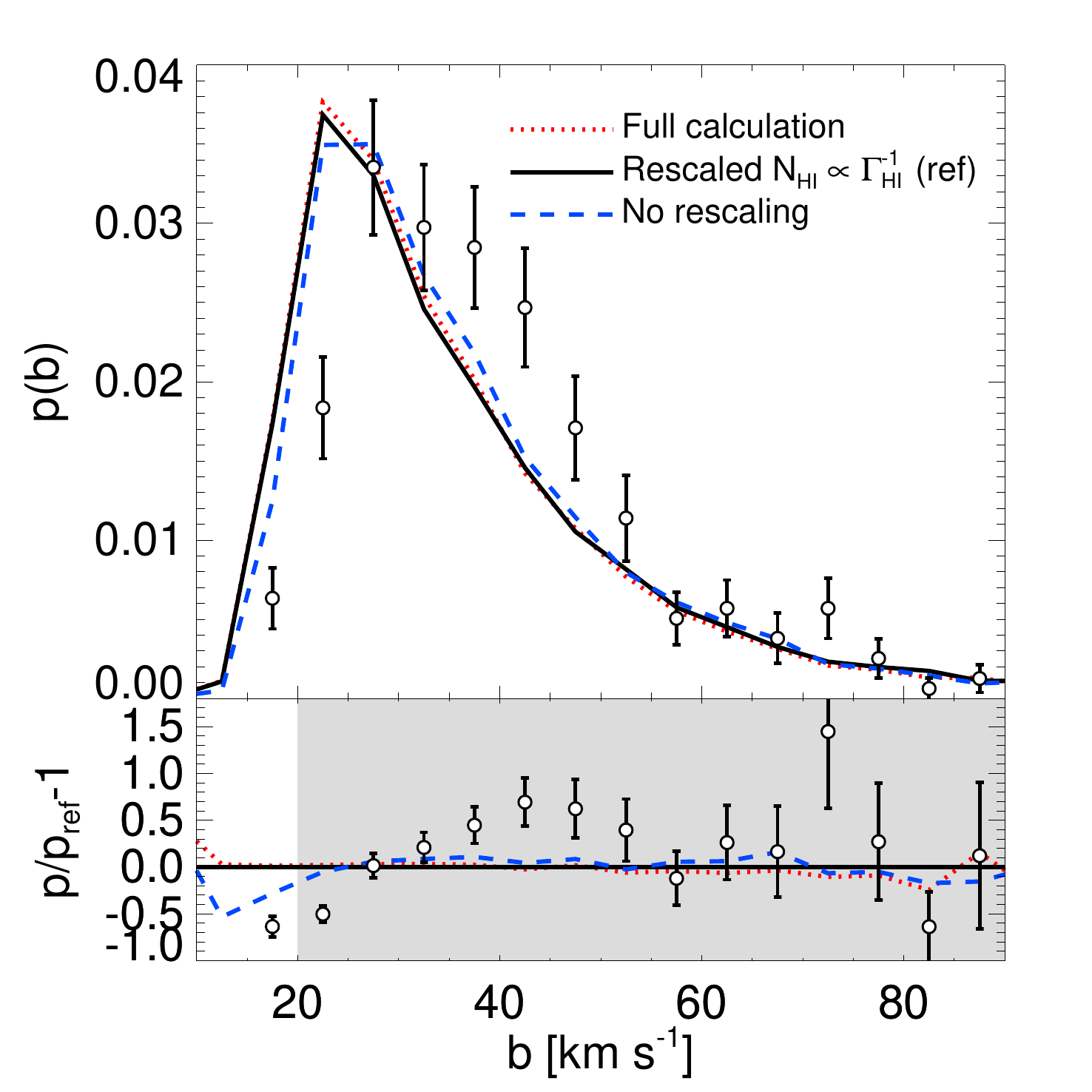}
  \end{minipage}
  \vspace{-0.3cm}
  \caption{The effect of rescaling the \Lya optical depths under the
    assumption of photoionisation equilibrium.  {\it Left:} The CDDF
    measured from COS data for \Lya absorbers with Doppler parameters
    $20\rm\,km\,s^{-1}\leq b \leq 90\rm\,km\,s^{-1}$, compared to the
    CDDF obtained from the H02 simulation for no rescaling (i.e. using
    the native UV background amplitude, red dotted), a linear
    rescaling of the pixel optical depths (or equivalently, column
    densities) in post-processing (black solid) and a full
    recalculation of the ionisation balance including the effect of
    collisional ionisation and free electrons from ionised helium (red
    dotted). {\it Right:} The corresponding Doppler parameter
    probability distribution for \Lya absorbers with column densities
    $10^{13.3}\rm\,cm^{-2}\leq N_{\rm HI} \leq
    10^{14.5}\rm\,cm^{-2}$. The figure is otherwise the same as
    Figure~\ref{fig:phystest}.}
  \label{fig:ionbaltest}  
\end{figure*}

The effect of the assumed signal-to-noise ratio and line spread
function (LSF) on our Voigt profile fits to simulated spectra are
shown in Fig.~\ref{fig:SNtest} and Fig.~\ref{fig:LSFtest},
respectively.

Our fiducial, flux independent signal-to-noise ratio of $\rm S/N=30$
per $19\rm\, km\,s^{-1}$ resolution element (black curves) is compared
to $\rm S/N=20$ (red dotted curves) and $\rm S/N=40$ (blue dashed
curves) in Fig.~\ref{fig:SNtest}.  The signal-to-noise ratio affects
the identification of narrow ($b<10\rm\,km\,s^{-1}$) absorbers, and
impacts on the completeness of low column density ($N_{\rm
  HI}<10^{13}\rm\,cm^{-2}$) absorption lines.  We have also tested
more complicated noise models that use a combination of flux
independent and flux dependent terms (not shown), but we find very
little difference between these and a flux independent noise model for
lines with $10^{13.3}\rm\,cm^{-2}\leq N_{\rm HI} \leq
10^{14.5}\rm\,cm^{-2}$ and $20\rm \,km\,s^{-1} \leq b \leq 90\rm\,
km\,s^{-1}$.

In Fig.~\ref{fig:LSFtest}, we show the effect of deconvolving the mock
spectra with the COS LSF (G130M/1327 LP1, black
curves)\footnote{\url{https://www.stsci.edu/hst/instrumentation/cos/performance/spectral-resolution}},
as well as the COS LSF at a different lifetime position, G130M/1327
LP3 (blue dashed curves) and a Gaussian with FWHM$=19\rm\,km\,s^{-1}$.
As was the case for the signal-to-noise, the narrow, low column
density lines are the most strongly affected if using an incorrect
model for the LSF.  This is most apparent for the Gaussian LSF, which
lacks the extended wings that are present in the COS LSF.

Based on the results shown in Fig.~\ref{fig:SNtest} and
Fig.~\ref{fig:LSFtest}, we judge that \Lya forest absorption lines
with $10^{13.3}\rm\,cm^{-2}\leq N_{\rm HI} \leq 10^{14.5}\rm\,cm^{-2}$
and $20\rm \,km\,s^{-1} \leq b \leq 90\rm\, km\,s^{-1}$ should be the
least affected by variations in the assumed signal-to-noise or
spectral resolution of the COS data.

Finally, as discussed in Section~\ref{sec:obs}, the \HI column
densities in our simulated spectra are rescaled by a constant to match
the amplitude of the observed CDDF at $10^{13.3}\rm\,cm^{-2}\leq
N_{\rm HI} \leq 10^{14.5}\rm\,cm^{-2}$ \citep[following][]{Viel2017}.
This is equivalent to rescaling the \HI photoionisation rate,
$\Gamma_{\rm HI}$, since $N_{\rm HI}\propto \Gamma_{\rm HI}^{-1}$ for
optically thin gas in photoionisation equilibrium with the UV
background.  However, as noted by \cite{Khaire2019pk}, it is possible
this assumption may break down if the gas responsible for the \Lya
forest absorption at $z\simeq 0.1$ is hot ($T>10^{5}\rm\,K$) and
collisionally ionised.  We test this explicitly in
Fig.~\ref{fig:ionbaltest} using three cases: no rescaling of the
column densities (blue dashed curves), the post-processed linear
scaling of the column densities that we use throughout this work
(black curves), and a full recalculation of the IGM ionisation balance
using a photoionisation rate that is scaled by the same factor used
in the post-processed case (red dotted curves).  For the column
density and Doppler parameter range we consider in this work, the
agreement between the approximate (black curves) and full calculation
(red dotted curves) is excellent, justifying our assumption.

\section{Comparison to Illustris-TNG} \label{app:TNG}

\begin{figure*}
  \begin{minipage}{1.00\textwidth}
    \includegraphics[width=0.49\textwidth]{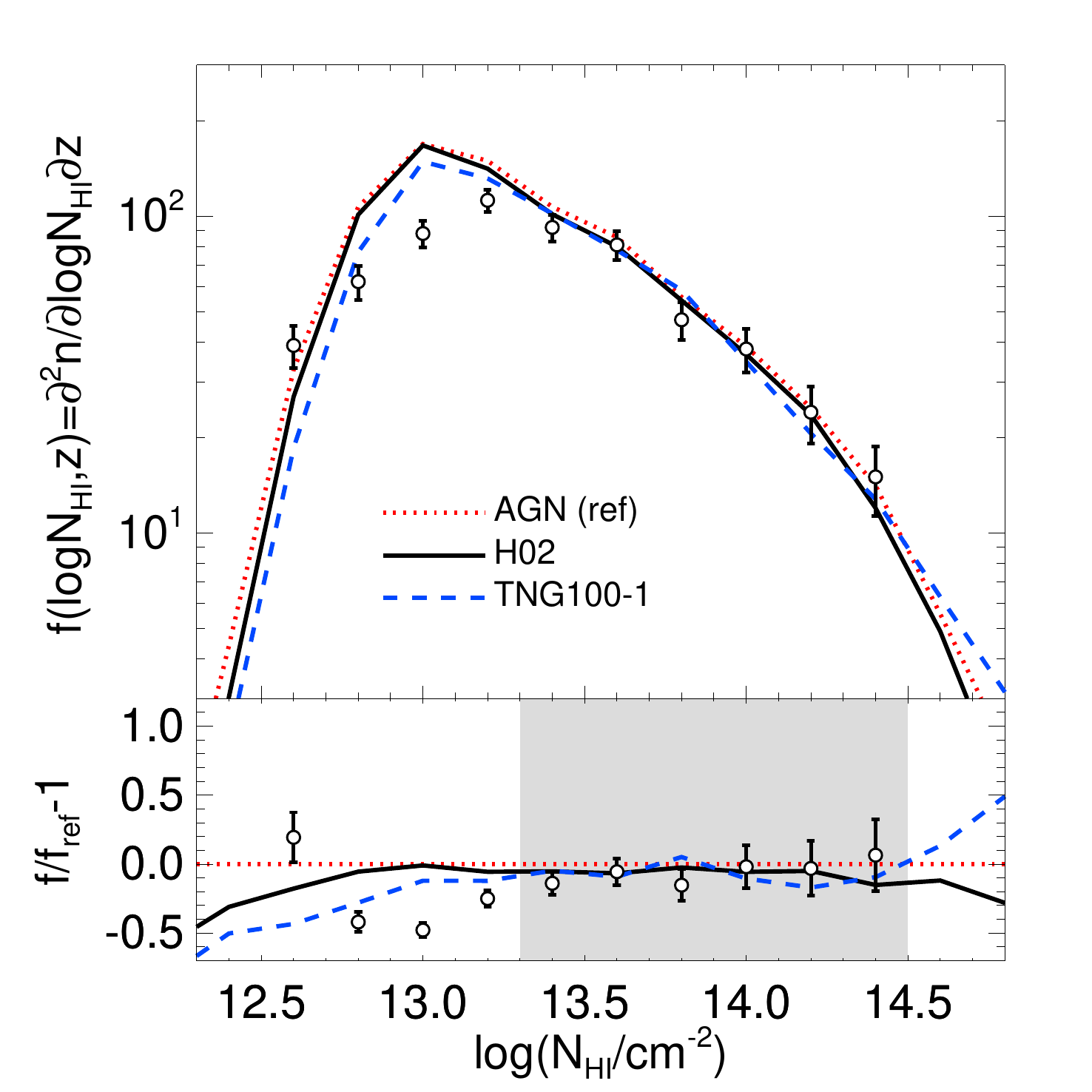}
    \includegraphics[width=0.49\textwidth]{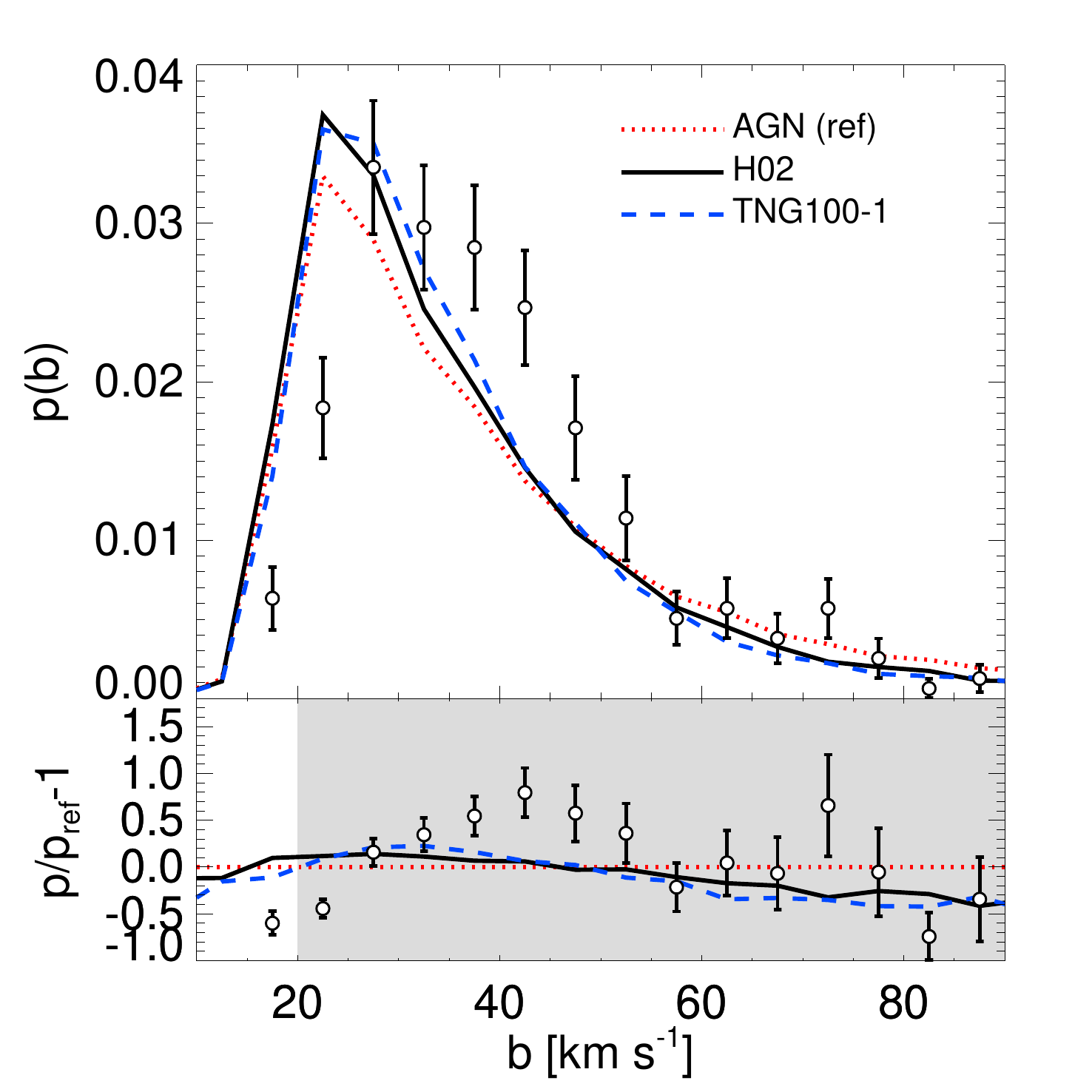}
  \end{minipage}
  \vspace{-0.3cm}
  \caption{Comparison of the TNG100-1 simulation to the CDDF and
    Doppler parameter distribution measured from COS data.  {\it
      Left:} The CDDF for \Lya absorbers with Doppler parameters
    $20\rm\,km\,s^{-1}\leq b \leq 90\rm\,km\,s^{-1}$, compared to the
    TNG100-1 simulation (blue dashed curve), AGN simulation (red
    dotted curve) and H02 simulation (solid black curve). Note that we
    do not apply a correction for mass resolution and box size to the
    CDDF from TNG100-1.  {\it Right:} The corresponding Doppler
    parameter probability distribution for \Lya absorbers with column
    densities $10^{13.3}\rm\,cm^{-2}\leq N_{\rm HI} \leq
    10^{14.5}\rm\,cm^{-2}$. The figure is otherwise the same as
    Figure~\ref{fig:phystest}.}
  \label{fig:TNG}  
\end{figure*}

In Figure~\ref{fig:TNG}, we compare the COS CDDF and Doppler parameter
distribution to Voigt profile fits obtained from \Lya forest spectra
extracted from the publicly available Illustris TNG100-1 simulation at
$z=0.1$
\citep{Marinacci2018,Naiman2018,Nelson2018,Springel2018,Pillepich2018,Nelson2019}.
This provides a test of whether or not the results of this study are
peculiar to our numerical implementation.

In contrast to the simulations used in this work, IllustrisTNG
includes metal line cooling and magneto-hydrodynamics, as well as
different implementations for AGN feedback, galactic winds, metal
enrichment and the UV background.  IllustrisTNG furthermore uses the
\textsc{AREPO} code \citep{Springel2010}, which employs a moving-mesh
hydrodynamics scheme instead of smoothed particle hydrodynamics.  The
cosmological parameters for the TNG100-1 simulation are very similar
to those used in this work, with $\Omega_{\rm m}=0.3089$,
$\Omega_{\Lambda}=0.6911$, $h=0.6774$, $\Omega_{\rm b}=0.0486$,
$\sigma_{8}=0.8159$ and $n=0.9667$.  The box size of TNG100-1 is
$75h^{-1}\rm cMpc$, with a gas particle mass of $M_{\rm gas}=9.4\times
10^{5}h^{-1}M_{\odot}$ (i.e. a factor $\sim 2$ larger volume and a
factor $\sim 7$ smaller gas particle mass compared to our fiducial AGN
simulation).  Photoionisation and photoheating is provided by the
$2011$ update of the \citet{FaucherGiguere2009} UV background
model.\footnote{\url{https://galaxies.northwestern.edu/uvb-fg09/}} At
$z=0.1$, this has $\log(\Gamma_{\rm HI}^{\rm FG09}/\rm
s^{-1})=-13.31$.

Adopting $\log(\Gamma_{\rm HI}^{\rm CDDF}/\rm s^{-1})=-13.08$ in the
TNG100-1 \Lya forest spectra provides a good match (within
$1$--$2\sigma$) to the amplitude and shape of the CDDF (this
corresponds to $\Gamma_{\rm HI}^{\rm CDDF}/\Gamma_{\rm HI}^{\rm
  HM12}=2.37$).  However, the line widths remain systematically
narrower than the COS data, with the number of lines at $b=22.5\pm
2.5\rm\,km\,s^{-1}$ over-predicted by $\sim 5.5\sigma$.  This is
consistent with the simulations used in this work.

\bsp	
\label{lastpage}
\end{document}